\documentclass[twocolumn, aps, pra, longbibliography,  
 unsortedaddress]{revtex4-1}
\usepackage[english]{babel}
\usepackage[dvipsnames]{xcolor}
\usepackage{tikz}
\usepackage{tikz-cd}
\usepackage{graphicx}
\usepackage{dcolumn}
\usepackage{bm}
\usepackage{amsmath}
\usepackage{amssymb}
\usepackage{dsfont}
\usepackage{lipsum}  
\usepackage{color}
\usepackage{multirow}
\usepackage{stmaryrd}
\usepackage{upgreek}
\usepackage{bbm}
\usepackage{xcolor}
\usepackage{ragged2e}
\usepackage{siunitx}
\usepackage[colorlinks=true,urlcolor=blue,citecolor=blue,linkcolor=blue]{hyperref}
\usepackage{braket}
\usepackage{mathtools}
\usepackage[mathlines]{lineno}
\usepackage[normalem]{ulem}
\usepackage{scalerel}
\usepackage{ bbold }
\usepackage{inputenc}
\usepackage[T1]{fontenc}
\usepackage{tikz}
\usepackage{cleveref}
\usepackage{amsfonts}
\usepackage{natbib}
\usepackage{booktabs}
\usepackage{algorithm} 
\usepackage{algpseudocode} 
\usepackage{blkarray}

\newcommand{\co}[1]{\texttt{#1}}
\newcommand\headercell[1]{
   \smash[b]{\begin{tabular}[t]{@{}c@{}} #1 \end{tabular}}}
\DeclareMathOperator{\Tr}{Tr}

\begin{document}

\preprint{APS/123-QED}

\title{Ancilla-free implementation of generalized measurements \\ for qubits embedded in a qudit space}

\author{Laurin E. Fischer} \email{aur@zurich.ibm.com}
\author{Daniel Miller}
\author{Francesco Tacchino}
\author{\\Panagiotis Kl. Barkoutsos}
\author{Daniel J. Egger}
\email{deg@zurich.ibm.com}
\author{Ivano Tavernelli}
\email{ita@zurich.ibm.com}
\affiliation{IBM Quantum, IBM Research Europe – Zurich, S\"aumerstrasse 4, CH-8803 R\"uschlikon, Switzerland}

\begin{abstract}
Informationally complete (IC) positive operator-valued measures (POVMs) are generalized quantum measurements that offer advantages over the standard computational basis readout of qubits. 
For instance, IC-POVMs enable efficient extraction of operator expectation values, a crucial step in many quantum algorithms. POVM measurements are typically implemented by coupling one additional ancilla qubit to each logical qubit, thus imposing high demands on the device size and connectivity. 
Here, we show how to implement a general class of IC-POVMs without ancilla qubits. 
We exploit the higher-dimensional Hilbert space of a qudit in which qubits are often encoded.
POVMs can then be realized by coupling each qubit to two of the available qudit states, followed by a projective measurement.
We develop the required control pulse sequences and numerically establish their feasibility for superconducting transmon qubits through pulse-level simulations. 
Finally, we present an experimental demonstration of a qudit-space POVM measurement on IBM Quantum hardware. 
This paves the way to making POVM measurements broadly available to quantum computing applications.
\end{abstract}

\maketitle

\section{Introduction}
\label{chap:introduction}
Steady progress in the field of quantum technology, attested by continuing improvements in both quantum algorithms~\cite{grinko2021iterative, egger2021warmstarting, sokolov2022orders} and hardware performance~\cite{place2021new, jurcevic2021demonstration}, suggests that quantum computers may soon provide significant advantages over their classical counterparts in fields such as optimization, machine learning, finance, quantum physics and chemistry.
In particular, \emph{ab initio} computational studies of molecular systems and materials represent natural areas of application for quantum computers~\cite{bauer2020quantum, mcardle2020quantum, motta2021emerging, ollitrault2020quantum, sokolov2020quantum, ollitrault2020hardware}. 
These prospects have also attracted interest from the material and drug design industries~\cite{kuhn2019accuracy, robert2021resourceefficient}. 

Proof-of-principle experiments for small molecular systems have been successfully demonstrated on various quantum computing platforms~\cite{kandala2017hardwareefficient, hempel2018quantum, lanyon2010quantum}.
Crucially, these applications should be extended to problem sizes of practical interest to reach the scale at which quantum advantage can be indisputably claimed.
On current noisy hardware without error correction, the realizable circuit depths are limited by finite gate fidelities and qubit coherence times.
Variational algorithms address these issues by leveraging classical resources in combination with, e.g., adaptive quantum protocols and effective sampling from parametrized quantum states~\cite{mcclean2016theory, cerezo2021variational}. 
For example, the \emph{variational quantum eigensolver} (VQE) can be used, among other applications, to obtain the ground state energy of molecules~\cite{peruzzo2014variational}.
This is achieved by measuring the expectation value of the Hamiltonian for a trial state prepared with a parameterized ansatz circuit. 
By updating the parameters with a classical optimizer, the energy is minimized to approach the true ground state, in the spirit of the variational principle.
A sufficiently good accuracy is only reached if the ansatz circuit is expressive enough to closely approximate the actual ground state. 
Moreover, the convergence of the classical optimizer can be obstructed by vanishing gradients and local minima, particularly under the influence of hardware noise~\cite{wang2021noiseinduced}.
Overcoming these issues~\cite{sokolov2020quantum,Eddins2022, holmes2022connecting,wang2021error} still leaves the large number of measurement shots needed to estimate the target observables 
as a major bottleneck of VQE~\cite{gonthier2020identifying}.
This is commonly referred to as the \emph{measurement problem}.
For example, the small-scale molecular calculations of $\text{H}_2$, LiH, and $\text{BeH}_2$ reported in Ref.~\cite{kandala2017hardwareefficient} required measuring $\mathcal{O}(10^9)$ quantum circuits. 
On larger problem instances, these requirements can grow unsustainably large, e.g., an estimate for the $\text{Fe}_2\text{S}_2$ complex predicts up to $\mathcal{O}(10^{13})$ required measurements per energy evaluation~\cite{Wecker2015}.
Even with the high sampling rate of superconducting quantum processors of up to 100 kHz, this task would take decades to complete.
Circuit execution speed~\cite{wack2021quality} and measurement number reduction are therefore crucial to variational algorithms.

Known strategies to alleviate the measurement problem include Pauli groupings~\cite{kandala2017hardwareefficient,  gokhale_ON3_Measurement_2020,  verteletskyi2020measurement, hamamura2020efficient, crawford2021efficient, miller2022hardwaretailored}, 
classical shadows~\cite{huang2020predicting,hadfield2020measurements,Zhao2021}, and 
machine learning~\cite{torlai2020precise}.
Recent work suggests that informationally complete positive operator-valued measures (IC-POVMs) can also efficiently estimate quantum states and observables, for example, they achieve a near optimal scaling in the number of measurements for the reconstruction of fermionic reduced density matrices~\cite{jiang2020optimal,Bonet2020}.
In the context of observable expectation value sampling, adapting the POVM to the target observable reduces the measurement overhead by one order of magnitude compared to a standard Pauli grouping in hydrogen chains with 14 qubits~\cite{garcia2021learning}. 
However, the experimental realization of IC-POVMs requires coupling each qubit representing the trial state to two additional quantum states~\cite{chen2007ancilla}. 
Traditionally, this is done by coupling each qubit to an ancillary one before readout~\cite{garcia2021learning, brida2012ancillaassisted}.
This approach doubles the number of necessary qubits during the measurement stage, and therefore halves the usable portion of a quantum chip. 
Moreover, the limited connectivity of most quantum architectures leads to a significant \textsc{Swap}-gate overhead~\cite{weidenfeller2022scaling}. 

In this work, we conceptualize and implement a measurement scheme for IC-POVMs, which does not require ancilla qubits. 
Many quantum computing architectures encode qubits in two levels of a larger Hilbert space, e.g., the energetically lowest states of a transmon or two long-lived states of an atom or ion~\cite{peterer2015coherence, low2020practical, shi2021quantum}.
We use two additional states in this surrounding qudit space to realize programmable single-qubit POVM measurements. 
This requires the ability to distinguish four qudit states through projective measurements and a short pulse sequence coupling to the qudit states at the very end of a quantum circuit. 
As a result, the coherence and gate fidelity requirements of these additional states are much less stringent than for the qubit states. 

Our paper is organized as follows. 
In Sec.~\ref{chap:theory}, we propose a practical implementation of POVM measurements for qubits embedded in a qudit space. 
In Sec.~\ref{chap:SC_qubits}, we demonstrate an experimental implementation of our scheme on a superconducting qubit in IBM Quantum hardware. 
Finally, in Sec.~\ref{chap:applications}, we show how qudit-based POVMs implemented in superconducting transmon hardware can sample operators with low variance through pulse-level numerical simulations.

\section{Theory}
\label{chap:theory}
The POVM formalism describes general measurements of a state $\rho_{\text{S}}$ on a system Hilbert space $\mathcal{H}_{\text{S}}$.
Formally, an $M$-outcome POVM is a set of $M$ positive semi-definite Hermitian operators $ \Pi^0, \dots, \Pi^{M-1}$ acting on $\mathcal{H}_{\text{S}}$ which satisfy the completeness relation $\sum_{m=0}^{M-1} \Pi^m = \mathbbm{1}$, where $\mathbbm{1}$ is the identity.
Each operator $\Pi^m$ represents one possible outcome of the measurement that occurs with a probability
\begin{align}
\label{eq:outcome_probability_POVM}
p_m = \text{Tr}(\rho_{\text{S}} \Pi^m).
\end{align}
Standard projective measurements of an orthonormal basis of pure states $\ket{\psi_m}$ form a special case of POVM measurements, where ${\Pi^m = \ket{\psi_m}\!\bra{\psi_m}}$.
A POVM measurement is \emph{informationally complete} (IC) if every Hermitian operator $\mathcal{O}$ can be written as
\begin{align}
\label{eq:O_decomposed_POVM}
\mathcal{O} & = \sum\limits_{m} c_{m} \Pi^{m}, \quad c_m \in \mathbb{R}.
\end{align}
In this case, the probability distribution $\{p_m\}$ in Eq.~\eqref{eq:outcome_probability_POVM} contains the full information about the state $\rho_\text{S}$. 
In particular, $\{p_m\}$ suffices to compute the expectation value of $\mathcal{O}$ as
\begin{align}
\label{eq:O_expectation_value_POVM}
\langle \mathcal{O} \rangle &= \text{Tr}(\rho_\text{S} \mathcal{O}) =  \sum\limits_{m} c_{m} p_m .
\end{align}
This expectation value can thus be estimated from $\mathcal{N}$ samples drawn from the POVM outcome distribution as $\widehat{\langle \mathcal{O} \rangle} = \sum_m c_m \mathcal{N}_m/\mathcal{N}$, where $\mathcal{N}_m$ denotes the number of times outcome $m$ was observed. 
The error $\epsilon$ on this estimator is the standard error of the mean
\begin{align}
\label{eq:O_error_POVM}
\epsilon^2\Big( \widehat{\langle \mathcal{O} \rangle}\Big) =
\text{Var}(\mathcal{O})/\mathcal{N} =  \Big(\sum\limits_m c_m^2 p_m  - \langle \mathcal{O} \rangle^2\Big)/\mathcal{N} .
\end{align}
Tailoring the POVM operators to the specific observable $\mathcal{O}$ and the state $\rho_\text{S}$ considerably reduces the corresponding variance $\text{Var}(\mathcal{O})$~\cite{garcia2021learning}. 

General POVMs on $\mathcal{H}_{\text{S}}$ can be implemented by coupling to an extended space $\mathcal{H}_{\text{ext}}$ either through a tensor product extension (TPE) $\mathcal{H}_{\text{ext}} = \mathcal{H}_{\text{S}} \otimes \mathcal{H}_{\text{A}}$ or a direct sum extension (DSE) $\mathcal{H}_{\text{ext}} = \mathcal{H}_{\text{S}} \oplus \mathcal{H}_{\text{A}}$ \cite{chen2007ancilla}.
To realize POVM measurements, a specific unitary $U$ is applied to $\mathcal{H}_{\text{ext}}$ such that the probability distribution of a subsequent $M$-outcome projective measurement on $\mathcal{H}_{\text{ext}}$ coincides with the POVM outcome distribution $\{p_m\}$ for the original state $\rho_\text{S}$. 
Before applying $U$, the initial state on $\mathcal{H}_{\text{ext}}$ is of the form $\rho = \rho_{\text{S}} \otimes \rho_{\text{A}}$ in a TPE
while in a DSE it has no support on $\mathcal{H}_{\text{A}}$.
In both cases, the existence of $U$ is guaranteed by Naimark's dilation theorem~\cite{gelfand1943imbedding}.

We consider IC-POVM measurements on $N$-qubit systems, specifically product POVMs where each global operator $\Pi^m$ is given by a tensor product of local single-qubit operators of rank one. 
Each such local POVM includes $M=4$ linearly independent operators~\cite{chen2007ancilla}. 
The global POVM then consists of $4^N$ product operators, the minimal number required for informational completeness.
Such POVMs are typically implemented in a TPE by coupling each of the $N$ qubits to an ancilla qubit. 
The single-qubit POVM operators then define a two-qubit unitary $U$ acting on the system and ancilla qubit.
This can be accomplished with three \textsc{Cnot}-gates and single-qubit gates through the KAK decomposition~\cite{vatan2004optimal, drury2008constructive}, which can also be improved by scaling pulses~\cite{earnest2021pulseefficient}.
The relation between $U$ and the POVM operators $\Pi^m$ is detailed in App.~\ref{app_sec:POVM_ancilla_implementation}.

\begin{figure}
\includegraphics[width=0.99\columnwidth]{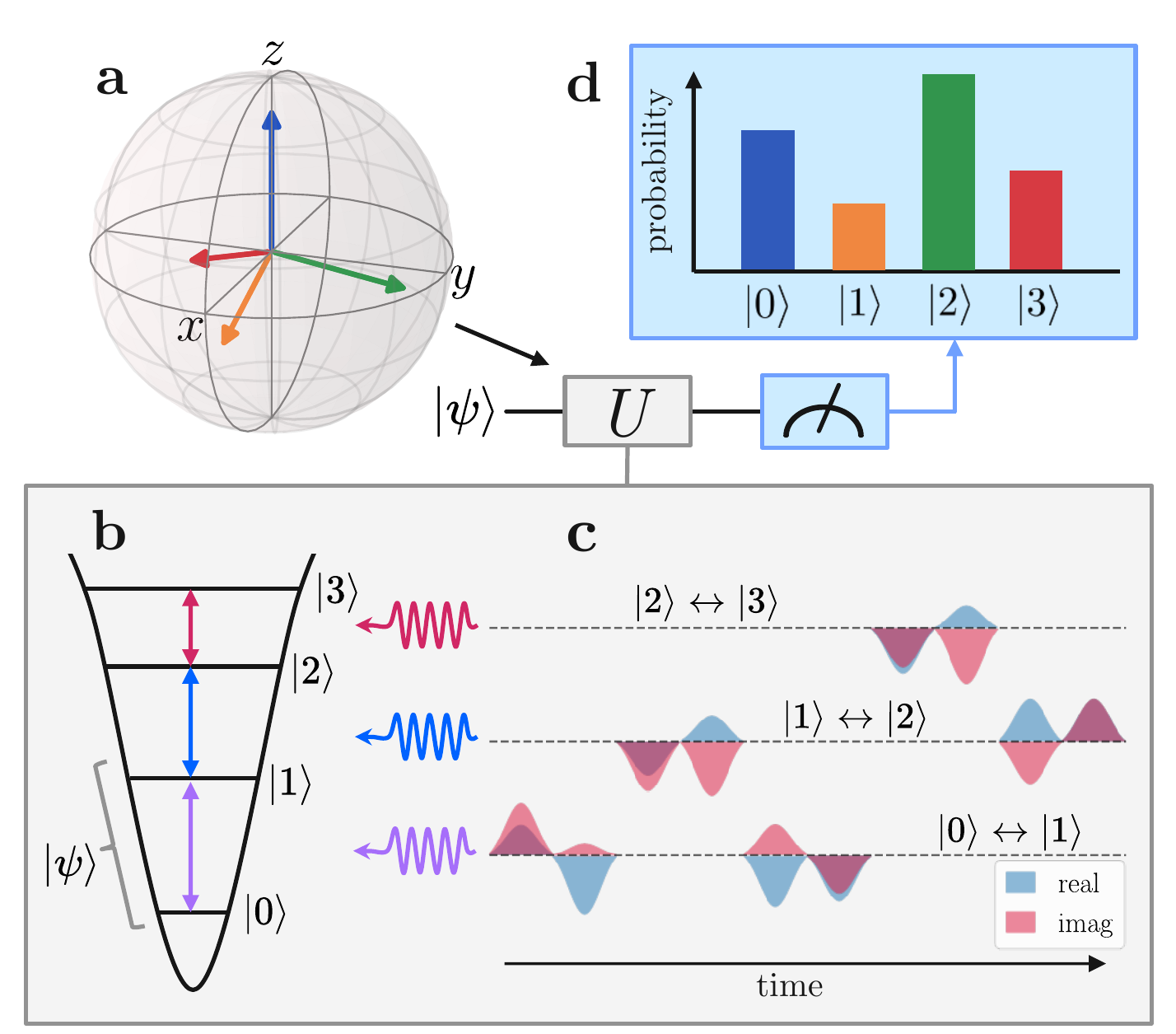}
\caption{Schematic of a POVM implementation in qudit space. 
\textbf{a)} The $M=4$ rank-one, single-qubit POVM operators, represented on a Bloch sphere, define a four-dimensional unitary $U$ which encodes the POVM operators. 
\textbf{b)} We realize this unitary on the qudit space in which the qubit state $\ket{\psi}$ is encoded.
\textbf{c)} This can be achieved by a sequence of ten $\pi/2$-pulses 
that couple adjacent levels.
\textbf{d)} Finally, a projective measurement of the four states yields the outcome probabilities  of the four POVM operators.
}
\label{fig:overview_schematic}
\end{figure}

The overhead of ancilla-based POVM implementations in a TPE, which doubles the qubit count, can be avoided if the qubit states $\ket{0}$ and $\ket{1}$ are encoded in the higher-dimensional Hilbert space of a qudit. 
Instead of an ancilla, we use two additional states of the qudit space, denoted $\ket{2}$ and $\ket{3}$, which are not populated during the quantum circuit to realize a single-qubit POVM through a DSE, see Fig.~\ref{fig:overview_schematic}.
The states $\ket{2}$ and $\ket{3}$ may be higher-excited states of a superconducting transmon qubit~\cite{peterer2015coherence} or additional states of the level structure in trapped ions~\cite{low2020practical} and neutral atoms~\cite{shi2021quantum}. 
We implement the POVM-encoding unitary $U$ on the qudit space through a sequence of pulses that couple adjacent levels.
This approach is suitable to architectures where an external drive with a dipole coupling is available, e.g., through microwave or laser pulses.

We now review the action of individual pulses and then decompose $U$ into rotations generated by such pulses.
Let $H_0^{\text{lf}} = \sum_{n=0}^{d-1} E_n \ket{n}\bra{n}$ denote the qudit Hamiltonian in its eigenbasis in the laboratory frame (lf).
An external drive 
\begin{align}
\label{eq_theo:pulse_drive_definition}
\epsilon(t) = \Omega(t)\cos\left(\omega_D t - \phi \right)
\end{align}
with envelope $\Omega(t)$, drive frequency $\omega_D$, and phase $\phi$ leads to an interaction Hamiltonian 
\begin{align}
\label{eq_theo:interaction_Ham_lab_frame}
  H_{\text{int}}^{\text{lf}}(t) = \epsilon(t)\sum_{n=0}^{d-2} g_n \ket{n}\bra{n+1} + \text{h.c.} \,.
\end{align}
Here, $g_n$ denotes the coupling strength to the $n \!\leftrightarrow\! n+1 $ transition and we set $\hbar=1$. 
By transforming into the rotating frame (rf) of the drive, and applying the rotating wave approximation (dropping terms rotating at $2\omega_D$), these Hamiltonians become
\begin{align}
    H_0^{\text{rf}} &= \sum_{n=0}^{d-1} \left(E_{n} - n \omega_D \right) \ket{n}\bra{n}  \label{eqn:theory_H0_rot_frame} \quad \text{and}\\
 H_{\text{int}}^{\text{rf}}(t) &= \frac{\Omega(t)}{2}\sum_{n=0}^{d-2} g_n  e^{i\phi}  \ket{n+1}\bra{n} + \text{h.c.} \, . \label{eqn:theory_Hint_rot_frame}
\end{align}
Setting $\omega_D= E_{n+1} - E_n$, i.e., on resonance with the $n \!\leftrightarrow\! n+1 $ transition, and evolving $H_0^{\text{rf}}+ H_\text{int}^{\text{rf}}(t)$ for a duration $T$ results in the qudit unitary
\begin{align}
\label{eq:givens_rotation_hardware}
\mathcal{R}_{n\leftrightarrow n+1}(\theta, \phi) &= \mathcal{G}_{n\leftrightarrow n+1}(\theta, \phi) \\
& \hspace{-4mm} \times
\text{diag}\left(e^{-i E_0 T}, \ldots, e^{-i (E_{d-1}- (d-1)\omega_D )T} \right).
\nonumber
\end{align}
Here, we assume that other transitions are far detuned.
The qudit operator $\mathcal{G}_{n\leftrightarrow n+1}(\theta, \phi)$ applies a \emph{Givens} rotation
\begin{align} 
\label{eq:theory_given_rotation_definition}
 G(\theta, \phi)
 &= \begin{pmatrix} 
    \cos(\theta/2) & -i \sin(\theta/2) e^{-i\phi}\\
     -i \sin(\theta/2) e^{i\phi}  & \cos(\theta/2)
    \end{pmatrix}
\end{align}
to the subspace spanned by $\ket{n}$ and $\ket{n+1}$ and
acts as the identity everywhere else. $G(\theta, \phi)$ is a rotation of angle $\theta \sim g_n \int_{0}^T \Omega(t) \text{d}t$ around an axis in the $xy$-plane with a polar angle given by the drive phase $\phi$.
The diagonal matrix in Eq.~\eqref{eq:givens_rotation_hardware} imprints phases on all non-resonant states.

We further define generalized $\mathcal{Z}_{n\leftrightarrow n+1}(\varphi)$-rotations, that act as $\text{diag}(e^{-i\frac{\varphi}{2}}, e^{i\frac{\varphi}{2}})$ on the states $\ket{n}$ and $\ket{n+1}$ and as the identity elsewhere.
Such generalized \mbox{$\mathcal{Z}$-gates} can be engineered from two Givens rotations~\cite{low2020practical}.
For qubits, it is common to implement $z$-rotations virtually by adjusting the phases $\phi$ of subsequent drive pulses~\cite{mckay2017efficient, murali2019full}. 
We generalize this concept to virtually implement qudit-space \mbox{$\mathcal{Z}$-gates}, as detailed in App.~\ref{app_sec:POVM_virtual_Z}.

We construct the POVM-encoding unitary $U$ 
from $\mathcal{R}$-rotations as in Eq.~\eqref{eq:givens_rotation_hardware}
by adapting an algorithm presented in Ref.~\cite{schirmer2002constructive} that decomposes $U$ (up to remaining phases on the diagonal) into a sequence of Givens rotations $\mathcal{G}_{n\leftrightarrow n+1}(\theta, \phi)$, following a strategy similar to a QR decomposition~\cite{golub1996matrix}. 
We extend this algorithm in two ways. 
First, we add $\mathcal{Z}$-gates to the sequence to fully decompose $U$ (including all relative phases) without increasing the number of pulses.
Second, we replace the inaccessible $\mathcal{G}$-rotations in the decomposition of $U$ with the realistic $\mathcal{R}$-rotations in Eq.~\eqref{eq:givens_rotation_hardware}, that include additional phases acquired by idle levels. 
We absorb these phases into the angles $\phi$ of the subsequent $\mathcal{R}$-pulses. 
The details of the decomposition algorithm of $U$ into $\mathcal{R}$-gates are given in App.~\ref{app:povm_decomposition}.
Here, we only quote our main result: The target unitary $U$ can always be realized as a sequence of five $\mathcal{R}$-rotations
\begin{align}
\label{eq:theo_final_sequence_U}
U &=
 \mathcal{R}_{1\leftrightarrow 2}(\theta_5, \phi_5) \mathcal{R}_{2\leftrightarrow 3}(\theta_4, \phi_4) \\
 &\hspace{10mm} \times \mathcal{R}_{0\leftrightarrow 1}(\theta_3, \phi_3) \mathcal{R}_{1\leftrightarrow 2}(\theta_2, \phi_2) \mathcal{R}_{0\leftrightarrow1 }(\theta_1, \phi_1).  \nonumber
\end{align}
The specific choice of the targeted POVM operators $\Pi^m$ enter through the angles
$\theta_i$ and $\phi_i$, while the order in which the transitions are driven is fixed and independent of the POVM. 

Finally, let $\sqrt{\mathcal{X}}_{n\leftrightarrow n+1}$ denote a $\pi/2$-pulse around the $x$-axis between the states $\ket{n}$ and $\ket{n+1}$. 
Any $\mathcal{R}$-rotation can be realized by two $\sqrt{\mathcal{X}}$-pulses and three virtual $\mathcal{Z}$-gates, see App.~\ref{app_sec:POVM_decomp_algorithm}.
This has the great practical benefit that only the three pulses $\sqrt{\mathcal{X}}_{0\leftrightarrow1}$, $\sqrt{\mathcal{X}}_{1\leftrightarrow2}$, and $\sqrt{\mathcal{X}}_{2\leftrightarrow3}$, rather than a parametrized family of pulses, require calibration.
It is thus helpful to decompose the pulse sequence in Eq.~\eqref{eq:theo_final_sequence_U} into $\sqrt{\mathcal{X}}$-gates, shifting all angular dependencies into near-perfect virtual $\mathcal{Z}$-gates. 
Common calibration techniques applicable to the qudit-space pulses are readily available~\cite{sheldon2016characterizing}.
The resulting pulse sequence for the implementation of $U$ requires a total of ten $\sqrt{\mathcal{X}}$-pulses, see Fig.~\ref{fig:overview_schematic}(c) for an example where each pulse is depicted with a Gaussian envelope.

\section{Implementation in superconducting qubits}
\label{chap:SC_qubits}
\begin{figure*}
\centering
\includegraphics[width=2\columnwidth]{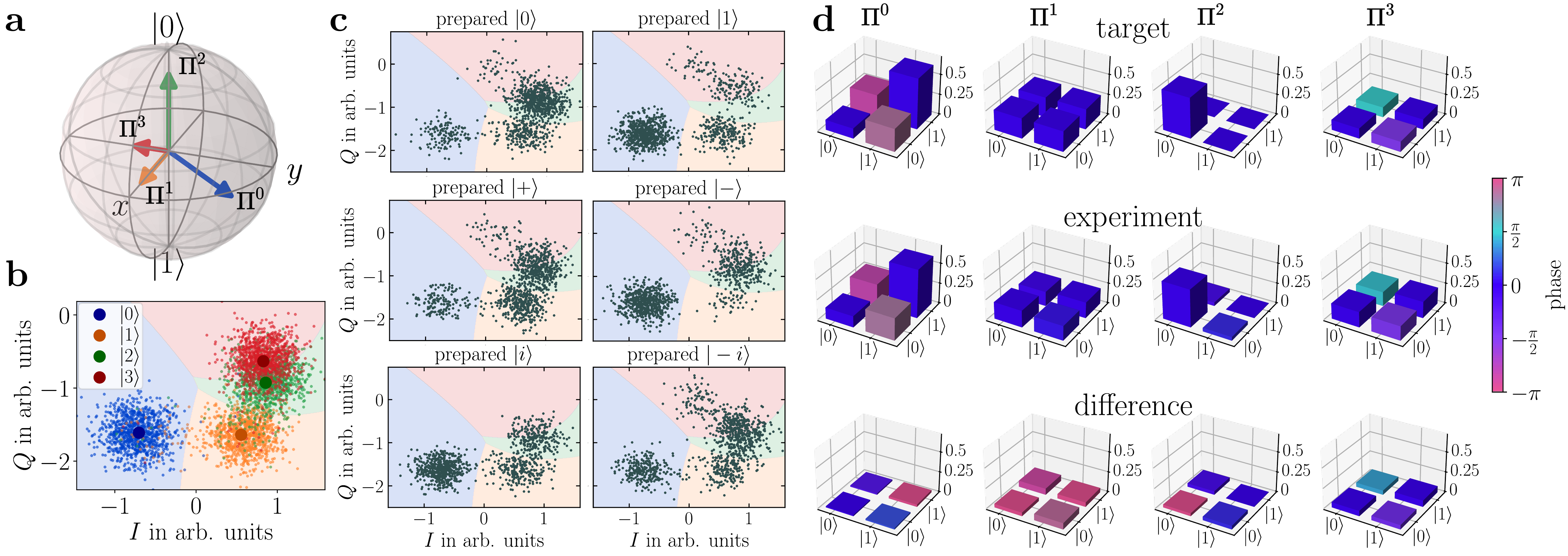}
\caption[]{Experimental realization of a single-qubit informationally complete POVM in the qudit space of a transmon qubit. 
\textbf{a)} Target POVM operators on the Bloch sphere. 
\textbf{b)} Calibration of measurement discrimination in the IQ-plane. Shaded regions show the decision boundaries of the classifier and large circles denote the average over all shots.
\textbf{c)} Raw data of POVM measurement outcomes for the six single-qubit stabilizer states. 
\textbf{d)} Characterization of the experimentally realized POVM operators $\Pi^i$ plotted as matrix histograms. The top row shows
the theoretical target operators, the middle row shows the POVM operators obtained from a maximum-likelihood detector tomography of the experimental data after applying readout error mitigation, and the bottom row shows their difference. Data taken on qubit 0 of \textit{ibmq\_lima} with $E_\text{J}/E_\text{C} \sim 45$.}
\label{fig:experiment_results_POVM}
\end{figure*}

We now present and discuss experimental results of a qudit-space POVM measurement in a superconducting transmon qubit.
Transmons are a popular qubit architecture as they enjoy long coherence times relative to the duration of their gates~\cite{place2021new} and can gather measurements at elevated trigger rates, typically around $1\,\text{--}\,100\,\text{kHz}$~\cite{wack2021quality}. 
They are built from a non-linear resonance circuit created by a Josephson junction shunted by a capacitor and are characterized by the ratio of the Josephson energy $E_{\text{J}}$ to the charging energy $E_{\text{C}}$, with $E_{\text{J}}/E_{\text{C}} \gg 1$
\cite{koch2007charge}. 
The spectrum of a transmon is described by an anharmonic oscillator, with the qubit encoded in the ground state $\ket{0}$ and the first excited state $\ket{1}$. 
For details on this architecture see App.~\ref{app:details_hardware}. 

\subsection{Qudit control of transmons}
We propose to use the energetically next-highest states $\ket{2}$ and $\ket{3}$ in addition to the qubit states $\ket{0}$ and $\ket{1}$ to implement qudit-based POVM measurements. 
With the decomposition in Eq.~\eqref{eq:theo_final_sequence_U}, we only need to drive transitions between adjacent states.
In existing experimental setups, these states are accessed by switching the carrier frequency of the microwave drive pulses. 
Current IBM Quantum systems employ qubits with $0 \!\leftrightarrow\! 1$ transition frequencies of $\sim\! 5\,\text{GHz}$ and anharmonicities of $\sim\!-300\,\text{MHz}$. 
Drive pulses are generated by an arbitrary waveform generator with a sampling rate of $4.5\times 10^9\,\text{s}^{-1}$~\cite{IBMQuantum}. 
We can thus apply modulations to the carrier frequency of up to approximately $\pm 1\,\text{GHz}$ (still oversampling by a factor of 4.5). 
The carrier frequencies of $\sim 4.7\,\text{GHz}$ and $\sim 4.3\,\text{GHz}$ required to address the $1\!\leftrightarrow\!2$ and $2\!\leftrightarrow\!3$ transitions, respectively, are thus well within the capabilities of our control hardware.
Coherent control of the $\ket{2}$ state following this procedure has already found applications in excited state promotion readout~\cite{elder2020highfidelity, jurcevic2021demonstration}, entanglement studies~\cite{cervera-lierta2021experimental}, gate decompositions~\cite{galda2021implementing}, fast resets~\cite{egger2018pulsed}, and entangling operations~\cite{egger2019entanglement}.

Qudit-based POVM measurements require sufficient lifetimes of the higher excited states.
On typical transmon qubits, we observe that the decay from $\ket{3}$ occurs predominantly sequentially as $\ket{3}\!\rightarrow\! \ket{2}\!\rightarrow\! \ket{1}\!\rightarrow\! \ket{0}$, while transitions such as $\ket{3}\!\rightarrow\! \ket{1}$ are strongly suppressed, see App.~\ref{sec_app:decay_higher_states}. 
This is in agreement with theory~\cite{catelani2012decoherence}, and previous experiments~\cite{peterer2015coherence}.
For our purposes, coherence in $\ket{2}$ and $\ket{3}$ is only required during the POVM pulse sequence, which lasts a total of $\mathcal{O}(100\,\text{ns})$ using at most ten $\sqrt{\mathcal{X}}$-pulses. 
With measured lifetimes of $>25\,\upmu\text{s}$ for the $\ket{3}$ and $\ket{2}$ states, we do not expect the decay of higher excited states to be a limiting factor. 

Transmons are dispersively measured by coupling them to a readout resonator~\cite{wallraff2005approaching}. 
The transmitted signal is typically down-converted and integrated, resulting in a point in the IQ-plane, which is then discriminated into $\ket{0}$ and $\ket{1}$.
Dispersive readout can be extended to distinguish between the four qudit states. Recently, separation of the lowest three states with fidelities >95\% has been demonstrated experimentally~\cite{blok2021quantum}.

A challenge for qudit control of transmons is the \emph{charge dispersion} of higher-excited states.
The exact eigenenergies of all transmon states fluctuate under charge noise of the environment, see App.~\ref{sec_sup:transmon_qubits}. 
This effect increases exponentially for the energetically higher states posing a threat for high-fidelity pulses on the $1\!\leftrightarrow\!2$ and especially on the $2\!\leftrightarrow\!3$ transition. 
As a result, transition frequencies fluctuate considerably from one experimental run to another. 
For IBM Quantum hardware with $E_{\text{J}}/E_{\text{C}} \sim 40$, we observe that the $2\!\leftrightarrow\!3$ transition frequency varies by $15$ to $20\;\text{MHz}$, see App.~\ref{sec_app:meas_charge_disp}.
To ensure a resonant driving of the transition, the corresponding drive pulses thus need to cover a broad spectral range.
This can be achieved by shortening the pulses, which typically increases phase errors and leakage to neighboring levels. 
Pulse shaping techniques such as DRAG and advanced optimal control help alleviate this issue~\cite{motzoi2009simple, gambetta2011analytic, werninghaus2021leakage}.
Furthermore, applying the POVM pulse sequence requires tracking the phases of idle levels. 
The acquired phases depend on the eigenenergies of each level, which are subject to charge dispersion. 
Conveniently, the unitary that encodes the POVM  requires a single drive of the $2 \!\leftrightarrow\! 3$ transition, see Eq.~\eqref{eq:theo_final_sequence_U}. 
Hence, the $\ket{3}$ state is only populated once during the sequence, so that any phase uncertainty after the $2\!\leftrightarrow\!3$ pulse becomes irrelevant upon measurement in the qudit basis. 
Thus, whereas full coherent control of the $\ket{3}$ state is difficult to achieve, the relatively simple pulse sequence required for the POVM measurement is particularly robust to phase uncertainties of this state. 

\subsection{Experimental demonstration}
\label{sec:experimental_demo}

As a proof-of-principle demonstration on IBM Quantum hardware, we implement a single-qubit IC-POVM which consists of the target POVM operators
\begin{equation}
\label{eq:experimental_POVM_operators}
\begin{aligned}
    \Pi^0 &= \frac{3}{4} \ket{\psi_0}\bra{\psi_0} ,& \quad 
    \Pi^1 &= \frac{1}{2} \ket{+}\bra{+} ,\\ 
    \Pi^2 &= \frac{1}{2} \ket{0}\bra{0} , 
    & \quad \Pi^3 &= \frac{1}{4} \ket{-i}\bra{-i}  \qquad
\end{aligned}
\end{equation}
with $\ket{\psi_0} = \left(\ket{0} + \left(i-2\right)\ket{1}\right)/\sqrt{6}$.
Three of the operators ($\Pi^1 , \Pi^2$, and $\Pi^3$) point along the Cartesian axes of the Bloch sphere, while $\Pi^0 $ points into the octant which lies opposite of all other vectors, see Fig.~\ref{fig:experiment_results_POVM}\textbf{a}. 
The unitary that encodes this POVM is realized with a sequence consisting of two $\sqrt{\mathcal{X}}_{0 \leftrightarrow1}$, two $\sqrt{\mathcal{X}}_{1 \leftrightarrow2}$ and one $\sqrt{\mathcal{X}}_{2 \leftrightarrow3}$ gates, see App.~\ref{sec:app_details_experiments_POVM}.
We use the standard single-qubit $SX$-gate that comes with a highly calibrated \textsc{Drag}-pulse exposed to the user by IBM Quantum systems as the $\sqrt{\mathcal{X}}_{0 \leftrightarrow 1}$-pulse.  
All further pulse-level calibrations and the POVM measurements are implemented through Qiskit's pulse module \cite{alexander2020qiskit, mckay2018qiskit}.
For the $1\!\leftrightarrow\!2$ and $2\!\leftrightarrow\!3$ transitions, we first calibrate the transition frequency with spectroscopy after preparing the initial states $\ket{1}$ and $\ket{2}$, respectively.  
For simplicity, we implement the $\sqrt{\mathcal{X}}$-gates on these transitions with Gaussian pulses. 
We choose a duration of $32\,\text{ns}$ for the $\sqrt{\mathcal{X}}_{1 \leftrightarrow2}$- and $14\,\text{ns}$ for the $\sqrt{\mathcal{X}}_{2 \leftrightarrow3}$-pulse. 
These durations are shorter than the $36\,\text{ns}$ standard single-qubit pulse to mitigate charge dispersion in higher-excited states by an increased spectral width. 
Simulations suggest that even shorter pulses are beneficial, see App.~\ref{app:details_simulation}. 
However, we find it more difficult to calibrate them. 
After fixing the pulse duration, we calibrate the angle of the rotations through sinusoidal fits to Rabi oscillations with varying pulse amplitudes.
To calibrate the readout, we prepare and measure the states $\ket{0}$, $\ket{1}$, $\ket{2}$, and $\ket{3}$ separately through a sequence of appropriate $\sqrt{\mathcal{X}}$-gates and use this data to train a classifier with a quadratic decision boundary, as shown in Fig.~\ref{fig:experiment_results_POVM}\textbf{b}.
For each state, we obtain a characteristic signal that clusters in different regions of the IQ-plane.

We investigate how well our pulse sequence along with the calibrated measurement implements the desired POVM with \emph{quantum detector tomography} (QDT) \cite{dariano2004quantum, lundeen2009tomography}, which characterizes the realized POVM operators.  
Hereby, a set of reference states is prepared and measured by our POVM implementation. 
We choose the set of single-qubit states $\ket{0}$, $\ket{1}$, $\ket{+}$, $\ket{-}$, $\ket{i}$, and $\ket{-i}$ for this purpose.
From the obtained outcome distributions, shown in Fig.~\ref{fig:experiment_results_POVM}\textbf{c}, the underlying experimental POVM operators can be estimated with a maximum-likelihood (ML) procedure, which guarantees that they form a valid POVM \cite{fiurasek2001maximumlikelihood}, see App. \ref{app:detector_tomography}.
Note that, on the Bloch sphere, the tomography states $\ket{-}$, $\ket{1}$, and $\ket{i}$ lie opposite the POVM operators $\Pi^1$, $\Pi^2$, and $\Pi^3$, respectively.
They should thus have zero measurement probability of the corresponding outcomes, which is attested by a noticeable lack of counts in the respective regions of the IQ-plane in the raw data of Fig.~\ref{fig:experiment_results_POVM}\textbf{c}.
As a result, the operators obtained from the maximum-likelihood detector tomography are in good qualitative agreement with the theoretical target operators, see Fig.~\ref{fig:experiment_results_POVM}\textbf{d}. 

\begin{table}[]
\caption{Measured readout assignment error probabilities when preparing the four qudit states of a transmon.}
\centering
\begin{ruledtabular}
\begin{tabular}{@{} c|rrrr @{}}
\headercell{Measured \\} & \multicolumn{4}{c@{}}{Prepared} \\
& $\ket{0}$ & $\ket{1}$ & $\ket{2}$ & $\ket{3}$ \\ 
\midrule
$\ket{0}$ & 98.3\,\% & 4.2\,\% & 0.6\,\% & 0.2\,\% \\
$\ket{1}$ & 0.5\,\% & 88.8\,\% &  8.8\,\% &  2.1\,\%  \\
$\ket{2}$ & 0.8\,\% &  6.9\,\% & 59.3\,\% & 22.8\,\% \\
$\ket{3}$ & 0.4\,\% &  0.1\,\% & 31.3\,\% & 74.9\,\% \\
\end{tabular}
\end{ruledtabular}
\label{tab:readout_assignments}
\end{table}

We quantify the fidelity through the \emph{operational distance} $D_\text{OD}$~\cite{maciejewski2020mitigation, puchala2018strategies}, a measure on the POVM space, between the experimentally realized and the target POVM with $0 \leq D_\text{OD} \leq 1$ and $D_\text{OD}=0$ for coinciding POVMs, see App.~\ref{app:operational_distance}.  
The raw measurement data presented in Fig.~\ref{fig:experiment_results_POVM}\textbf{c} yields $D_\text{OD} = 0.22$.
We identify the overlap of the detection regions in the IQ-plane between $\ket{1}$ and $\ket{2}$ and especially $\ket{2}$ and $\ket{3}$ as the main experimental limitation for qudit-based POVM measurements.
Specifically, in our experiments, around one quarter of the prepared states in $\ket{3}$ are identified as $\ket{2}$ and vice versa, see Tab.~\ref{tab:readout_assignments}.
To mitigate misassignment errors, we apply readout error mitigation based on the inversion of the misassignment matrix, constrained to non-negative probability vectors~\cite{maciejewski2020mitigation}.
Thereby, we can partially correct the measured raw data 
and achieve an improved $D_\text{OD}$ of $0.15$ between the theoretical and the ML-estimated experimental POVM.

The difficulty to reliably distinguish the states $\ket{2}$ and $\ket{3}$ complicates the calibration of the average $2\!\leftrightarrow\!3$ transition frequency. 
At the moment, this renders the implementation of POVMs that require virtual $\mathcal{Z}_{2\leftrightarrow3}$-gates infeasible. 
This motivates the choice of the POVM operators in Eq.~\eqref{eq:experimental_POVM_operators} for our experiments, which are achievable with a slightly simplified pulse sequence, compared to the most general case of Eq.~\eqref{eq:theo_final_sequence_U}, see App.~\ref{sec:app_details_experiments_POVM}.
The measurement pulses used in our experiment are the default pulses provided by the backend, which are optimized for maximal separation of the $\ket{0}$ and $\ket{1}$ states.
A large-scale implementation of qudit-space POVM measurements would require a more careful calibration of the readout pulses, which optimizes the separation of all four involved basis states. 
This would make the virtual $\mathcal{Z}_{2\leftrightarrow3}$-gates feasible and improve the $\sqrt{\mathcal{X}}_{2\leftrightarrow3}$-gate.

\subsection{Optimal transmon parameter regime}
\label{sec:optimal_transmon_regime}

\begin{figure}
\includegraphics[width=0.99\columnwidth]{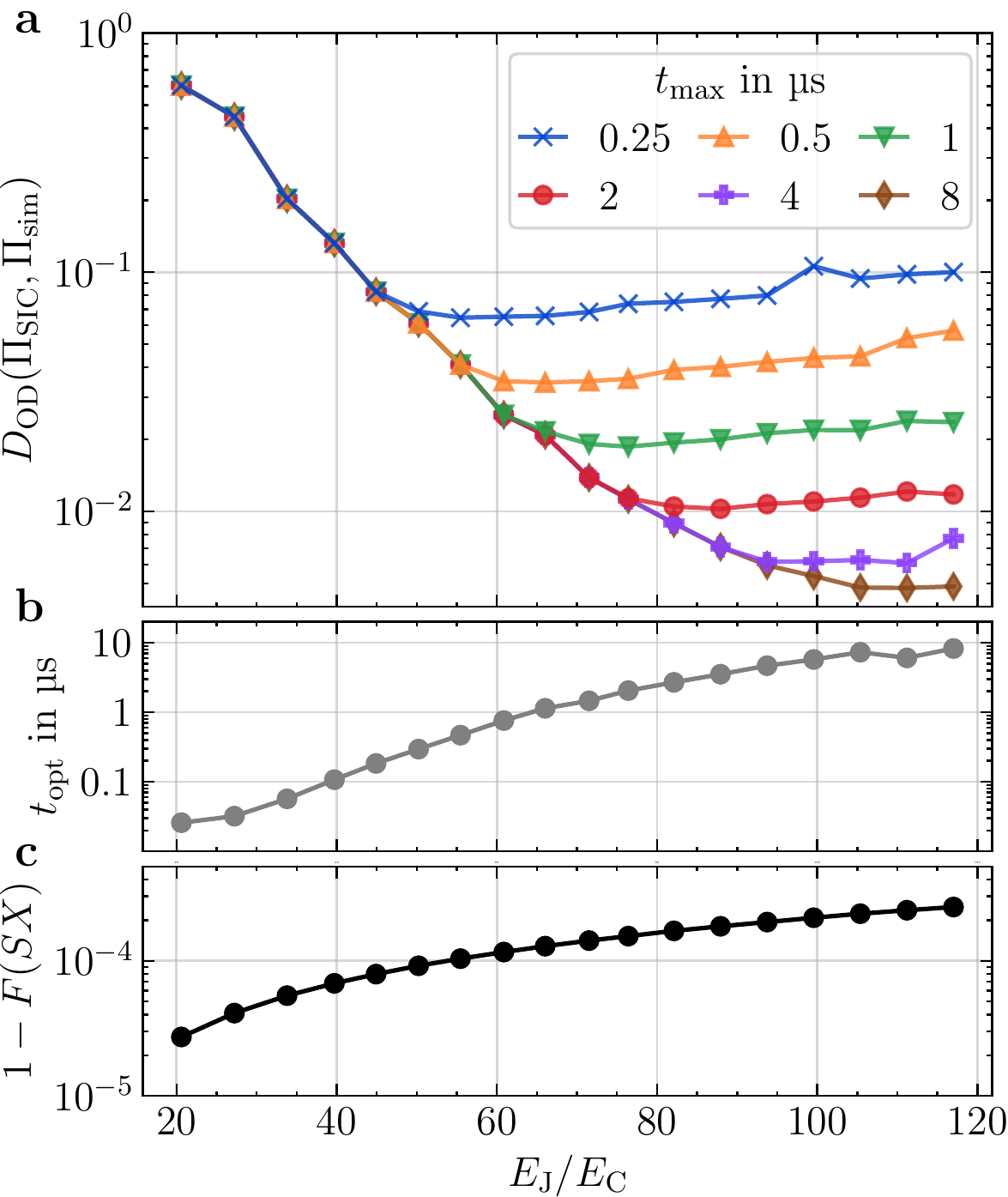}
\caption[]{Simulations of a pulse schedule implementing a SIC-POVM under charge noise for different ratios $E_{\text{J}}/E_{\text{C}}$ of a transmon qubit with a frequency of $5\,\text{GHz}$.
\textbf{a)} Operational distance $D_\text{OD}$ between theory and simulated POVMs for different maximal durations $t_\text{max}$ of the pulse schedules. 
\textbf{b)} Optimal total durations $t_{\text{opt}}$ of the POVM pulse schedules that reach the best operational distance. 
\textbf{c)} Average gate error of a single-qubit $SX$-gate realized through Gaussian pulses with a fixed duration of $36\,\text{ns}$. 
}
\label{fig:transmon_parameter_regime}
\end{figure}

In the previous section, we demonstrated a qudit-based POVM measurement on a quantum device with an $E_{\text{J}}/E_{\text{C}}$-ratio of $\sim 45$. This value was chosen for optimal qubit operation.
However, the substantial charge dispersion in states $\ket{2}$ and $\ket{3}$ of the transmon suggests that larger $E_{\text{J}}/E_{\text{C}}$-ratios may be advantageous for qudit POVMs. 
This would sacrifice some anharmonicity to decrease the charge noise.
We now quantitatively assess this trade-off through numerical pulse-level simulations, which account for both leakage errors due to finite anharmonicity and phase errors due to charge noise, but neglect readout misassignment errors. 

We start by probing how the achievable $D_\text{OD}$ depends on $E_{\text{J}}/E_{\text{C}}$, using a single-qubit symmetric, informationally complete (SIC) POVM $\boldsymbol{\Pi}_{\mathrm{SIC}}$ as an example of a generic POVM.
It consists of four operators $\Pi_{\text{SIC}}^m = \frac{1}{2} \ket{\psi_m}\bra{\psi_m}$ with $\ket{\psi_0} = \ket{0}$ and $\ket{\psi_m} = \left( \ket{0} + \sqrt{2} e^{2\pi i (m-1)/3} \ket{1} \right)/\sqrt{3}$ with $ m\in \{1, 2, 3 \}$ that point towards the corners of a regular tetrahedron, see Fig.~\ref{fig:overview_schematic}\textbf{a}. 
In contrast to the experimentally demonstrated POVM in Eq.~\eqref{eq:experimental_POVM_operators}, $\boldsymbol{\Pi}_{\mathrm{SIC}}$ requires implementing the pulse sequence from Eq.~\eqref{eq:theo_final_sequence_U} in its full generality.
We simulate this sequence with Gaussian pulse envelopes on a single transmon by numerically integrating the time-dependent Schr{\"o}dinger equation.
For details on how we model charge dispersion and calibrate pulses see App.~\ref{app:details_simulation}.
As the $E_{\text{J}}/E_{\text{C}}$-ratio increases and charge noise becomes less prevalent, 
$D_\text{OD}(\Pi_{\text{SIC}}, \Pi_{\text{sim}})$
decreases, see Fig.~\ref{fig:transmon_parameter_regime}\textbf{a}. 
While the $D_\text{OD}$ is limited to $0.1$ for $E_{\text{J}}/E_{\text{C}} \sim 40$, it improves to $0.01$ for $E_{\text{J}}/E_{\text{C}} \sim 80$.
The change in anharmonicity with $E_{\text{J}}/E_{\text{C}}$ affects the duration of the pulse sequence that achieves the optimal $D_\text{OD}$, as plotted in Fig.~\ref{fig:transmon_parameter_regime}\textbf{b}.
In the low $E_{\text{J}}/E_{\text{C}}$-regime, short pulses are favored as a broad spectral width is required to cover the large spread of the charge noise, and leakage is minimal due to the large anharmonicity.
Conversely, with increasing $E_{\text{J}}/E_{\text{C}}$, the anharmonicity of the transmon is reduced, which amplifies leakage.
The optimal pulse durations thus increase with the ratio $E_{\text{J}}/E_{\text{C}}$. 

The longer the pulse sequence, the more it is subject to non-unitary processes like decoherence, which are not considered in our simulation. 
Consequently, there is a trade-off between the optimal durations of the pulses under unitary dynamics and noise induced by finite coherence times. 
We therefore limit the total duration of the POVM-encoding pulse sequence to different maximally allowed durations $t_{\text{max}}$, see Fig.~\ref{fig:transmon_parameter_regime}\textbf{a}.
We find that, for fixed $t_{\text{max}}$, the $D_\text{OD}$ improves with increasing $E_{\text{J}}/E_{\text{C}}$ until an optimal ratio is reached after which the $D_\text{OD}$ gradually increases. 
In the parameter regime of current IBM Quantum hardware ($E_{\text{J}}/E_{\text{C}} \sim 35\,\text{--}\,45$), the optimal POVM pulse sequence time is $\sim\! 100\,\text{ns}$. 
On this timescale, we do not expect decoherence to be significant, see App.~\ref{sec_app:decay_higher_states}.
For reference, single-qubit gates typically last $36\,\text{ns}$~\cite{IBMQuantum}.
Finally, changing the transmon parameters also affects the conventional gates run in the quantum circuit prior to the POVM measurement. 
This is exemplified by the average gate fidelity $F$ of a single-qubit $36\,\text{ns}$ $SX$-gate, which is shown in Fig.~\ref{fig:transmon_parameter_regime}\textbf{c}. 
As $E_{\text{J}}/E_{\text{C}}$ increases from 20 to 120 the gate fidelity decreases by roughly one order of magnitude due to the reduced anharmonicity. 

The trade-off between anharmonicity and charge noise in a transmon qubit is a complex interplay of many factors, including coherence times, gate fidelities and gate speed~\cite{koch2007charge}.
Our simulations suggest that, when taking qudit POVM fidelities into account, the optimal hardware regime shifts towards higher $E_{\text{J}}/E_{\text{C}}$-ratios.
While this improves the quality of qudit-space POVM measurements, it comes at the expense of either slightly worse gate fidelities or slightly slower gate speeds, whose severity ultimately depend on the available coherence times.
Optimal control methods may alleviate such issues~\cite{werninghaus2021leakage}.

\section{Application to operator sampling}
\label{chap:applications}
Our experimental realization is currently limited by misassignment errors in the readout due to insufficient separation in the IQ-plane.
However, even with perfect readout fidelities, the considerable charge noise of current-generation transmon qubits still raises the question whether qudit POVMs with ODs of $\sim\!0.1$ are sufficient for practical applications. 
Here, we address this question through numerical simulations of optimized IC-POVMs for estimating the expectation value of an observable $\mathcal{O}$ as developed in Ref.~\cite{garcia2021learning}. 

\subsection{Device noise mitigation through detector tomography}
\label{sec:bias_mitigation_tomography}

\begin{figure}
\centering
\includegraphics[width=0.99\columnwidth]{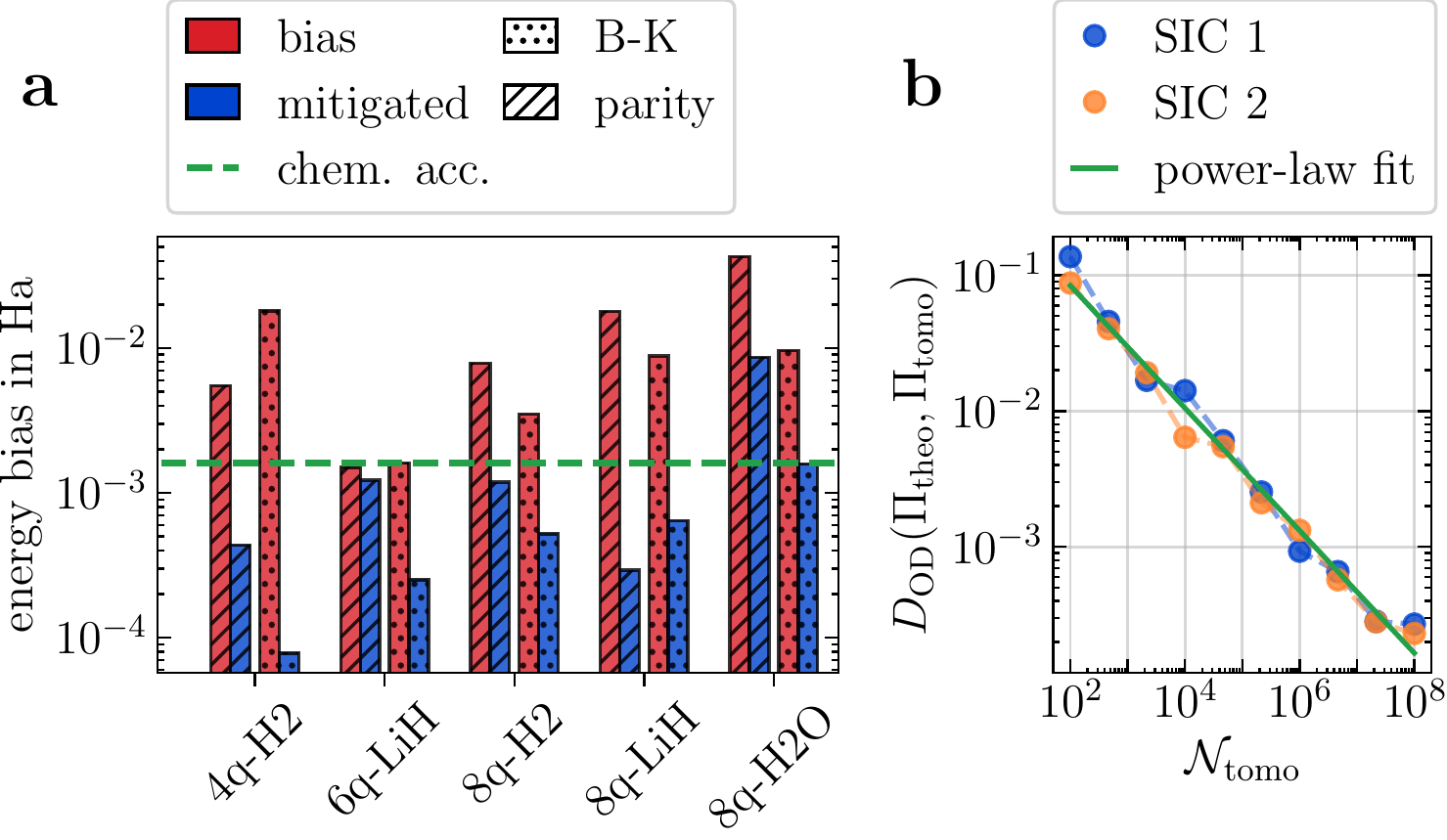}
\caption[]{Error Mitigation through detector tomography on qudit-space POVM measurements. \textbf{a)} Simulations of energy measurements with optimized POVMs for trained VQE states of small molecular Hamiltonians ($\text{H}_2$, $\text{LiH}$, and $\text{H}_2\text{O}$) obtained from parity and Bravyi-Kitaev (B-K) fermion-to-qubit mappings. Charge noise leads to a bias in the POVM estimator (red). Detector tomography with a total of $10^5$ shots reduces this bias (blue). \textbf{b)} Operational distance between the experimetal POVM $\boldsymbol{\Pi}_{\text{exp}}$
simulated under charge noise and its tomographic reconstruction $\boldsymbol{\Pi}_{\text{tomo}}$ for the standard SIC-POVM (blue) and a SIC-POVM defined in Ref.~\cite{jiang2020optimal} (yellow) as a function of the total shots used for the detector tomography $\mathcal{N}_{\text{tomo}}$. A power-law fit yields a scaling of $D_{\text{OD}} \sim \mathcal{N}_{\text{tomo}}^{-0.45}$.}
\label{fig:error_mitigation_tomography}
\end{figure}

\begin{figure*}
\centering
\includegraphics[width=1.99\columnwidth]{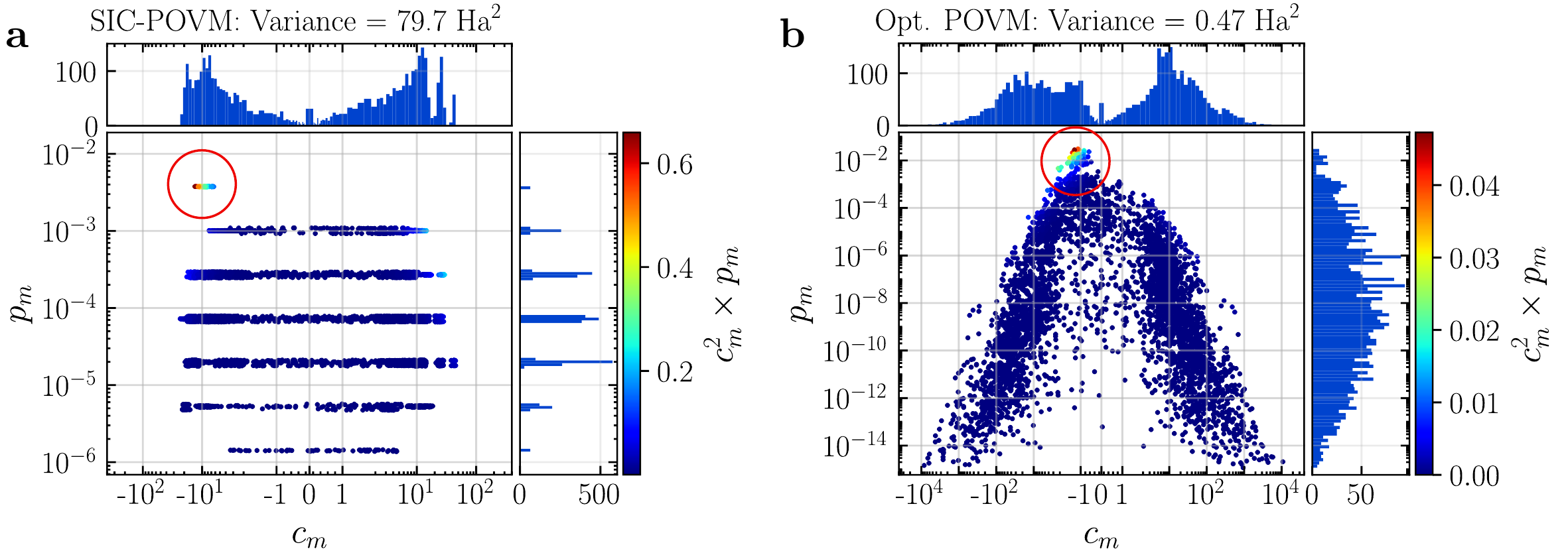}
\caption[]{Variance reduction through an optimized POVM for a six-qubit LiH Hamiltonian $\mathcal{O}_{\text{LiH}}$ with a trained VQE state $\ket{\psi}_\text{VQE}$. 
Each scatter point represents the probability $p_m$ of obtaining outcome $m$ for the state $\ket{\psi}_\text{VQE}$ and the corresponding coefficient $c_m$ of the observable $\mathcal{O}_{\text{LiH}}$ in the global IC-POVM basis,  
see Eqs.~\eqref{eq:outcome_probability_POVM}$\,$--$\,$\eqref{eq:O_decomposed_POVM}. 
There are $M=4^6$ POVM outcomes and $m\in\{0, \dots, M-1\}$.
The histograms show the marginal distributions of $c$ and $p$, while color indicates how much each data point $(c_m, p_m)$ contributes to the variance.
Only the few outcomes circled in red contribute significantly. 
\textbf{a)} Exact theoretical distribution where each qubit is measured in a SIC-POVM.
\textbf{b)} Distribution of a POVM which was optimized to reduce the variance for this observable and state. 
The probabilities $p_m$ are obtained from a simulation of the qudit-space POVM scheme under noisy conditions in transmon hardware with $E_\text{J}/E_\text{C} = 45$. 
The coefficients $c_m$ are computed following the error mitigation strategy with $\mathcal{N}_\text{tomo} = 10^5$ tomography shots.
}
\label{fig:povm_scatter_plots}
\end{figure*}

We denote the optimized (theoretical) target POVM by $\boldsymbol{\Pi}_{\text{theo}}$, which defines a target unitary in the qudit space of each transmon with corresponding outcome probabilities $p^{\text{theo}}_m$ according to Eq.~\eqref{eq:outcome_probability_POVM}.
However, due to device noise, the effective (experimental) channel that is applied to the qudits encodes a different POVM, denoted by $\boldsymbol{\Pi}_{\text{exp}}$, which slightly deviates from the theoretical one. 
In practice, $\boldsymbol{\Pi}_{\text{exp}}$ defines the experimental measurement probabilities of the outcomes $p^{\text{exp}}_m$, while $\boldsymbol{\Pi}_{\text{theo}}$ is used to obtain the decomposition of $\mathcal{O}$ with coefficients $c^{\text{theo}}_m$, as defined in Eq.~\eqref{eq:O_decomposed_POVM}.
The combined estimator converges to $\widehat{\langle \mathcal{O} \rangle} =  \sum_m c^{\text{theo}}_m p^{\text{exp}}_m$, which differs from the theoretical expectation value due to the imperfections in the device, leading to a bias $\sum_m c^{\text{theo}}_m ( p^{\text{exp}}_m - p^{\text{theo}}_m )$.

To estimate the impact of this bias on practical applications,
we study its effects on energy measurements of trained VQE ansatz states for small molecular Hamiltonians mapped onto four to eight qubits. 
As the target operators $\boldsymbol{\Pi}_{\text{theo}}$, we use POVMs that minimize the variance for the respective Hamiltonians over the trial states as reported in Ref.~\cite{garcia2021learning}.
These POVMs are simulated under charge noise for a device with $E_\text{J}/E_\text{C}=45$,
see App.~\ref{app:details_simulation}. 
The biases that arise from the device noise are shown in Fig.~\ref{fig:error_mitigation_tomography}\textbf{a} (red bars). 
In most cases, we observe that charge noise creates biases that prevent energy estimations down to chemical accuracy.

To attenuate the large biases induced by the hardware noise, we propose an efficient error mitigation strategy in which the mismatch between $\boldsymbol{\Pi}_{\text{theo}}$ and $\boldsymbol{\Pi}_{\text{exp}}$ is reduced by means of quantum detector tomography~\cite{dariano2004quantum, lundeen2009tomography}. 
This process allows an accurate estimation of the POVM operators that are actually implemented in the device, denoted by $\boldsymbol{\Pi}_{\text{tomo}}$. 
With this procedure, we first compute the decomposition of $\mathcal{O}$ into the operators of $\boldsymbol{\Pi}_{\text{tomo}}$, i.e., ${\mathcal{O} = \sum_m c^{\text{tomo}}_m \Pi_{\text{tomo}}^m}$ and then use the new coefficients $c^{\text{tomo}}_m$ to estimate the expectation value as $\widehat{ \langle \mathcal{O} \rangle} = \sum_m c^{\text{tomo}}_m p^{\text{exp}}_m$. 
With an increasing number of tomography shots, the OD between $\boldsymbol{\Pi}_{\text{exp}}$ and $\boldsymbol{\Pi}_{\text{tomo}}$ can be arbitrarily decreased, see Fig.~\ref{fig:error_mitigation_tomography}. 
In turn, the systematic bias $\sum_m c^{\text{tomo}}_m ( p^{\text{exp}}_m - p^{\text{tomo}}_m )$ converges to zero for infinitely many tomography shots.
The desired accuracy in a given application thus defines
how many measurements should be dedicated to the detector tomography. 
Crucially, since the POVMs we consider are always products of single-qubit POVMs, the tomographic reconstruction can be carried out on all qubits in parallel. 
Thus, the overhead in the shot budget is constant, and we do not expect this process to hamper the scalability of qudit-based POVMs. 
Our simulations indicate that, even for current transmon hardware with $E_\text{J}/E_\text{C} \sim 45$, qudit-space POVM measurements characterized through detector tomography are sufficiently accurate for quantum chemistry applications. 

\subsection{Qudit-based POVMs for variance reduction}

Finally, we discuss whether the qudit POVM measurements in noisy conditions can be utilized to reduce the variance of an estimator of $\langle \mathcal{O} \rangle$. 
As an example, we consider the 6-qubit Hamiltonian $\mathcal{O}_\text{LiH}$ of a LiH molecule in the STO-3G basis obtained from the Bravyi-Kitaev mapping and investigate the number of shots needed to estimate the energy of a trained VQE state $\ket{\psi}_\text{VQE}$ within chemical accuracy (in the chosen basis set). 
We compare two situations where firstly, each qubit is measured using a SIC-POVM and secondly, the qubits are measured by means of a product POVM optimized to minimize the variance of $\mathcal{O}_\text{LiH}$ in the state $\ket{\psi}_\text{VQE}$~\cite{garcia2021learning}.
For a given POVM, the variance of a specific observable is determined by its decomposition coefficients $c_m$ and the measurement probability distribution $p_m$ of the state, see Eq.~\eqref{eq:O_error_POVM}. 
Namely, the second moment $\sum_m c_m^2 p_m$ determines the accuracy $\epsilon$ of the POVM-based estimator. 
In particular the outcomes $m$ with both high absolute value of $c_m$ and high measurement probability $p_m$ contribute to $\epsilon$. 
For the outcome distribution of the SIC-POVM, due to the symmetry of the POVM operators, the data is highly structured, see Fig.~\ref{fig:povm_scatter_plots}\textbf{a}.
The outcomes with highest probability attain high values of $\left|c_m\right|$, which results in a large second moment of $80.86\,\text{Ha}^2$. 
By measuring in an optimized POVM, even under charge noise, the second moment is considerably reduced to $1.59\,\text{Ha}^2$.
This approaches the optimum set by the squared first moment $\langle \mathcal{O} \rangle^2 = 1.12\,\text{Ha}^2$.
This effect can be explained by inspecting the shape of the the distribution in Fig.~\ref{fig:povm_scatter_plots}\textbf{b}, which shows a ``squeezing'' such that the most probable outcomes are associated with low absolute values of $c_m$. 
This in turn leads to very large absolute coefficients for other outcomes, which, in contrast, have negligible measurement probability and thus hardly contribute to the variance. 

We observe that with the generic SIC-POVM scheme about $3.5\times10^7$ shots are required to estimate $\langle \mathcal{O}_{\text{LiH}} \rangle$ to within chemical accuracy. 
In contrast, only $3.1 \times 10^5$ shots are required when using the optimized POVM in a qudit-based scheme using a transmon affected by state-of-the-art charge noise. 
This number already includes $10^5$ shots devoted solely to the detector tomography used for the bias mitigation discussed in Sec.~\ref{sec:bias_mitigation_tomography}. 
With a circuit execution rate of $10\,\text{kHz}$, the optimized POVM reduces the measurement time from 1 hour down to 30 seconds.
It is important to note that in this application the mitigated bias lies well within chemical accuracy, as shown in Fig.~\ref{fig:error_mitigation_tomography}. 
Based on this example, we conclude that qudit-space POVM measurements constitute a valid, shot-efficient approach to estimate observables with high precision.

\section{Discussions \& Conclusions}
\label{chap:discussion}
We introduced a method to perform general POVM measurements for qubits via a Naimark dilation construction, 
which extends the qubit space into a qudit space through the addition of two extra levels, 
rather than coupling to an additional ancilla qubit. 
Our strategy makes optimal use of the available quantum resources in a system without requiring full qudit control -- a challenging task in general. 
We couple the qubit states to the two additional levels of the surrounding qudit for only a short duration at the measurement stage of the quantum circuit. 
Therefore, only modest coherence and pulse fidelities are required.
Compared to ancilla-based POVM implementations, we circumvent the doubling of the quantum register size and thus save half of the qubits on the chip, while also avoiding a considerable \textsc{Swap}-gate overhead in case of limited device connectivity.
The result is a protocol that is applicable to various qubit architectures including super- and semiconducting qubits, trapped ions, and cold atoms. 

For a superconducting transmon qubit, we detailed an implementation of qudit-space POVM measurements, including a description of the decomposition into suitable elementary pulses between adjacent levels, and of the required calibrations. 
Specifically, we proposed ways to operate the necessary frame changes by tracking advances in relative phases, as well as 
generalizing the concept of virtual $\mathcal{Z}$-gates to the qudit space. 
Compared to the standard qubit setting, our proposal admittedly requires further calibrations involving the additional states. 
However, these calibrations can be performed on all qudits in parallel and are typically faster than two-qubit gate calibrations.

Exploiting the functionalities of Qiskit Pulse~\cite{alexander2020qiskit}, we successfully performed a proof-of-principle experiment using the four lowest levels of a transmon in IBM Quantum hardware. 
We found that measurement misassignments are currently the main limitation of the proposed qudit-based POVMs, which prevents the scaling up to multi-qubit implementations.  
This calls for a more thorough design and optimization of the shape and frequency of measurement pulses with the aim of obtaining a sufficient dispersive shift for all four qudit levels. 
Moreover, the importance of choosing the readout resonator frequency appropriately, such that no transitions between higher excited states are accidentally resonant to the resonator frequency, has also been pointed out~\cite{peterer2015coherence}.

From preliminary pulse-level simulations, we conclude that tuning the qubits deeper into the transmon regime would be beneficial to achieve optimal POVM fidelities, as this limits the impact of charge noise in the higher-excited states. 
Nonetheless, our results indicate that the implementation of qudit-based POVMs in state-of-the-art IBM Quantum hardware can significantly reduce the number of measurements required to estimate expectation values. 
To achieve this goal, we designed a shot-efficient strategy based on detector tomography to mitigate systematic errors arising from experimental imperfections.

In addition to operator averaging, informationally complete POVMs can be employed for other paradigmatic quantum information tasks, including  state tomography~\cite{carrasquilla2019reconstructing} and the extraction of classical shadows~\cite{acharya2021shadow}. 
In all these cases, our strategy offers a resource-effective route towards their implementation in state-of-the-art quantum processors. 
On a broader perspective, our results open up new opportunities to exploit the multi-level structure available on many different qubit architectures, thus contributing to the development of a richer operational toolbox, and extending the native capabilities of current quantum computing architectures.

\section{Acknowledgements}

This research is part of two projects that have received funding from the European Union’s Horizon 2020 research and innovation programme under the Marie Skłodowska-Curie grant agreements No.~847471 and No.~955479.
This work was supported as a part of NCCR SPIN, a National Centre of Competence in Research, funded by the Swiss National Science Foundation (grant number 51NF40-180604).
We acknowledge the use of IBM Quantum services for this work.
IBM, the IBM logo, and ibm.com are trademarks of International Business Machines Corp., registered in many jurisdictions worldwide. Other product and service names might be trademarks of IBM or other companies. 
The current list of IBM trademarks is available at \url{https://www.ibm.com/legal/copytrade}.

\bibliography{main}

\begin{thebibliography}{83}%
\makeatletter
\providecommand \@ifxundefined [1]{%
 \@ifx{#1\undefined}
}%
\providecommand \@ifnum [1]{%
 \ifnum #1\expandafter \@firstoftwo
 \else \expandafter \@secondoftwo
 \fi
}%
\providecommand \@ifx [1]{%
 \ifx #1\expandafter \@firstoftwo
 \else \expandafter \@secondoftwo
 \fi
}%
\providecommand \natexlab [1]{#1}%
\providecommand \enquote  [1]{``#1''}%
\providecommand \bibnamefont  [1]{#1}%
\providecommand \bibfnamefont [1]{#1}%
\providecommand \citenamefont [1]{#1}%
\providecommand \href@noop [0]{\@secondoftwo}%
\providecommand \href [0]{\begingroup \@sanitize@url \@href}%
\providecommand \@href[1]{\@@startlink{#1}\@@href}%
\providecommand \@@href[1]{\endgroup#1\@@endlink}%
\providecommand \@sanitize@url [0]{\catcode `\\12\catcode `\$12\catcode
  `\&12\catcode `\#12\catcode `\^12\catcode `\_12\catcode `\%12\relax}%
\providecommand \@@startlink[1]{}%
\providecommand \@@endlink[0]{}%
\providecommand \url  [0]{\begingroup\@sanitize@url \@url }%
\providecommand \@url [1]{\endgroup\@href {#1}{\urlprefix }}%
\providecommand \urlprefix  [0]{URL }%
\providecommand \Eprint [0]{\href }%
\providecommand \doibase [0]{http://dx.doi.org/}%
\providecommand \selectlanguage [0]{\@gobble}%
\providecommand \bibinfo  [0]{\@secondoftwo}%
\providecommand \bibfield  [0]{\@secondoftwo}%
\providecommand \translation [1]{[#1]}%
\providecommand \BibitemOpen [0]{}%
\providecommand \bibitemStop [0]{}%
\providecommand \bibitemNoStop [0]{.\EOS\space}%
\providecommand \EOS [0]{\spacefactor3000\relax}%
\providecommand \BibitemShut  [1]{\csname bibitem#1\endcsname}%
\let\auto@bib@innerbib\@empty
\bibitem [{\citenamefont {Grinko}\ \emph {et~al.}(2021)\citenamefont {Grinko},
  \citenamefont {Gacon}, \citenamefont {Zoufal},\ and\ \citenamefont
  {Woerner}}]{grinko2021iterative}%
  \BibitemOpen
  \bibfield  {author} {\bibinfo {author} {\bibfnamefont {Dmitry}\ \bibnamefont
  {Grinko}}, \bibinfo {author} {\bibfnamefont {Julien}\ \bibnamefont {Gacon}},
  \bibinfo {author} {\bibfnamefont {Christa}\ \bibnamefont {Zoufal}}, \ and\
  \bibinfo {author} {\bibfnamefont {Stefan}\ \bibnamefont {Woerner}},\
  }\bibfield  {title} {\enquote {\bibinfo {title} {Iterative quantum amplitude
  estimation},}\ }\href {\doibase 10.1038/s41534-021-00379-1} {\bibfield
  {journal} {\bibinfo  {journal} {npj Quantum Information}\ }\textbf {\bibinfo
  {volume} {7}},\ \bibinfo {pages} {52} (\bibinfo {year} {2021})}\BibitemShut
  {NoStop}%
\bibitem [{\citenamefont {Egger}\ \emph {et~al.}(2021)\citenamefont {Egger},
  \citenamefont {Mareček},\ and\ \citenamefont
  {Woerner}}]{egger2021warmstarting}%
  \BibitemOpen
  \bibfield  {author} {\bibinfo {author} {\bibfnamefont {Daniel~J.}\
  \bibnamefont {Egger}}, \bibinfo {author} {\bibfnamefont {Jakub}\ \bibnamefont
  {Mareček}}, \ and\ \bibinfo {author} {\bibfnamefont {Stefan}\ \bibnamefont
  {Woerner}},\ }\bibfield  {title} {\enquote {\bibinfo {title} {Warm-starting
  quantum optimization},}\ }\href {\doibase 10.22331/q-2021-06-17-479}
  {\bibfield  {journal} {\bibinfo  {journal} {Quantum}\ }\textbf {\bibinfo
  {volume} {5}},\ \bibinfo {pages} {479} (\bibinfo {year} {2021})}\BibitemShut
  {NoStop}%
\bibitem [{\citenamefont {Sokolov}\ \emph {et~al.}(2022)\citenamefont
  {Sokolov}, \citenamefont {Dobrautz}, \citenamefont {Luo}, \citenamefont
  {Alavi},\ and\ \citenamefont {Tavernelli}}]{sokolov2022orders}%
  \BibitemOpen
  \bibfield  {author} {\bibinfo {author} {\bibfnamefont {Igor~O.}\ \bibnamefont
  {Sokolov}}, \bibinfo {author} {\bibfnamefont {Werner}\ \bibnamefont
  {Dobrautz}}, \bibinfo {author} {\bibfnamefont {Hongjun}\ \bibnamefont {Luo}},
  \bibinfo {author} {\bibfnamefont {Ali}\ \bibnamefont {Alavi}}, \ and\
  \bibinfo {author} {\bibfnamefont {Ivano}\ \bibnamefont {Tavernelli}},\
  }\bibfield  {title} {\enquote {\bibinfo {title} {Orders of magnitude
  reduction in the computational overhead for quantum many-body problems on
  quantum computers via an exact transcorrelated method},}\ }\href
  {http://arxiv.org/abs/2201.03049} {\bibfield  {journal} {\bibinfo  {journal}
  {arXiv:2201.03049 [quant-ph]}\ } (\bibinfo {year} {2022})}\BibitemShut
  {NoStop}%
\bibitem [{\citenamefont {Place}\ \emph {et~al.}(2021)\citenamefont {Place},
  \citenamefont {Rodgers}, \citenamefont {Mundada}, \citenamefont {Smitham},
  \citenamefont {Fitzpatrick}, \citenamefont {Leng}, \citenamefont {Premkumar},
  \citenamefont {Bryon}, \citenamefont {Vrajitoarea}, \citenamefont {Sussman},
  \citenamefont {Cheng}, \citenamefont {Madhavan}, \citenamefont {Babla},
  \citenamefont {Le}, \citenamefont {Gang}, \citenamefont {Jäck},
  \citenamefont {Gyenis}, \citenamefont {Yao}, \citenamefont {Cava},
  \citenamefont {de~Leon},\ and\ \citenamefont {Houck}}]{place2021new}%
  \BibitemOpen
  \bibfield  {author} {\bibinfo {author} {\bibfnamefont {Alexander P.~M.}\
  \bibnamefont {Place}}, \bibinfo {author} {\bibfnamefont {Lila V.~H.}\
  \bibnamefont {Rodgers}}, \bibinfo {author} {\bibfnamefont {Pranav}\
  \bibnamefont {Mundada}}, \bibinfo {author} {\bibfnamefont {Basil~M.}\
  \bibnamefont {Smitham}}, \bibinfo {author} {\bibfnamefont {Mattias}\
  \bibnamefont {Fitzpatrick}}, \bibinfo {author} {\bibfnamefont {Zhaoqi}\
  \bibnamefont {Leng}}, \bibinfo {author} {\bibfnamefont {Anjali}\ \bibnamefont
  {Premkumar}}, \bibinfo {author} {\bibfnamefont {Jacob}\ \bibnamefont
  {Bryon}}, \bibinfo {author} {\bibfnamefont {Andrei}\ \bibnamefont
  {Vrajitoarea}}, \bibinfo {author} {\bibfnamefont {Sara}\ \bibnamefont
  {Sussman}}, \bibinfo {author} {\bibfnamefont {Guangming}\ \bibnamefont
  {Cheng}}, \bibinfo {author} {\bibfnamefont {Trisha}\ \bibnamefont
  {Madhavan}}, \bibinfo {author} {\bibfnamefont {Harshvardhan~K.}\ \bibnamefont
  {Babla}}, \bibinfo {author} {\bibfnamefont {Xuan~Hoang}\ \bibnamefont {Le}},
  \bibinfo {author} {\bibfnamefont {Youqi}\ \bibnamefont {Gang}}, \bibinfo
  {author} {\bibfnamefont {Berthold}\ \bibnamefont {Jäck}}, \bibinfo {author}
  {\bibfnamefont {András}\ \bibnamefont {Gyenis}}, \bibinfo {author}
  {\bibfnamefont {Nan}\ \bibnamefont {Yao}}, \bibinfo {author} {\bibfnamefont
  {Robert~J.}\ \bibnamefont {Cava}}, \bibinfo {author} {\bibfnamefont
  {Nathalie~P.}\ \bibnamefont {de~Leon}}, \ and\ \bibinfo {author}
  {\bibfnamefont {Andrew~A.}\ \bibnamefont {Houck}},\ }\bibfield  {title}
  {\enquote {\bibinfo {title} {New material platform for superconducting
  transmon qubits with coherence times exceeding 0.3 milliseconds},}\ }\href
  {\doibase 10.1038/s41467-021-22030-5} {\bibfield  {journal} {\bibinfo
  {journal} {Nature Communications}\ }\textbf {\bibinfo {volume} {12}},\
  \bibinfo {pages} {1779} (\bibinfo {year} {2021})}\BibitemShut {NoStop}%
\bibitem [{\citenamefont {Jurcevic}\ \emph {et~al.}(2021)\citenamefont
  {Jurcevic}, \citenamefont {Javadi-Abhari}, \citenamefont {Bishop},
  \citenamefont {Lauer}, \citenamefont {Bogorin}, \citenamefont {Brink},
  \citenamefont {Capelluto}, \citenamefont {Günlük}, \citenamefont {Itoko},
  \citenamefont {Kanazawa}, \citenamefont {Kandala}, \citenamefont {Keefe},
  \citenamefont {Krsulich}, \citenamefont {Landers}, \citenamefont
  {Lewandowski}, \citenamefont {McClure}, \citenamefont {Nannicini},
  \citenamefont {Narasgond}, \citenamefont {Nayfeh}, \citenamefont {Pritchett},
  \citenamefont {Rothwell}, \citenamefont {Srinivasan}, \citenamefont
  {Sundaresan}, \citenamefont {Wang}, \citenamefont {Wei}, \citenamefont
  {Wood}, \citenamefont {Yau}, \citenamefont {Zhang}, \citenamefont {Dial},
  \citenamefont {Chow},\ and\ \citenamefont
  {Gambetta}}]{jurcevic2021demonstration}%
  \BibitemOpen
  \bibfield  {author} {\bibinfo {author} {\bibfnamefont {Petar}\ \bibnamefont
  {Jurcevic}}, \bibinfo {author} {\bibfnamefont {Ali}\ \bibnamefont
  {Javadi-Abhari}}, \bibinfo {author} {\bibfnamefont {Lev~S.}\ \bibnamefont
  {Bishop}}, \bibinfo {author} {\bibfnamefont {Isaac}\ \bibnamefont {Lauer}},
  \bibinfo {author} {\bibfnamefont {Daniela~F.}\ \bibnamefont {Bogorin}},
  \bibinfo {author} {\bibfnamefont {Markus}\ \bibnamefont {Brink}}, \bibinfo
  {author} {\bibfnamefont {Lauren}\ \bibnamefont {Capelluto}}, \bibinfo
  {author} {\bibfnamefont {Oktay}\ \bibnamefont {Günlük}}, \bibinfo {author}
  {\bibfnamefont {Toshinari}\ \bibnamefont {Itoko}}, \bibinfo {author}
  {\bibfnamefont {Naoki}\ \bibnamefont {Kanazawa}}, \bibinfo {author}
  {\bibfnamefont {Abhinav}\ \bibnamefont {Kandala}}, \bibinfo {author}
  {\bibfnamefont {George~A.}\ \bibnamefont {Keefe}}, \bibinfo {author}
  {\bibfnamefont {Kevin}\ \bibnamefont {Krsulich}}, \bibinfo {author}
  {\bibfnamefont {William}\ \bibnamefont {Landers}}, \bibinfo {author}
  {\bibfnamefont {Eric~P.}\ \bibnamefont {Lewandowski}}, \bibinfo {author}
  {\bibfnamefont {Douglas~T}\ \bibnamefont {McClure}}, \bibinfo {author}
  {\bibfnamefont {Giacomo}\ \bibnamefont {Nannicini}}, \bibinfo {author}
  {\bibfnamefont {Adinath}\ \bibnamefont {Narasgond}}, \bibinfo {author}
  {\bibfnamefont {Hasan~M.}\ \bibnamefont {Nayfeh}}, \bibinfo {author}
  {\bibfnamefont {Emily}\ \bibnamefont {Pritchett}}, \bibinfo {author}
  {\bibfnamefont {Mary~Beth}\ \bibnamefont {Rothwell}}, \bibinfo {author}
  {\bibfnamefont {Srikanth}\ \bibnamefont {Srinivasan}}, \bibinfo {author}
  {\bibfnamefont {Neereja}\ \bibnamefont {Sundaresan}}, \bibinfo {author}
  {\bibfnamefont {Cindy}\ \bibnamefont {Wang}}, \bibinfo {author}
  {\bibfnamefont {Ken~X.}\ \bibnamefont {Wei}}, \bibinfo {author}
  {\bibfnamefont {Christopher~J.}\ \bibnamefont {Wood}}, \bibinfo {author}
  {\bibfnamefont {Jeng-Bang}\ \bibnamefont {Yau}}, \bibinfo {author}
  {\bibfnamefont {Eric~J.}\ \bibnamefont {Zhang}}, \bibinfo {author}
  {\bibfnamefont {Oliver~E.}\ \bibnamefont {Dial}}, \bibinfo {author}
  {\bibfnamefont {Jerry~M.}\ \bibnamefont {Chow}}, \ and\ \bibinfo {author}
  {\bibfnamefont {Jay~M.}\ \bibnamefont {Gambetta}},\ }\bibfield  {title}
  {\enquote {\bibinfo {title} {Demonstration of quantum volume 64 on a
  superconducting quantum computing system},}\ }\href {\doibase
  10.1088/2058-9565/abe519} {\bibfield  {journal} {\bibinfo  {journal} {Quantum
  Science and Technology}\ }\textbf {\bibinfo {volume} {6}},\ \bibinfo {pages}
  {025020} (\bibinfo {year} {2021})}\BibitemShut {NoStop}%
\bibitem [{\citenamefont {Bauer}\ \emph {et~al.}(2020)\citenamefont {Bauer},
  \citenamefont {Bravyi}, \citenamefont {Motta},\ and\ \citenamefont
  {Chan}}]{bauer2020quantum}%
  \BibitemOpen
  \bibfield  {author} {\bibinfo {author} {\bibfnamefont {Bela}\ \bibnamefont
  {Bauer}}, \bibinfo {author} {\bibfnamefont {Sergey}\ \bibnamefont {Bravyi}},
  \bibinfo {author} {\bibfnamefont {Mario}\ \bibnamefont {Motta}}, \ and\
  \bibinfo {author} {\bibfnamefont {Garnet Kin-Lic}\ \bibnamefont {Chan}},\
  }\bibfield  {title} {\enquote {\bibinfo {title} {Quantum {algorithms} for
  {quantum} {chemistry} and {quantum} {materials} {science}},}\ }\href
  {\doibase 10.1021/acs.chemrev.9b00829} {\bibfield  {journal} {\bibinfo
  {journal} {Chemical Reviews}\ }\textbf {\bibinfo {volume} {120}},\ \bibinfo
  {pages} {12685--12717} (\bibinfo {year} {2020})}\BibitemShut {NoStop}%
\bibitem [{\citenamefont {McArdle}\ \emph {et~al.}(2020)\citenamefont
  {McArdle}, \citenamefont {Endo}, \citenamefont {Aspuru-Guzik}, \citenamefont
  {Benjamin},\ and\ \citenamefont {Yuan}}]{mcardle2020quantum}%
  \BibitemOpen
  \bibfield  {author} {\bibinfo {author} {\bibfnamefont {Sam}\ \bibnamefont
  {McArdle}}, \bibinfo {author} {\bibfnamefont {Suguru}\ \bibnamefont {Endo}},
  \bibinfo {author} {\bibfnamefont {Alán}\ \bibnamefont {Aspuru-Guzik}},
  \bibinfo {author} {\bibfnamefont {Simon~C.}\ \bibnamefont {Benjamin}}, \ and\
  \bibinfo {author} {\bibfnamefont {Xiao}\ \bibnamefont {Yuan}},\ }\bibfield
  {title} {\enquote {\bibinfo {title} {Quantum computational chemistry},}\
  }\href {\doibase 10.1103/RevModPhys.92.015003} {\bibfield  {journal}
  {\bibinfo  {journal} {Reviews of Modern Physics}\ }\textbf {\bibinfo {volume}
  {92}},\ \bibinfo {pages} {015003} (\bibinfo {year} {2020})}\BibitemShut
  {NoStop}%
\bibitem [{\citenamefont {Motta}\ and\ \citenamefont
  {Rice}(2021)}]{motta2021emerging}%
  \BibitemOpen
  \bibfield  {author} {\bibinfo {author} {\bibfnamefont {Mario}\ \bibnamefont
  {Motta}}\ and\ \bibinfo {author} {\bibfnamefont {Julia~E.}\ \bibnamefont
  {Rice}},\ }\bibfield  {title} {\enquote {\bibinfo {title} {Emerging quantum
  computing algorithms for quantum chemistry},}\ }\href
  {https://onlinelibrary.wiley.com/doi/10.1002/wcms.1580} {\bibfield  {journal}
  {\bibinfo  {journal} {WIREs Computational Molecular Science; e1580}\ }
  (\bibinfo {year} {2021})}\BibitemShut {NoStop}%
\bibitem [{\citenamefont {Ollitrault}\ \emph
  {et~al.}(2020{\natexlab{a}})\citenamefont {Ollitrault}, \citenamefont
  {Kandala}, \citenamefont {Chen}, \citenamefont {Barkoutsos}, \citenamefont
  {Mezzacapo}, \citenamefont {Pistoia}, \citenamefont {Sheldon}, \citenamefont
  {Woerner}, \citenamefont {Gambetta},\ and\ \citenamefont
  {Tavernelli}}]{ollitrault2020quantum}%
  \BibitemOpen
  \bibfield  {author} {\bibinfo {author} {\bibfnamefont {Pauline~J.}\
  \bibnamefont {Ollitrault}}, \bibinfo {author} {\bibfnamefont {Abhinav}\
  \bibnamefont {Kandala}}, \bibinfo {author} {\bibfnamefont {Chun-Fu}\
  \bibnamefont {Chen}}, \bibinfo {author} {\bibfnamefont {Panagiotis~Kl.}\
  \bibnamefont {Barkoutsos}}, \bibinfo {author} {\bibfnamefont {Antonio}\
  \bibnamefont {Mezzacapo}}, \bibinfo {author} {\bibfnamefont {Marco}\
  \bibnamefont {Pistoia}}, \bibinfo {author} {\bibfnamefont {Sarah}\
  \bibnamefont {Sheldon}}, \bibinfo {author} {\bibfnamefont {Stefan}\
  \bibnamefont {Woerner}}, \bibinfo {author} {\bibfnamefont {Jay~M.}\
  \bibnamefont {Gambetta}}, \ and\ \bibinfo {author} {\bibfnamefont {Ivano}\
  \bibnamefont {Tavernelli}},\ }\bibfield  {title} {\enquote {\bibinfo {title}
  {Quantum equation of motion for computing molecular excitation energies on a
  noisy quantum processor},}\ }\href {\doibase
  10.1103/PhysRevResearch.2.043140} {\bibfield  {journal} {\bibinfo  {journal}
  {Physical Review Research}\ }\textbf {\bibinfo {volume} {2}},\ \bibinfo
  {pages} {043140} (\bibinfo {year} {2020}{\natexlab{a}})}\BibitemShut
  {NoStop}%
\bibitem [{\citenamefont {Sokolov}\ \emph {et~al.}(2020)\citenamefont
  {Sokolov}, \citenamefont {Barkoutsos}, \citenamefont {Ollitrault},
  \citenamefont {Greenberg}, \citenamefont {Rice}, \citenamefont {Pistoia},\
  and\ \citenamefont {Tavernelli}}]{sokolov2020quantum}%
  \BibitemOpen
  \bibfield  {author} {\bibinfo {author} {\bibfnamefont {Igor~O.}\ \bibnamefont
  {Sokolov}}, \bibinfo {author} {\bibfnamefont {Panagiotis~Kl.}\ \bibnamefont
  {Barkoutsos}}, \bibinfo {author} {\bibfnamefont {Pauline~J.}\ \bibnamefont
  {Ollitrault}}, \bibinfo {author} {\bibfnamefont {Donny}\ \bibnamefont
  {Greenberg}}, \bibinfo {author} {\bibfnamefont {Julia}\ \bibnamefont {Rice}},
  \bibinfo {author} {\bibfnamefont {Marco}\ \bibnamefont {Pistoia}}, \ and\
  \bibinfo {author} {\bibfnamefont {Ivano}\ \bibnamefont {Tavernelli}},\
  }\bibfield  {title} {\enquote {\bibinfo {title} {Quantum orbital-optimized
  unitary coupled cluster methods in the strongly correlated regime: {Can}
  quantum algorithms outperform their classical equivalents?}}\ }\href
  {\doibase 10.1063/1.5141835} {\bibfield  {journal} {\bibinfo  {journal} {The
  Journal of Chemical Physics}\ }\textbf {\bibinfo {volume} {152}},\ \bibinfo
  {pages} {124107} (\bibinfo {year} {2020})}\BibitemShut {NoStop}%
\bibitem [{\citenamefont {Ollitrault}\ \emph
  {et~al.}(2020{\natexlab{b}})\citenamefont {Ollitrault}, \citenamefont
  {Baiardi}, \citenamefont {Reiher},\ and\ \citenamefont
  {Tavernelli}}]{ollitrault2020hardware}%
  \BibitemOpen
  \bibfield  {author} {\bibinfo {author} {\bibfnamefont {Pauline~J.}\
  \bibnamefont {Ollitrault}}, \bibinfo {author} {\bibfnamefont {Alberto}\
  \bibnamefont {Baiardi}}, \bibinfo {author} {\bibfnamefont {Markus}\
  \bibnamefont {Reiher}}, \ and\ \bibinfo {author} {\bibfnamefont {Ivano}\
  \bibnamefont {Tavernelli}},\ }\bibfield  {title} {\enquote {\bibinfo {title}
  {Hardware efficient quantum algorithms for vibrational structure
  calculations},}\ }\href {\doibase 10.1039/D0SC01908A} {\bibfield  {journal}
  {\bibinfo  {journal} {Chemical Science}\ }\textbf {\bibinfo {volume} {11}},\
  \bibinfo {pages} {6842--6855} (\bibinfo {year}
  {2020}{\natexlab{b}})}\BibitemShut {NoStop}%
\bibitem [{\citenamefont {Kühn}\ \emph {et~al.}(2019)\citenamefont {Kühn},
  \citenamefont {Zanker}, \citenamefont {Deglmann}, \citenamefont {Marthaler},\
  and\ \citenamefont {Weiß}}]{kuhn2019accuracy}%
  \BibitemOpen
  \bibfield  {author} {\bibinfo {author} {\bibfnamefont {Michael}\ \bibnamefont
  {Kühn}}, \bibinfo {author} {\bibfnamefont {Sebastian}\ \bibnamefont
  {Zanker}}, \bibinfo {author} {\bibfnamefont {Peter}\ \bibnamefont
  {Deglmann}}, \bibinfo {author} {\bibfnamefont {Michael}\ \bibnamefont
  {Marthaler}}, \ and\ \bibinfo {author} {\bibfnamefont {Horst}\ \bibnamefont
  {Weiß}},\ }\bibfield  {title} {\enquote {\bibinfo {title} {Accuracy and
  resource estimations for quantum chemistry on a near-term quantum
  computer},}\ }\href {\doibase 10.1021/acs.jctc.9b00236} {\bibfield  {journal}
  {\bibinfo  {journal} {Journal of Chemical Theory and Computation}\ }\textbf
  {\bibinfo {volume} {15}},\ \bibinfo {pages} {4764--4780} (\bibinfo {year}
  {2019})}\BibitemShut {NoStop}%
\bibitem [{\citenamefont {Robert}\ \emph {et~al.}(2021)\citenamefont {Robert},
  \citenamefont {Barkoutsos}, \citenamefont {Woerner},\ and\ \citenamefont
  {Tavernelli}}]{robert2021resourceefficient}%
  \BibitemOpen
  \bibfield  {author} {\bibinfo {author} {\bibfnamefont {Anton}\ \bibnamefont
  {Robert}}, \bibinfo {author} {\bibfnamefont {Panagiotis~Kl.}\ \bibnamefont
  {Barkoutsos}}, \bibinfo {author} {\bibfnamefont {Stefan}\ \bibnamefont
  {Woerner}}, \ and\ \bibinfo {author} {\bibfnamefont {Ivano}\ \bibnamefont
  {Tavernelli}},\ }\bibfield  {title} {\enquote {\bibinfo {title}
  {Resource-efficient quantum algorithm for protein folding},}\ }\href
  {\doibase 10.1038/s41534-021-00368-4} {\bibfield  {journal} {\bibinfo
  {journal} {npj Quantum Information}\ }\textbf {\bibinfo {volume} {7}},\
  \bibinfo {pages} {38} (\bibinfo {year} {2021})}\BibitemShut {NoStop}%
\bibitem [{\citenamefont {Kandala}\ \emph {et~al.}(2017)\citenamefont
  {Kandala}, \citenamefont {Mezzacapo}, \citenamefont {Temme}, \citenamefont
  {Takita}, \citenamefont {Brink}, \citenamefont {Chow},\ and\ \citenamefont
  {Gambetta}}]{kandala2017hardwareefficient}%
  \BibitemOpen
  \bibfield  {author} {\bibinfo {author} {\bibfnamefont {Abhinav}\ \bibnamefont
  {Kandala}}, \bibinfo {author} {\bibfnamefont {Antonio}\ \bibnamefont
  {Mezzacapo}}, \bibinfo {author} {\bibfnamefont {Kristan}\ \bibnamefont
  {Temme}}, \bibinfo {author} {\bibfnamefont {Maika}\ \bibnamefont {Takita}},
  \bibinfo {author} {\bibfnamefont {Markus}\ \bibnamefont {Brink}}, \bibinfo
  {author} {\bibfnamefont {Jerry~M.}\ \bibnamefont {Chow}}, \ and\ \bibinfo
  {author} {\bibfnamefont {Jay~M.}\ \bibnamefont {Gambetta}},\ }\bibfield
  {title} {\enquote {\bibinfo {title} {Hardware-efficient variational quantum
  eigensolver for small molecules and quantum magnets},}\ }\href {\doibase
  10.1038/nature23879} {\bibfield  {journal} {\bibinfo  {journal} {Nature}\
  }\textbf {\bibinfo {volume} {549}},\ \bibinfo {pages} {242--246} (\bibinfo
  {year} {2017})}\BibitemShut {NoStop}%
\bibitem [{\citenamefont {Hempel}\ \emph {et~al.}(2018)\citenamefont {Hempel},
  \citenamefont {Maier}, \citenamefont {Romero}, \citenamefont {McClean},
  \citenamefont {Monz}, \citenamefont {Shen}, \citenamefont {Jurcevic},
  \citenamefont {Lanyon}, \citenamefont {Love}, \citenamefont {Babbush},
  \citenamefont {Aspuru-Guzik}, \citenamefont {Blatt},\ and\ \citenamefont
  {Roos}}]{hempel2018quantum}%
  \BibitemOpen
  \bibfield  {author} {\bibinfo {author} {\bibfnamefont {Cornelius}\
  \bibnamefont {Hempel}}, \bibinfo {author} {\bibfnamefont {Christine}\
  \bibnamefont {Maier}}, \bibinfo {author} {\bibfnamefont {Jonathan}\
  \bibnamefont {Romero}}, \bibinfo {author} {\bibfnamefont {Jarrod}\
  \bibnamefont {McClean}}, \bibinfo {author} {\bibfnamefont {Thomas}\
  \bibnamefont {Monz}}, \bibinfo {author} {\bibfnamefont {Heng}\ \bibnamefont
  {Shen}}, \bibinfo {author} {\bibfnamefont {Petar}\ \bibnamefont {Jurcevic}},
  \bibinfo {author} {\bibfnamefont {Ben~P.}\ \bibnamefont {Lanyon}}, \bibinfo
  {author} {\bibfnamefont {Peter}\ \bibnamefont {Love}}, \bibinfo {author}
  {\bibfnamefont {Ryan}\ \bibnamefont {Babbush}}, \bibinfo {author}
  {\bibfnamefont {Alán}\ \bibnamefont {Aspuru-Guzik}}, \bibinfo {author}
  {\bibfnamefont {Rainer}\ \bibnamefont {Blatt}}, \ and\ \bibinfo {author}
  {\bibfnamefont {Christian~F.}\ \bibnamefont {Roos}},\ }\bibfield  {title}
  {\enquote {\bibinfo {title} {Quantum chemistry calculations on a trapped-ion
  quantum simulator},}\ }\href {\doibase 10.1103/PhysRevX.8.031022} {\bibfield
  {journal} {\bibinfo  {journal} {Physical Review X}\ }\textbf {\bibinfo
  {volume} {8}},\ \bibinfo {pages} {031022} (\bibinfo {year}
  {2018})}\BibitemShut {NoStop}%
\bibitem [{\citenamefont {Lanyon}\ \emph {et~al.}(2010)\citenamefont {Lanyon},
  \citenamefont {Whitfield}, \citenamefont {Gillett}, \citenamefont {Goggin},
  \citenamefont {Almeida}, \citenamefont {Kassal}, \citenamefont {Biamonte},
  \citenamefont {Mohseni}, \citenamefont {Powell}, \citenamefont {Barbieri},
  \citenamefont {Aspuru-Guzik},\ and\ \citenamefont
  {White}}]{lanyon2010quantum}%
  \BibitemOpen
  \bibfield  {author} {\bibinfo {author} {\bibfnamefont {Benjamin~P.}\
  \bibnamefont {Lanyon}}, \bibinfo {author} {\bibfnamefont {James~D.}\
  \bibnamefont {Whitfield}}, \bibinfo {author} {\bibfnamefont {Geoff~G.}\
  \bibnamefont {Gillett}}, \bibinfo {author} {\bibfnamefont {Michael~E.}\
  \bibnamefont {Goggin}}, \bibinfo {author} {\bibfnamefont {Marcelo~P.}\
  \bibnamefont {Almeida}}, \bibinfo {author} {\bibfnamefont {Ivan}\
  \bibnamefont {Kassal}}, \bibinfo {author} {\bibfnamefont {Jacob~D.}\
  \bibnamefont {Biamonte}}, \bibinfo {author} {\bibfnamefont {Masoud}\
  \bibnamefont {Mohseni}}, \bibinfo {author} {\bibfnamefont {Ben~J.}\
  \bibnamefont {Powell}}, \bibinfo {author} {\bibfnamefont {Marco}\
  \bibnamefont {Barbieri}}, \bibinfo {author} {\bibfnamefont {Alán}\
  \bibnamefont {Aspuru-Guzik}}, \ and\ \bibinfo {author} {\bibfnamefont
  {Andrew~G.}\ \bibnamefont {White}},\ }\bibfield  {title} {\enquote {\bibinfo
  {title} {Towards quantum chemistry on a quantum computer},}\ }\href {\doibase
  10.1038/nchem.483} {\bibfield  {journal} {\bibinfo  {journal} {Nature
  Chemistry}\ }\textbf {\bibinfo {volume} {2}},\ \bibinfo {pages} {106--111}
  (\bibinfo {year} {2010})}\BibitemShut {NoStop}%
\bibitem [{\citenamefont {McClean}\ \emph {et~al.}(2016)\citenamefont
  {McClean}, \citenamefont {Romero}, \citenamefont {Babbush},\ and\
  \citenamefont {Aspuru-Guzik}}]{mcclean2016theory}%
  \BibitemOpen
  \bibfield  {author} {\bibinfo {author} {\bibfnamefont {Jarrod~R.}\
  \bibnamefont {McClean}}, \bibinfo {author} {\bibfnamefont {Jonathan}\
  \bibnamefont {Romero}}, \bibinfo {author} {\bibfnamefont {Ryan}\ \bibnamefont
  {Babbush}}, \ and\ \bibinfo {author} {\bibfnamefont {Alán}\ \bibnamefont
  {Aspuru-Guzik}},\ }\bibfield  {title} {\enquote {\bibinfo {title} {The theory
  of variational hybrid quantum-classical algorithms},}\ }\href {\doibase
  10.1088/1367-2630/18/2/023023} {\bibfield  {journal} {\bibinfo  {journal}
  {New Journal of Physics}\ }\textbf {\bibinfo {volume} {18}},\ \bibinfo
  {pages} {023023} (\bibinfo {year} {2016})}\BibitemShut {NoStop}%
\bibitem [{\citenamefont {Cerezo}\ \emph {et~al.}(2021)\citenamefont {Cerezo},
  \citenamefont {Arrasmith}, \citenamefont {Babbush}, \citenamefont {Benjamin},
  \citenamefont {Endo}, \citenamefont {Fujii}, \citenamefont {McClean},
  \citenamefont {Mitarai}, \citenamefont {Yuan}, \citenamefont {Cincio},\ and\
  \citenamefont {Coles}}]{cerezo2021variational}%
  \BibitemOpen
  \bibfield  {author} {\bibinfo {author} {\bibfnamefont {Marco}\ \bibnamefont
  {Cerezo}}, \bibinfo {author} {\bibfnamefont {Andrew}\ \bibnamefont
  {Arrasmith}}, \bibinfo {author} {\bibfnamefont {Ryan}\ \bibnamefont
  {Babbush}}, \bibinfo {author} {\bibfnamefont {Simon~C.}\ \bibnamefont
  {Benjamin}}, \bibinfo {author} {\bibfnamefont {Suguru}\ \bibnamefont {Endo}},
  \bibinfo {author} {\bibfnamefont {Keisuke}\ \bibnamefont {Fujii}}, \bibinfo
  {author} {\bibfnamefont {Jarrod~R.}\ \bibnamefont {McClean}}, \bibinfo
  {author} {\bibfnamefont {Kosuke}\ \bibnamefont {Mitarai}}, \bibinfo {author}
  {\bibfnamefont {Xiao}\ \bibnamefont {Yuan}}, \bibinfo {author} {\bibfnamefont
  {Lukasz}\ \bibnamefont {Cincio}}, \ and\ \bibinfo {author} {\bibfnamefont
  {Patrick~J.}\ \bibnamefont {Coles}},\ }\bibfield  {title} {\enquote {\bibinfo
  {title} {Variational quantum algorithms},}\ }\href {\doibase
  10.1038/s42254-021-00348-9} {\bibfield  {journal} {\bibinfo  {journal}
  {Nature Reviews Physics}\ }\textbf {\bibinfo {volume} {3}},\ \bibinfo {pages}
  {625--644} (\bibinfo {year} {2021})}\BibitemShut {NoStop}%
\bibitem [{\citenamefont {Peruzzo}\ \emph {et~al.}(2014)\citenamefont
  {Peruzzo}, \citenamefont {McClean}, \citenamefont {Shadbolt}, \citenamefont
  {Yung}, \citenamefont {Zhou}, \citenamefont {Love}, \citenamefont
  {Aspuru-Guzik},\ and\ \citenamefont {O’Brien}}]{peruzzo2014variational}%
  \BibitemOpen
  \bibfield  {author} {\bibinfo {author} {\bibfnamefont {Alberto}\ \bibnamefont
  {Peruzzo}}, \bibinfo {author} {\bibfnamefont {Jarrod}\ \bibnamefont
  {McClean}}, \bibinfo {author} {\bibfnamefont {Peter}\ \bibnamefont
  {Shadbolt}}, \bibinfo {author} {\bibfnamefont {Man-Hong}\ \bibnamefont
  {Yung}}, \bibinfo {author} {\bibfnamefont {Xiao-Qi}\ \bibnamefont {Zhou}},
  \bibinfo {author} {\bibfnamefont {Peter~J.}\ \bibnamefont {Love}}, \bibinfo
  {author} {\bibfnamefont {Alán}\ \bibnamefont {Aspuru-Guzik}}, \ and\
  \bibinfo {author} {\bibfnamefont {Jeremy~L.}\ \bibnamefont {O’Brien}},\
  }\bibfield  {title} {\enquote {\bibinfo {title} {A variational eigenvalue
  solver on a photonic quantum processor},}\ }\href {\doibase
  10.1038/ncomms5213} {\bibfield  {journal} {\bibinfo  {journal} {Nature
  Communications}\ }\textbf {\bibinfo {volume} {5}},\ \bibinfo {pages} {4213}
  (\bibinfo {year} {2014})}\BibitemShut {NoStop}%
\bibitem [{\citenamefont {Wang}\ \emph
  {et~al.}(2021{\natexlab{a}})\citenamefont {Wang}, \citenamefont {Fontana},
  \citenamefont {Cerezo}, \citenamefont {Sharma}, \citenamefont {Sone},
  \citenamefont {Cincio},\ and\ \citenamefont {Coles}}]{wang2021noiseinduced}%
  \BibitemOpen
  \bibfield  {author} {\bibinfo {author} {\bibfnamefont {Samson}\ \bibnamefont
  {Wang}}, \bibinfo {author} {\bibfnamefont {Enrico}\ \bibnamefont {Fontana}},
  \bibinfo {author} {\bibfnamefont {Marco}\ \bibnamefont {Cerezo}}, \bibinfo
  {author} {\bibfnamefont {Kunal}\ \bibnamefont {Sharma}}, \bibinfo {author}
  {\bibfnamefont {Akira}\ \bibnamefont {Sone}}, \bibinfo {author}
  {\bibfnamefont {Lukasz}\ \bibnamefont {Cincio}}, \ and\ \bibinfo {author}
  {\bibfnamefont {Patrick~J.}\ \bibnamefont {Coles}},\ }\bibfield  {title}
  {\enquote {\bibinfo {title} {Noise-induced barren plateaus in variational
  quantum algorithms},}\ }\href {\doibase 10.1038/s41467-021-27045-6}
  {\bibfield  {journal} {\bibinfo  {journal} {Nature Communications}\ }\textbf
  {\bibinfo {volume} {12}},\ \bibinfo {pages} {6961} (\bibinfo {year}
  {2021}{\natexlab{a}})}\BibitemShut {NoStop}%
\bibitem [{\citenamefont {Eddins}\ \emph {et~al.}(2022)\citenamefont {Eddins},
  \citenamefont {Motta}, \citenamefont {Gujarati}, \citenamefont {Bravyi},
  \citenamefont {Mezzacapo}, \citenamefont {Hadfield},\ and\ \citenamefont
  {Sheldon}}]{Eddins2022}%
  \BibitemOpen
  \bibfield  {author} {\bibinfo {author} {\bibfnamefont {Andrew}\ \bibnamefont
  {Eddins}}, \bibinfo {author} {\bibfnamefont {Mario}\ \bibnamefont {Motta}},
  \bibinfo {author} {\bibfnamefont {Tanvi~P.}\ \bibnamefont {Gujarati}},
  \bibinfo {author} {\bibfnamefont {Sergey}\ \bibnamefont {Bravyi}}, \bibinfo
  {author} {\bibfnamefont {Antonio}\ \bibnamefont {Mezzacapo}}, \bibinfo
  {author} {\bibfnamefont {Charles}\ \bibnamefont {Hadfield}}, \ and\ \bibinfo
  {author} {\bibfnamefont {Sarah}\ \bibnamefont {Sheldon}},\ }\bibfield
  {title} {\enquote {\bibinfo {title} {Doubling the size of quantum simulators
  by entanglement forging},}\ }\href {\doibase 10.1103/PRXQuantum.3.010309}
  {\bibfield  {journal} {\bibinfo  {journal} {PRX Quantum}\ }\textbf {\bibinfo
  {volume} {3}},\ \bibinfo {pages} {010309} (\bibinfo {year}
  {2022})}\BibitemShut {NoStop}%
\bibitem [{\citenamefont {Holmes}\ \emph {et~al.}(2022)\citenamefont {Holmes},
  \citenamefont {Sharma}, \citenamefont {Cerezo},\ and\ \citenamefont
  {Coles}}]{holmes2022connecting}%
  \BibitemOpen
  \bibfield  {author} {\bibinfo {author} {\bibfnamefont {Zoë}\ \bibnamefont
  {Holmes}}, \bibinfo {author} {\bibfnamefont {Kunal}\ \bibnamefont {Sharma}},
  \bibinfo {author} {\bibfnamefont {Marco}\ \bibnamefont {Cerezo}}, \ and\
  \bibinfo {author} {\bibfnamefont {Patrick~J.}\ \bibnamefont {Coles}},\
  }\bibfield  {title} {\enquote {\bibinfo {title} {Connecting ansatz
  expressibility to gradient magnitudes and barren plateaus},}\ }\href
  {\doibase 10.1103/PRXQuantum.3.010313} {\bibfield  {journal} {\bibinfo
  {journal} {PRX Quantum}\ }\textbf {\bibinfo {volume} {3}},\ \bibinfo {pages}
  {010313} (\bibinfo {year} {2022})}\BibitemShut {NoStop}%
\bibitem [{\citenamefont {Wang}\ \emph
  {et~al.}(2021{\natexlab{b}})\citenamefont {Wang}, \citenamefont {Czarnik},
  \citenamefont {Arrasmith}, \citenamefont {Cerezo}, \citenamefont {Cincio},\
  and\ \citenamefont {Coles}}]{wang2021error}%
  \BibitemOpen
  \bibfield  {author} {\bibinfo {author} {\bibfnamefont {Samson}\ \bibnamefont
  {Wang}}, \bibinfo {author} {\bibfnamefont {Piotr}\ \bibnamefont {Czarnik}},
  \bibinfo {author} {\bibfnamefont {Andrew}\ \bibnamefont {Arrasmith}},
  \bibinfo {author} {\bibfnamefont {M.}~\bibnamefont {Cerezo}}, \bibinfo
  {author} {\bibfnamefont {Lukasz}\ \bibnamefont {Cincio}}, \ and\ \bibinfo
  {author} {\bibfnamefont {Patrick~J.}\ \bibnamefont {Coles}},\ }\href@noop {}
  {\enquote {\bibinfo {title} {Can error mitigation improve trainability of
  noisy variational quantum algorithms?}}\ } (\bibinfo {year}
  {2021}{\natexlab{b}}),\ \Eprint {http://arxiv.org/abs/2109.01051}
  {arXiv:2109.01051 [quant-ph]} \BibitemShut {NoStop}%
\bibitem [{\citenamefont {Gonthier}\ \emph {et~al.}(2020)\citenamefont
  {Gonthier}, \citenamefont {Radin}, \citenamefont {Buda}, \citenamefont
  {Doskocil}, \citenamefont {Abuan},\ and\ \citenamefont
  {Romero}}]{gonthier2020identifying}%
  \BibitemOpen
  \bibfield  {author} {\bibinfo {author} {\bibfnamefont {Jérôme~F.}\
  \bibnamefont {Gonthier}}, \bibinfo {author} {\bibfnamefont {Maxwell~D.}\
  \bibnamefont {Radin}}, \bibinfo {author} {\bibfnamefont {Corneliu}\
  \bibnamefont {Buda}}, \bibinfo {author} {\bibfnamefont {Eric~J.}\
  \bibnamefont {Doskocil}}, \bibinfo {author} {\bibfnamefont {Clena~M.}\
  \bibnamefont {Abuan}}, \ and\ \bibinfo {author} {\bibfnamefont {Jhonathan}\
  \bibnamefont {Romero}},\ }\bibfield  {title} {\enquote {\bibinfo {title}
  {Identifying challenges towards practical quantum advantage through resource
  estimation: the measurement roadblock in the variational quantum
  eigensolver},}\ }\href {http://arxiv.org/abs/2012.04001} {\bibfield
  {journal} {\bibinfo  {journal} {arXiv:2012.04001 [quant-ph]}\ } (\bibinfo
  {year} {2020})}\BibitemShut {NoStop}%
\bibitem [{\citenamefont {Wecker}\ \emph {et~al.}(2015)\citenamefont {Wecker},
  \citenamefont {Hastings},\ and\ \citenamefont {Troyer}}]{Wecker2015}%
  \BibitemOpen
  \bibfield  {author} {\bibinfo {author} {\bibfnamefont {Dave}\ \bibnamefont
  {Wecker}}, \bibinfo {author} {\bibfnamefont {Matthew~B.}\ \bibnamefont
  {Hastings}}, \ and\ \bibinfo {author} {\bibfnamefont {Matthias}\ \bibnamefont
  {Troyer}},\ }\bibfield  {title} {\enquote {\bibinfo {title} {Progress towards
  practical quantum variational algorithms},}\ }\href {\doibase
  10.1103/PhysRevA.92.042303} {\bibfield  {journal} {\bibinfo  {journal}
  {Physical Review A}\ }\textbf {\bibinfo {volume} {92}},\ \bibinfo {pages}
  {042303} (\bibinfo {year} {2015})}\BibitemShut {NoStop}%
\bibitem [{\citenamefont {Wack}\ \emph {et~al.}(2021)\citenamefont {Wack},
  \citenamefont {Paik}, \citenamefont {Javadi-Abhari}, \citenamefont
  {Jurcevic}, \citenamefont {Faro}, \citenamefont {Gambetta},\ and\
  \citenamefont {Johnson}}]{wack2021quality}%
  \BibitemOpen
  \bibfield  {author} {\bibinfo {author} {\bibfnamefont {Andrew}\ \bibnamefont
  {Wack}}, \bibinfo {author} {\bibfnamefont {Hanhee}\ \bibnamefont {Paik}},
  \bibinfo {author} {\bibfnamefont {Ali}\ \bibnamefont {Javadi-Abhari}},
  \bibinfo {author} {\bibfnamefont {Petar}\ \bibnamefont {Jurcevic}}, \bibinfo
  {author} {\bibfnamefont {Ismael}\ \bibnamefont {Faro}}, \bibinfo {author}
  {\bibfnamefont {Jay~M.}\ \bibnamefont {Gambetta}}, \ and\ \bibinfo {author}
  {\bibfnamefont {Blake~R.}\ \bibnamefont {Johnson}},\ }\bibfield  {title}
  {\enquote {\bibinfo {title} {Quality, {Speed}, and {Scale}: three key
  attributes to measure the performance of near-term quantum computers},}\
  }\href {http://arxiv.org/abs/2110.14108} {\bibfield  {journal} {\bibinfo
  {journal} {arXiv:2110.14108 [quant-ph]}\ } (\bibinfo {year}
  {2021})}\BibitemShut {NoStop}%
\bibitem [{\citenamefont {Gokhale}\ \emph {et~al.}(2020)\citenamefont
  {Gokhale}, \citenamefont {Angiuli}, \citenamefont {Ding}, \citenamefont
  {Gui}, \citenamefont {Tomesh}, \citenamefont {Suchara}, \citenamefont
  {Martonosi},\ and\ \citenamefont {Chong}}]{gokhale_ON3_Measurement_2020}%
  \BibitemOpen
  \bibfield  {author} {\bibinfo {author} {\bibfnamefont {Pranav}\ \bibnamefont
  {Gokhale}}, \bibinfo {author} {\bibfnamefont {Olivia}\ \bibnamefont
  {Angiuli}}, \bibinfo {author} {\bibfnamefont {Yongshan}\ \bibnamefont
  {Ding}}, \bibinfo {author} {\bibfnamefont {Kaiwen}\ \bibnamefont {Gui}},
  \bibinfo {author} {\bibfnamefont {Teague}\ \bibnamefont {Tomesh}}, \bibinfo
  {author} {\bibfnamefont {Martin}\ \bibnamefont {Suchara}}, \bibinfo {author}
  {\bibfnamefont {Margaret}\ \bibnamefont {Martonosi}}, \ and\ \bibinfo
  {author} {\bibfnamefont {Frederic~T.}\ \bibnamefont {Chong}},\ }\bibfield
  {title} {\enquote {\bibinfo {title} {{$O(N^3)$ Measurement Cost for
  Variational Quantum Eigensolver on Molecular Hamiltonians}},}\ }\href
  {\doibase 10.1109/TQE.2020.3035814} {\bibfield  {journal} {\bibinfo
  {journal} {IEEE Transactions on Quantum Engineering}\ }\textbf {\bibinfo
  {volume} {1}},\ \bibinfo {pages} {1--24} (\bibinfo {year}
  {2020})}\BibitemShut {NoStop}%
\bibitem [{\citenamefont {Verteletskyi}\ \emph {et~al.}(2020)\citenamefont
  {Verteletskyi}, \citenamefont {Yen},\ and\ \citenamefont
  {Izmaylov}}]{verteletskyi2020measurement}%
  \BibitemOpen
  \bibfield  {author} {\bibinfo {author} {\bibfnamefont {Vladyslav}\
  \bibnamefont {Verteletskyi}}, \bibinfo {author} {\bibfnamefont {Tzu-Ching}\
  \bibnamefont {Yen}}, \ and\ \bibinfo {author} {\bibfnamefont {Artur~F.}\
  \bibnamefont {Izmaylov}},\ }\bibfield  {title} {\enquote {\bibinfo {title}
  {Measurement optimization in the variational quantum eigensolver using a
  minimum clique cover},}\ }\href {\doibase 10.1063/1.5141458} {\bibfield
  {journal} {\bibinfo  {journal} {The Journal of Chemical Physics}\ }\textbf
  {\bibinfo {volume} {152}},\ \bibinfo {pages} {124114} (\bibinfo {year}
  {2020})}\BibitemShut {NoStop}%
\bibitem [{\citenamefont {Hamamura}\ and\ \citenamefont
  {Imamichi}(2020)}]{hamamura2020efficient}%
  \BibitemOpen
  \bibfield  {author} {\bibinfo {author} {\bibfnamefont {Ikko}\ \bibnamefont
  {Hamamura}}\ and\ \bibinfo {author} {\bibfnamefont {Takashi}\ \bibnamefont
  {Imamichi}},\ }\bibfield  {title} {\enquote {\bibinfo {title} {Efficient
  evaluation of quantum observables using entangled measurements},}\ }\href
  {\doibase 10.1038/s41534-020-0284-2} {\bibfield  {journal} {\bibinfo
  {journal} {npj Quantum Information}\ }\textbf {\bibinfo {volume} {6}},\
  \bibinfo {pages} {56} (\bibinfo {year} {2020})}\BibitemShut {NoStop}%
\bibitem [{\citenamefont {Crawford}\ \emph {et~al.}(2021)\citenamefont
  {Crawford}, \citenamefont {Straaten}, \citenamefont {Wang}, \citenamefont
  {Parks}, \citenamefont {Campbell},\ and\ \citenamefont
  {Brierley}}]{crawford2021efficient}%
  \BibitemOpen
  \bibfield  {author} {\bibinfo {author} {\bibfnamefont {Ophelia}\ \bibnamefont
  {Crawford}}, \bibinfo {author} {\bibfnamefont {Barnaby~van}\ \bibnamefont
  {Straaten}}, \bibinfo {author} {\bibfnamefont {Daochen}\ \bibnamefont
  {Wang}}, \bibinfo {author} {\bibfnamefont {Thomas}\ \bibnamefont {Parks}},
  \bibinfo {author} {\bibfnamefont {Earl}\ \bibnamefont {Campbell}}, \ and\
  \bibinfo {author} {\bibfnamefont {Stephen}\ \bibnamefont {Brierley}},\
  }\bibfield  {title} {\enquote {\bibinfo {title} {Efficient quantum
  measurement of {Pauli} operators in the presence of finite sampling error},}\
  }\href {\doibase 10.22331/q-2021-01-20-385} {\bibfield  {journal} {\bibinfo
  {journal} {Quantum}\ }\textbf {\bibinfo {volume} {5}},\ \bibinfo {pages}
  {385} (\bibinfo {year} {2021})}\BibitemShut {NoStop}%
\bibitem [{\citenamefont {Miller}\ \emph {et~al.}(2022)\citenamefont {Miller},
  \citenamefont {Fischer}, \citenamefont {Sokolov}, \citenamefont
  {Barkoutsos},\ and\ \citenamefont {Tavernelli}}]{miller2022hardwaretailored}%
  \BibitemOpen
  \bibfield  {author} {\bibinfo {author} {\bibfnamefont {Daniel}\ \bibnamefont
  {Miller}}, \bibinfo {author} {\bibfnamefont {Laurin~E.}\ \bibnamefont
  {Fischer}}, \bibinfo {author} {\bibfnamefont {Igor~O.}\ \bibnamefont
  {Sokolov}}, \bibinfo {author} {\bibfnamefont {Panagiotis~Kl.}\ \bibnamefont
  {Barkoutsos}}, \ and\ \bibinfo {author} {\bibfnamefont {Ivano}\ \bibnamefont
  {Tavernelli}},\ }\bibfield  {title} {\enquote {\bibinfo {title}
  {Hardware-{Tailored} {Diagonalization} {Circuits}},}\ }\href
  {http://arxiv.org/abs/2203.03646} {\bibfield  {journal} {\bibinfo  {journal}
  {arXiv:2203.03646 [quant-ph]}\ } (\bibinfo {year} {2022})}\BibitemShut
  {NoStop}%
\bibitem [{\citenamefont {Huang}\ \emph {et~al.}(2020)\citenamefont {Huang},
  \citenamefont {Kueng},\ and\ \citenamefont {Preskill}}]{huang2020predicting}%
  \BibitemOpen
  \bibfield  {author} {\bibinfo {author} {\bibfnamefont {Hsin-Yuan}\
  \bibnamefont {Huang}}, \bibinfo {author} {\bibfnamefont {Richard}\
  \bibnamefont {Kueng}}, \ and\ \bibinfo {author} {\bibfnamefont {John}\
  \bibnamefont {Preskill}},\ }\bibfield  {title} {\enquote {\bibinfo {title}
  {Predicting many properties of a quantum system from very few
  measurements},}\ }\href {\doibase 10.1038/s41567-020-0932-7} {\bibfield
  {journal} {\bibinfo  {journal} {Nature Physics}\ }\textbf {\bibinfo {volume}
  {16}},\ \bibinfo {pages} {1050--1057} (\bibinfo {year} {2020})}\BibitemShut
  {NoStop}%
\bibitem [{\citenamefont {Hadfield}\ \emph {et~al.}(2020)\citenamefont
  {Hadfield}, \citenamefont {Bravyi}, \citenamefont {Raymond},\ and\
  \citenamefont {Mezzacapo}}]{hadfield2020measurements}%
  \BibitemOpen
  \bibfield  {author} {\bibinfo {author} {\bibfnamefont {Charles}\ \bibnamefont
  {Hadfield}}, \bibinfo {author} {\bibfnamefont {Sergey}\ \bibnamefont
  {Bravyi}}, \bibinfo {author} {\bibfnamefont {Rudy}\ \bibnamefont {Raymond}},
  \ and\ \bibinfo {author} {\bibfnamefont {Antonio}\ \bibnamefont
  {Mezzacapo}},\ }\href@noop {} {\enquote {\bibinfo {title} {Measurements of
  quantum hamiltonians with locally-biased classical shadows},}\ } (\bibinfo
  {year} {2020}),\ \Eprint {http://arxiv.org/abs/2006.15788} {arXiv:2006.15788
  [quant-ph]} \BibitemShut {NoStop}%
\bibitem [{\citenamefont {Zhao}\ \emph {et~al.}(2021)\citenamefont {Zhao},
  \citenamefont {Rubin},\ and\ \citenamefont {Miyake}}]{Zhao2021}%
  \BibitemOpen
  \bibfield  {author} {\bibinfo {author} {\bibfnamefont {Andrew}\ \bibnamefont
  {Zhao}}, \bibinfo {author} {\bibfnamefont {Nicholas~C.}\ \bibnamefont
  {Rubin}}, \ and\ \bibinfo {author} {\bibfnamefont {Akimasa}\ \bibnamefont
  {Miyake}},\ }\bibfield  {title} {\enquote {\bibinfo {title} {Fermionic
  partial tomography via classical shadows},}\ }\href {\doibase
  10.1103/PhysRevLett.127.110504} {\bibfield  {journal} {\bibinfo  {journal}
  {Physical Review Letters}\ }\textbf {\bibinfo {volume} {127}},\ \bibinfo
  {pages} {110504} (\bibinfo {year} {2021})}\BibitemShut {NoStop}%
\bibitem [{\citenamefont {Torlai}\ \emph {et~al.}(2020)\citenamefont {Torlai},
  \citenamefont {Mazzola}, \citenamefont {Carleo},\ and\ \citenamefont
  {Mezzacapo}}]{torlai2020precise}%
  \BibitemOpen
  \bibfield  {author} {\bibinfo {author} {\bibfnamefont {Giacomo}\ \bibnamefont
  {Torlai}}, \bibinfo {author} {\bibfnamefont {Guglielmo}\ \bibnamefont
  {Mazzola}}, \bibinfo {author} {\bibfnamefont {Giuseppe}\ \bibnamefont
  {Carleo}}, \ and\ \bibinfo {author} {\bibfnamefont {Antonio}\ \bibnamefont
  {Mezzacapo}},\ }\bibfield  {title} {\enquote {\bibinfo {title} {Precise
  measurement of quantum observables with neural-network estimators},}\ }\href
  {\doibase 10.1103/PhysRevResearch.2.022060} {\bibfield  {journal} {\bibinfo
  {journal} {Physical Review Research}\ }\textbf {\bibinfo {volume} {2}},\
  \bibinfo {pages} {022060} (\bibinfo {year} {2020})}\BibitemShut {NoStop}%
\bibitem [{\citenamefont {Jiang}\ \emph {et~al.}(2020)\citenamefont {Jiang},
  \citenamefont {Kalev}, \citenamefont {Mruczkiewicz},\ and\ \citenamefont
  {Neven}}]{jiang2020optimal}%
  \BibitemOpen
  \bibfield  {author} {\bibinfo {author} {\bibfnamefont {Zhang}\ \bibnamefont
  {Jiang}}, \bibinfo {author} {\bibfnamefont {Amir}\ \bibnamefont {Kalev}},
  \bibinfo {author} {\bibfnamefont {Wojciech}\ \bibnamefont {Mruczkiewicz}}, \
  and\ \bibinfo {author} {\bibfnamefont {Hartmut}\ \bibnamefont {Neven}},\
  }\bibfield  {title} {\enquote {\bibinfo {title} {Optimal fermion-to-qubit
  mapping via ternary trees with applications to reduced quantum states
  learning},}\ }\href {\doibase 10.22331/q-2020-06-04-276} {\bibfield
  {journal} {\bibinfo  {journal} {Quantum}\ }\textbf {\bibinfo {volume} {4}},\
  \bibinfo {pages} {276} (\bibinfo {year} {2020})}\BibitemShut {NoStop}%
\bibitem [{\citenamefont {Bonet-Monroig}\ \emph {et~al.}(2020)\citenamefont
  {Bonet-Monroig}, \citenamefont {Babbush},\ and\ \citenamefont
  {O'Brien}}]{Bonet2020}%
  \BibitemOpen
  \bibfield  {author} {\bibinfo {author} {\bibfnamefont {Xavier}\ \bibnamefont
  {Bonet-Monroig}}, \bibinfo {author} {\bibfnamefont {Ryan}\ \bibnamefont
  {Babbush}}, \ and\ \bibinfo {author} {\bibfnamefont {Thomas~E.}\ \bibnamefont
  {O'Brien}},\ }\bibfield  {title} {\enquote {\bibinfo {title} {Nearly optimal
  measurement scheduling for partial tomography of quantum states},}\ }\href
  {\doibase 10.1103/PhysRevX.10.031064} {\bibfield  {journal} {\bibinfo
  {journal} {Physical Review X}\ }\textbf {\bibinfo {volume} {10}},\ \bibinfo
  {pages} {031064} (\bibinfo {year} {2020})}\BibitemShut {NoStop}%
\bibitem [{\citenamefont {García-Pérez}\ \emph {et~al.}(2021)\citenamefont
  {García-Pérez}, \citenamefont {Rossi}, \citenamefont {Sokolov},
  \citenamefont {Tacchino}, \citenamefont {Barkoutsos}, \citenamefont
  {Mazzola}, \citenamefont {Tavernelli},\ and\ \citenamefont
  {Maniscalco}}]{garcia2021learning}%
  \BibitemOpen
  \bibfield  {author} {\bibinfo {author} {\bibfnamefont {Guillermo}\
  \bibnamefont {García-Pérez}}, \bibinfo {author} {\bibfnamefont
  {Matteo~A.C.}\ \bibnamefont {Rossi}}, \bibinfo {author} {\bibfnamefont
  {Boris}\ \bibnamefont {Sokolov}}, \bibinfo {author} {\bibfnamefont
  {Francesco}\ \bibnamefont {Tacchino}}, \bibinfo {author} {\bibfnamefont
  {Panagiotis~Kl.}\ \bibnamefont {Barkoutsos}}, \bibinfo {author}
  {\bibfnamefont {Guglielmo}\ \bibnamefont {Mazzola}}, \bibinfo {author}
  {\bibfnamefont {Ivano}\ \bibnamefont {Tavernelli}}, \ and\ \bibinfo {author}
  {\bibfnamefont {Sabrina}\ \bibnamefont {Maniscalco}},\ }\bibfield  {title}
  {\enquote {\bibinfo {title} {Learning to {measure}: {adaptive}
  {informationally} {complete} {generalized} {measurements} for {quantum}
  {algorithms}},}\ }\href {\doibase 10.1103/PRXQuantum.2.040342} {\bibfield
  {journal} {\bibinfo  {journal} {PRX Quantum}\ }\textbf {\bibinfo {volume}
  {2}},\ \bibinfo {pages} {040342} (\bibinfo {year} {2021})}\BibitemShut
  {NoStop}%
\bibitem [{\citenamefont {Chen}\ \emph {et~al.}(2007)\citenamefont {Chen},
  \citenamefont {Bergou}, \citenamefont {Zhu},\ and\ \citenamefont
  {Guo}}]{chen2007ancilla}%
  \BibitemOpen
  \bibfield  {author} {\bibinfo {author} {\bibfnamefont {Ping-Xing}\
  \bibnamefont {Chen}}, \bibinfo {author} {\bibfnamefont {János~A.}\
  \bibnamefont {Bergou}}, \bibinfo {author} {\bibfnamefont {Shi-Yao}\
  \bibnamefont {Zhu}}, \ and\ \bibinfo {author} {\bibfnamefont {Guang-Can}\
  \bibnamefont {Guo}},\ }\bibfield  {title} {\enquote {\bibinfo {title}
  {Ancilla dimensions needed to carry out positive-operator-valued
  measurement},}\ }\href {\doibase 10.1103/PhysRevA.76.060303} {\bibfield
  {journal} {\bibinfo  {journal} {Physical Review A}\ }\textbf {\bibinfo
  {volume} {76}},\ \bibinfo {pages} {060303} (\bibinfo {year}
  {2007})}\BibitemShut {NoStop}%
\bibitem [{\citenamefont {Brida}\ \emph {et~al.}(2012)\citenamefont {Brida},
  \citenamefont {Ciavarella}, \citenamefont {Degiovanni}, \citenamefont
  {Genovese}, \citenamefont {Migdall}, \citenamefont {Mingolla}, \citenamefont
  {Paris}, \citenamefont {Piacentini},\ and\ \citenamefont
  {Polyakov}}]{brida2012ancillaassisted}%
  \BibitemOpen
  \bibfield  {author} {\bibinfo {author} {\bibfnamefont {Giorgio}\ \bibnamefont
  {Brida}}, \bibinfo {author} {\bibfnamefont {Luigi}\ \bibnamefont
  {Ciavarella}}, \bibinfo {author} {\bibfnamefont {Ivo~P.}\ \bibnamefont
  {Degiovanni}}, \bibinfo {author} {\bibfnamefont {Marco}\ \bibnamefont
  {Genovese}}, \bibinfo {author} {\bibfnamefont {Alan}\ \bibnamefont
  {Migdall}}, \bibinfo {author} {\bibfnamefont {Maria~G.}\ \bibnamefont
  {Mingolla}}, \bibinfo {author} {\bibfnamefont {Matteo G.~A.}\ \bibnamefont
  {Paris}}, \bibinfo {author} {\bibfnamefont {Fabrizio}\ \bibnamefont
  {Piacentini}}, \ and\ \bibinfo {author} {\bibfnamefont {Sergey~V.}\
  \bibnamefont {Polyakov}},\ }\bibfield  {title} {\enquote {\bibinfo {title}
  {Ancilla-assisted calibration of a measuring apparatus},}\ }\href {\doibase
  10.1103/PhysRevLett.108.253601} {\bibfield  {journal} {\bibinfo  {journal}
  {Physical Review Letters}\ }\textbf {\bibinfo {volume} {108}},\ \bibinfo
  {pages} {253601} (\bibinfo {year} {2012})}\BibitemShut {NoStop}%
\bibitem [{\citenamefont {Weidenfeller}\ \emph {et~al.}(2022)\citenamefont
  {Weidenfeller}, \citenamefont {Valor}, \citenamefont {Gacon}, \citenamefont
  {Tornow}, \citenamefont {Bello}, \citenamefont {Woerner},\ and\ \citenamefont
  {Egger}}]{weidenfeller2022scaling}%
  \BibitemOpen
  \bibfield  {author} {\bibinfo {author} {\bibfnamefont {Johannes}\
  \bibnamefont {Weidenfeller}}, \bibinfo {author} {\bibfnamefont {Lucia~C.}\
  \bibnamefont {Valor}}, \bibinfo {author} {\bibfnamefont {Julien}\
  \bibnamefont {Gacon}}, \bibinfo {author} {\bibfnamefont {Caroline}\
  \bibnamefont {Tornow}}, \bibinfo {author} {\bibfnamefont {Luciano}\
  \bibnamefont {Bello}}, \bibinfo {author} {\bibfnamefont {Stefan}\
  \bibnamefont {Woerner}}, \ and\ \bibinfo {author} {\bibfnamefont {Daniel~J.}\
  \bibnamefont {Egger}},\ }\bibfield  {title} {\enquote {\bibinfo {title}
  {Scaling of the quantum approximate optimization algorithm on superconducting
  qubit based hardware},}\ }\href {http://arxiv.org/abs/2202.03459} {\bibfield
  {journal} {\bibinfo  {journal} {arXiv:2202.03459 [quant-ph]}\ } (\bibinfo
  {year} {2022})}\BibitemShut {NoStop}%
\bibitem [{\citenamefont {Peterer}\ \emph {et~al.}(2015)\citenamefont
  {Peterer}, \citenamefont {Bader}, \citenamefont {Jin}, \citenamefont {Yan},
  \citenamefont {Kamal}, \citenamefont {Gudmundsen}, \citenamefont {Leek},
  \citenamefont {Orlando}, \citenamefont {Oliver},\ and\ \citenamefont
  {Gustavsson}}]{peterer2015coherence}%
  \BibitemOpen
  \bibfield  {author} {\bibinfo {author} {\bibfnamefont {Michael~J.}\
  \bibnamefont {Peterer}}, \bibinfo {author} {\bibfnamefont {Samuel~J.}\
  \bibnamefont {Bader}}, \bibinfo {author} {\bibfnamefont {Xiaoyue}\
  \bibnamefont {Jin}}, \bibinfo {author} {\bibfnamefont {Fei}\ \bibnamefont
  {Yan}}, \bibinfo {author} {\bibfnamefont {Archana}\ \bibnamefont {Kamal}},
  \bibinfo {author} {\bibfnamefont {Theodore~J.}\ \bibnamefont {Gudmundsen}},
  \bibinfo {author} {\bibfnamefont {Peter~J.}\ \bibnamefont {Leek}}, \bibinfo
  {author} {\bibfnamefont {Terry~P.}\ \bibnamefont {Orlando}}, \bibinfo
  {author} {\bibfnamefont {William~D.}\ \bibnamefont {Oliver}}, \ and\ \bibinfo
  {author} {\bibfnamefont {Simon}\ \bibnamefont {Gustavsson}},\ }\bibfield
  {title} {\enquote {\bibinfo {title} {Coherence and {decay} of {higher}
  {energy} {levels} of a {superconducting} {transmon} {qubit}},}\ }\href
  {\doibase 10.1103/PhysRevLett.114.010501} {\bibfield  {journal} {\bibinfo
  {journal} {Physical Review Letters}\ }\textbf {\bibinfo {volume} {114}},\
  \bibinfo {pages} {010501} (\bibinfo {year} {2015})}\BibitemShut {NoStop}%
\bibitem [{\citenamefont {Low}\ \emph {et~al.}(2020)\citenamefont {Low},
  \citenamefont {White}, \citenamefont {Cox}, \citenamefont {Day},\ and\
  \citenamefont {Senko}}]{low2020practical}%
  \BibitemOpen
  \bibfield  {author} {\bibinfo {author} {\bibfnamefont {Pei~Jiang}\
  \bibnamefont {Low}}, \bibinfo {author} {\bibfnamefont {Brendan~M.}\
  \bibnamefont {White}}, \bibinfo {author} {\bibfnamefont {Andrew~A.}\
  \bibnamefont {Cox}}, \bibinfo {author} {\bibfnamefont {Matthew~L.}\
  \bibnamefont {Day}}, \ and\ \bibinfo {author} {\bibfnamefont {Crystal}\
  \bibnamefont {Senko}},\ }\bibfield  {title} {\enquote {\bibinfo {title}
  {Practical trapped-ion protocols for universal qudit-based quantum
  computing},}\ }\href {\doibase 10.1103/PhysRevResearch.2.033128} {\bibfield
  {journal} {\bibinfo  {journal} {Physical Review Research}\ }\textbf {\bibinfo
  {volume} {2}},\ \bibinfo {pages} {033128} (\bibinfo {year}
  {2020})}\BibitemShut {NoStop}%
\bibitem [{\citenamefont {Shi}(2021)}]{shi2021quantum}%
  \BibitemOpen
  \bibfield  {author} {\bibinfo {author} {\bibfnamefont {Xiaofeng}\
  \bibnamefont {Shi}},\ }\bibfield  {title} {\enquote {\bibinfo {title}
  {Quantum logic and entanglement by neutral {Rydberg} atoms: methods and
  fidelity},}\ }\href
  {https://iopscience.iop.org/article/10.1088/2058-9565/ac18b8} {\bibfield
  {journal} {\bibinfo  {journal} {Quantum Science and Technology}\ }\textbf
  {\bibinfo {volume} {10.1088}} (\bibinfo {year} {2021})}\BibitemShut {NoStop}%
\bibitem [{\citenamefont {Gelfand}\ and\ \citenamefont
  {Neumark}(1943)}]{gelfand1943imbedding}%
  \BibitemOpen
  \bibfield  {author} {\bibinfo {author} {\bibfnamefont {Israel}\ \bibnamefont
  {Gelfand}}\ and\ \bibinfo {author} {\bibfnamefont {Mark}\ \bibnamefont
  {Neumark}},\ }\bibfield  {title} {\enquote {\bibinfo {title} {On the
  imbedding of normed rings into the ring of operators in {Hilbert} space},}\
  }\href@noop {} {\bibfield  {journal} {\bibinfo  {journal} {Matematicheskii
  Sbornik}\ }\textbf {\bibinfo {volume} {12}},\ \bibinfo {pages} {197--217}
  (\bibinfo {year} {1943})}\BibitemShut {NoStop}%
\bibitem [{\citenamefont {Vatan}\ and\ \citenamefont
  {Williams}(2004)}]{vatan2004optimal}%
  \BibitemOpen
  \bibfield  {author} {\bibinfo {author} {\bibfnamefont {Farrokh}\ \bibnamefont
  {Vatan}}\ and\ \bibinfo {author} {\bibfnamefont {Colin}\ \bibnamefont
  {Williams}},\ }\bibfield  {title} {\enquote {\bibinfo {title} {Optimal
  quantum circuits for general two-qubit gates},}\ }\href {\doibase
  10.1103/PhysRevA.69.032315} {\bibfield  {journal} {\bibinfo  {journal}
  {Physical Review A}\ }\textbf {\bibinfo {volume} {69}},\ \bibinfo {pages}
  {032315} (\bibinfo {year} {2004})}\BibitemShut {NoStop}%
\bibitem [{\citenamefont {Drury}\ and\ \citenamefont
  {Love}(2008)}]{drury2008constructive}%
  \BibitemOpen
  \bibfield  {author} {\bibinfo {author} {\bibfnamefont {Byron}\ \bibnamefont
  {Drury}}\ and\ \bibinfo {author} {\bibfnamefont {Peter}\ \bibnamefont
  {Love}},\ }\bibfield  {title} {\enquote {\bibinfo {title} {Constructive
  quantum {Shannon} decomposition from {Cartan} involutions},}\ }\href
  {\doibase 10.1088/1751-8113/41/39/395305} {\bibfield  {journal} {\bibinfo
  {journal} {Journal of Physics A: Mathematical and Theoretical}\ }\textbf
  {\bibinfo {volume} {41}},\ \bibinfo {pages} {395305} (\bibinfo {year}
  {2008})}\BibitemShut {NoStop}%
\bibitem [{\citenamefont {Earnest}\ \emph {et~al.}(2021)\citenamefont
  {Earnest}, \citenamefont {Tornow},\ and\ \citenamefont
  {Egger}}]{earnest2021pulseefficient}%
  \BibitemOpen
  \bibfield  {author} {\bibinfo {author} {\bibfnamefont {Nathan}\ \bibnamefont
  {Earnest}}, \bibinfo {author} {\bibfnamefont {Caroline}\ \bibnamefont
  {Tornow}}, \ and\ \bibinfo {author} {\bibfnamefont {Daniel~J.}\ \bibnamefont
  {Egger}},\ }\bibfield  {title} {\enquote {\bibinfo {title} {Pulse-efficient
  circuit transpilation for quantum applications on cross-resonance-based
  hardware},}\ }\href {\doibase 10.1103/PhysRevResearch.3.043088} {\bibfield
  {journal} {\bibinfo  {journal} {Physical Review Research}\ }\textbf {\bibinfo
  {volume} {3}},\ \bibinfo {pages} {043088} (\bibinfo {year}
  {2021})}\BibitemShut {NoStop}%
\bibitem [{\citenamefont {McKay}\ \emph {et~al.}(2017)\citenamefont {McKay},
  \citenamefont {Wood}, \citenamefont {Sheldon}, \citenamefont {Chow},\ and\
  \citenamefont {Gambetta}}]{mckay2017efficient}%
  \BibitemOpen
  \bibfield  {author} {\bibinfo {author} {\bibfnamefont {David~C.}\
  \bibnamefont {McKay}}, \bibinfo {author} {\bibfnamefont {Christopher~J.}\
  \bibnamefont {Wood}}, \bibinfo {author} {\bibfnamefont {Sarah}\ \bibnamefont
  {Sheldon}}, \bibinfo {author} {\bibfnamefont {Jerry~M.}\ \bibnamefont
  {Chow}}, \ and\ \bibinfo {author} {\bibfnamefont {Jay~M.}\ \bibnamefont
  {Gambetta}},\ }\bibfield  {title} {\enquote {\bibinfo {title} {Efficient {Z}
  gates for quantum computing},}\ }\href {\doibase 10.1103/PhysRevA.96.022330}
  {\bibfield  {journal} {\bibinfo  {journal} {Physical Review A}\ }\textbf
  {\bibinfo {volume} {96}},\ \bibinfo {pages} {022330} (\bibinfo {year}
  {2017})}\BibitemShut {NoStop}%
\bibitem [{\citenamefont {Murali}\ \emph {et~al.}(2019)\citenamefont {Murali},
  \citenamefont {Linke}, \citenamefont {Martonosi}, \citenamefont {Abhari},
  \citenamefont {Nguyen},\ and\ \citenamefont {Alderete}}]{murali2019full}%
  \BibitemOpen
  \bibfield  {author} {\bibinfo {author} {\bibfnamefont {Prakash}\ \bibnamefont
  {Murali}}, \bibinfo {author} {\bibfnamefont {Norbert~Matthias}\ \bibnamefont
  {Linke}}, \bibinfo {author} {\bibfnamefont {Margaret}\ \bibnamefont
  {Martonosi}}, \bibinfo {author} {\bibfnamefont {Ali~Javadi}\ \bibnamefont
  {Abhari}}, \bibinfo {author} {\bibfnamefont {Nhung~Hong}\ \bibnamefont
  {Nguyen}}, \ and\ \bibinfo {author} {\bibfnamefont {Cinthia~Huerta}\
  \bibnamefont {Alderete}},\ }\bibfield  {title} {\enquote {\bibinfo {title}
  {Full-stack, real-system quantum computer studies: architectural comparisons
  and design insights},}\ }in\ \href {\doibase 10.1145/3307650.3322273} {\emph
  {\bibinfo {booktitle} {Proceedings of the 46th {International} {Symposium} on
  {Computer} {Architecture}}}}\ (\bibinfo  {publisher} {ACM},\ \bibinfo
  {address} {Phoenix Arizona},\ \bibinfo {year} {2019})\ pp.\ \bibinfo {pages}
  {527--540}\BibitemShut {NoStop}%
\bibitem [{\citenamefont {Schirmer}\ \emph {et~al.}(2002)\citenamefont
  {Schirmer}, \citenamefont {Greentree}, \citenamefont {Ramakrishna},\ and\
  \citenamefont {Rabitz}}]{schirmer2002constructive}%
  \BibitemOpen
  \bibfield  {author} {\bibinfo {author} {\bibfnamefont {Sophie~G.}\
  \bibnamefont {Schirmer}}, \bibinfo {author} {\bibfnamefont {Andrew~D.}\
  \bibnamefont {Greentree}}, \bibinfo {author} {\bibfnamefont {Viswanath}\
  \bibnamefont {Ramakrishna}}, \ and\ \bibinfo {author} {\bibfnamefont
  {Herschel}\ \bibnamefont {Rabitz}},\ }\bibfield  {title} {\enquote {\bibinfo
  {title} {Constructive control of quantum systems using factorization of
  unitary operators},}\ }\href {\doibase 10.1088/0305-4470/35/39/313}
  {\bibfield  {journal} {\bibinfo  {journal} {Journal of Physics A:
  Mathematical and General}\ }\textbf {\bibinfo {volume} {35}},\ \bibinfo
  {pages} {8315--8339} (\bibinfo {year} {2002})}\BibitemShut {NoStop}%
\bibitem [{\citenamefont {Golub}\ and\ \citenamefont
  {Van~Loan}(1996)}]{golub1996matrix}%
  \BibitemOpen
  \bibfield  {author} {\bibinfo {author} {\bibfnamefont {Gene~H.}\ \bibnamefont
  {Golub}}\ and\ \bibinfo {author} {\bibfnamefont {Charles~F.}\ \bibnamefont
  {Van~Loan}},\ }\href@noop {} {\emph {\bibinfo {title} {Matrix
  computations}}},\ \bibinfo {edition} {3rd}\ ed.,\ Johns {Hopkins} studies in
  the mathematical sciences\ (\bibinfo  {publisher} {Johns Hopkins University
  Press},\ \bibinfo {address} {Baltimore},\ \bibinfo {year} {1996})\BibitemShut
  {NoStop}%
\bibitem [{\citenamefont {Sheldon}\ \emph {et~al.}(2016)\citenamefont
  {Sheldon}, \citenamefont {Bishop}, \citenamefont {Magesan}, \citenamefont
  {Filipp}, \citenamefont {Chow},\ and\ \citenamefont
  {Gambetta}}]{sheldon2016characterizing}%
  \BibitemOpen
  \bibfield  {author} {\bibinfo {author} {\bibfnamefont {Sarah}\ \bibnamefont
  {Sheldon}}, \bibinfo {author} {\bibfnamefont {Lev~S.}\ \bibnamefont
  {Bishop}}, \bibinfo {author} {\bibfnamefont {Easwar}\ \bibnamefont
  {Magesan}}, \bibinfo {author} {\bibfnamefont {Stefan}\ \bibnamefont
  {Filipp}}, \bibinfo {author} {\bibfnamefont {Jerry~M.}\ \bibnamefont {Chow}},
  \ and\ \bibinfo {author} {\bibfnamefont {Jay~M.}\ \bibnamefont {Gambetta}},\
  }\bibfield  {title} {\enquote {\bibinfo {title} {Characterizing errors on
  qubit operations via iterative randomized benchmarking},}\ }\href {\doibase
  10.1103/PhysRevA.93.012301} {\bibfield  {journal} {\bibinfo  {journal}
  {Physical Review A}\ }\textbf {\bibinfo {volume} {93}},\ \bibinfo {pages}
  {012301} (\bibinfo {year} {2016})}\BibitemShut {NoStop}%
\bibitem [{\citenamefont {Koch}\ \emph {et~al.}(2007)\citenamefont {Koch},
  \citenamefont {Yu}, \citenamefont {Gambetta}, \citenamefont {Houck},
  \citenamefont {Schuster}, \citenamefont {Majer}, \citenamefont {Blais},
  \citenamefont {Devoret}, \citenamefont {Girvin},\ and\ \citenamefont
  {Schoelkopf}}]{koch2007charge}%
  \BibitemOpen
  \bibfield  {author} {\bibinfo {author} {\bibfnamefont {Jens}\ \bibnamefont
  {Koch}}, \bibinfo {author} {\bibfnamefont {Terri~M.}\ \bibnamefont {Yu}},
  \bibinfo {author} {\bibfnamefont {Jay}\ \bibnamefont {Gambetta}}, \bibinfo
  {author} {\bibfnamefont {Andrew~A.}\ \bibnamefont {Houck}}, \bibinfo {author}
  {\bibfnamefont {David~I.}\ \bibnamefont {Schuster}}, \bibinfo {author}
  {\bibfnamefont {Johannes}\ \bibnamefont {Majer}}, \bibinfo {author}
  {\bibfnamefont {Alexandre}\ \bibnamefont {Blais}}, \bibinfo {author}
  {\bibfnamefont {Michel~H.}\ \bibnamefont {Devoret}}, \bibinfo {author}
  {\bibfnamefont {Steven~M.}\ \bibnamefont {Girvin}}, \ and\ \bibinfo {author}
  {\bibfnamefont {Robert~J.}\ \bibnamefont {Schoelkopf}},\ }\bibfield  {title}
  {\enquote {\bibinfo {title} {Charge-insensitive qubit design derived from the
  {Cooper} pair box},}\ }\href {\doibase 10.1103/PhysRevA.76.042319} {\bibfield
   {journal} {\bibinfo  {journal} {Physical Review A}\ }\textbf {\bibinfo
  {volume} {76}},\ \bibinfo {pages} {042319} (\bibinfo {year}
  {2007})}\BibitemShut {NoStop}%
\bibitem [{\citenamefont {Quantum}()}]{IBMQuantum}%
  \BibitemOpen
  \bibfield  {author} {\bibinfo {author} {\bibfnamefont {IBM}\ \bibnamefont
  {Quantum}},\ }\href@noop {} {}\bibinfo {howpublished}
  {\url{https://quantum-computing.ibm.com/} (accessed Dec. 2021)}\BibitemShut
  {NoStop}%
\bibitem [{\citenamefont {Elder}\ \emph {et~al.}(2020)\citenamefont {Elder},
  \citenamefont {Wang}, \citenamefont {Reinhold}, \citenamefont {Hann},
  \citenamefont {Chou}, \citenamefont {Lester}, \citenamefont {Rosenblum},
  \citenamefont {Frunzio}, \citenamefont {Jiang},\ and\ \citenamefont
  {Schoelkopf}}]{elder2020highfidelity}%
  \BibitemOpen
  \bibfield  {author} {\bibinfo {author} {\bibfnamefont {Salvatore~S.}\
  \bibnamefont {Elder}}, \bibinfo {author} {\bibfnamefont {Christopher~S.}\
  \bibnamefont {Wang}}, \bibinfo {author} {\bibfnamefont {Philip}\ \bibnamefont
  {Reinhold}}, \bibinfo {author} {\bibfnamefont {Connor~T.}\ \bibnamefont
  {Hann}}, \bibinfo {author} {\bibfnamefont {Kevin~S.}\ \bibnamefont {Chou}},
  \bibinfo {author} {\bibfnamefont {Brian~J.}\ \bibnamefont {Lester}}, \bibinfo
  {author} {\bibfnamefont {Serge}\ \bibnamefont {Rosenblum}}, \bibinfo {author}
  {\bibfnamefont {Luigi}\ \bibnamefont {Frunzio}}, \bibinfo {author}
  {\bibfnamefont {Liang}\ \bibnamefont {Jiang}}, \ and\ \bibinfo {author}
  {\bibfnamefont {Robert~J.}\ \bibnamefont {Schoelkopf}},\ }\bibfield  {title}
  {\enquote {\bibinfo {title} {High-{fidelity} {measurement} of {qubits}
  {encoded} in {multilevel} {superconducting} {circuits}},}\ }\href {\doibase
  10.1103/PhysRevX.10.011001} {\bibfield  {journal} {\bibinfo  {journal}
  {Physical Review X}\ }\textbf {\bibinfo {volume} {10}},\ \bibinfo {pages}
  {011001} (\bibinfo {year} {2020})}\BibitemShut {NoStop}%
\bibitem [{\citenamefont {Cervera-Lierta}\ \emph {et~al.}(2022)\citenamefont
  {Cervera-Lierta}, \citenamefont {Krenn}, \citenamefont {Aspuru-Guzik},\ and\
  \citenamefont {Galda}}]{cervera-lierta2021experimental}%
  \BibitemOpen
  \bibfield  {author} {\bibinfo {author} {\bibfnamefont {Alba}\ \bibnamefont
  {Cervera-Lierta}}, \bibinfo {author} {\bibfnamefont {Mario}\ \bibnamefont
  {Krenn}}, \bibinfo {author} {\bibfnamefont {Alán}\ \bibnamefont
  {Aspuru-Guzik}}, \ and\ \bibinfo {author} {\bibfnamefont {Alexey}\
  \bibnamefont {Galda}},\ }\bibfield  {title} {\enquote {\bibinfo {title}
  {Experimental {High}-{Dimensional} {Greenberger}-{Horne}-{Zeilinger}
  {Entanglement} with {Superconducting} {Transmon} {Qutrits}},}\ }\href
  {\doibase 10.1103/PhysRevApplied.17.024062} {\bibfield  {journal} {\bibinfo
  {journal} {Physical Review Applied}\ }\textbf {\bibinfo {volume} {17}},\
  \bibinfo {pages} {024062} (\bibinfo {year} {2022})}\BibitemShut {NoStop}%
\bibitem [{\citenamefont {Galda}\ \emph {et~al.}(2021)\citenamefont {Galda},
  \citenamefont {Cubeddu}, \citenamefont {Kanazawa}, \citenamefont {Narang},\
  and\ \citenamefont {Earnest-Noble}}]{galda2021implementing}%
  \BibitemOpen
  \bibfield  {author} {\bibinfo {author} {\bibfnamefont {Alexey}\ \bibnamefont
  {Galda}}, \bibinfo {author} {\bibfnamefont {Michael}\ \bibnamefont
  {Cubeddu}}, \bibinfo {author} {\bibfnamefont {Naoki}\ \bibnamefont
  {Kanazawa}}, \bibinfo {author} {\bibfnamefont {Prineha}\ \bibnamefont
  {Narang}}, \ and\ \bibinfo {author} {\bibfnamefont {Nathan}\ \bibnamefont
  {Earnest-Noble}},\ }\bibfield  {title} {\enquote {\bibinfo {title}
  {Implementing a ternary decomposition of the {Toffoli} gate on
  fixed-frequency transmon qutrits},}\ }\href {http://arxiv.org/abs/2109.00558}
  {\bibfield  {journal} {\bibinfo  {journal} {arXiv:2109.00558 [quant-ph]}\ }
  (\bibinfo {year} {2021})}\BibitemShut {NoStop}%
\bibitem [{\citenamefont {Egger}\ \emph {et~al.}(2018)\citenamefont {Egger},
  \citenamefont {Werninghaus}, \citenamefont {Ganzhorn}, \citenamefont {Salis},
  \citenamefont {Fuhrer}, \citenamefont {Müller},\ and\ \citenamefont
  {Filipp}}]{egger2018pulsed}%
  \BibitemOpen
  \bibfield  {author} {\bibinfo {author} {\bibfnamefont {Daniel~J.}\
  \bibnamefont {Egger}}, \bibinfo {author} {\bibfnamefont {Max}\ \bibnamefont
  {Werninghaus}}, \bibinfo {author} {\bibfnamefont {Marc}\ \bibnamefont
  {Ganzhorn}}, \bibinfo {author} {\bibfnamefont {Gian}\ \bibnamefont {Salis}},
  \bibinfo {author} {\bibfnamefont {Andreas}\ \bibnamefont {Fuhrer}}, \bibinfo
  {author} {\bibfnamefont {Peter}\ \bibnamefont {Müller}}, \ and\ \bibinfo
  {author} {\bibfnamefont {Stefan}\ \bibnamefont {Filipp}},\ }\bibfield
  {title} {\enquote {\bibinfo {title} {Pulsed reset protocol for
  fixed-frequency superconducting qubits},}\ }\href {\doibase
  10.1103/PhysRevApplied.10.044030} {\bibfield  {journal} {\bibinfo  {journal}
  {Physical Review Applied}\ }\textbf {\bibinfo {volume} {10}},\ \bibinfo
  {pages} {044030} (\bibinfo {year} {2018})}\BibitemShut {NoStop}%
\bibitem [{\citenamefont {Egger}\ \emph {et~al.}(2019)\citenamefont {Egger},
  \citenamefont {Ganzhorn}, \citenamefont {Salis}, \citenamefont {Fuhrer},
  \citenamefont {Müller}, \citenamefont {Barkoutsos}, \citenamefont {Moll},
  \citenamefont {Tavernelli},\ and\ \citenamefont
  {Filipp}}]{egger2019entanglement}%
  \BibitemOpen
  \bibfield  {author} {\bibinfo {author} {\bibfnamefont {Daniel~J.}\
  \bibnamefont {Egger}}, \bibinfo {author} {\bibfnamefont {Max}\ \bibnamefont
  {Ganzhorn}}, \bibinfo {author} {\bibfnamefont {Gian}\ \bibnamefont {Salis}},
  \bibinfo {author} {\bibfnamefont {Andreas}\ \bibnamefont {Fuhrer}}, \bibinfo
  {author} {\bibfnamefont {Peter}\ \bibnamefont {Müller}}, \bibinfo {author}
  {\bibfnamefont {Panagiotis~Kl.}\ \bibnamefont {Barkoutsos}}, \bibinfo
  {author} {\bibfnamefont {Nikolaj}\ \bibnamefont {Moll}}, \bibinfo {author}
  {\bibfnamefont {Ivano}\ \bibnamefont {Tavernelli}}, \ and\ \bibinfo {author}
  {\bibfnamefont {Stefan}\ \bibnamefont {Filipp}},\ }\bibfield  {title}
  {\enquote {\bibinfo {title} {Entanglement generation in superconducting
  qubits using holonomic operations},}\ }\href {\doibase
  10.1103/PhysRevApplied.11.014017} {\bibfield  {journal} {\bibinfo  {journal}
  {Physical Review Applied}\ }\textbf {\bibinfo {volume} {11}},\ \bibinfo
  {pages} {014017} (\bibinfo {year} {2019})}\BibitemShut {NoStop}%
\bibitem [{\citenamefont {Catelani}\ \emph {et~al.}(2012)\citenamefont
  {Catelani}, \citenamefont {Nigg}, \citenamefont {Girvin}, \citenamefont
  {Schoelkopf},\ and\ \citenamefont {Glazman}}]{catelani2012decoherence}%
  \BibitemOpen
  \bibfield  {author} {\bibinfo {author} {\bibfnamefont {Gianluigi}\
  \bibnamefont {Catelani}}, \bibinfo {author} {\bibfnamefont {Simon~E.}\
  \bibnamefont {Nigg}}, \bibinfo {author} {\bibfnamefont {Steven~M.}\
  \bibnamefont {Girvin}}, \bibinfo {author} {\bibfnamefont {Robert~J.}\
  \bibnamefont {Schoelkopf}}, \ and\ \bibinfo {author} {\bibfnamefont
  {Leonid~I.}\ \bibnamefont {Glazman}},\ }\bibfield  {title} {\enquote
  {\bibinfo {title} {Decoherence of superconducting qubits caused by
  quasiparticle tunneling},}\ }\href {\doibase 10.1103/PhysRevB.86.184514}
  {\bibfield  {journal} {\bibinfo  {journal} {Physical Review B}\ }\textbf
  {\bibinfo {volume} {86}},\ \bibinfo {pages} {184514} (\bibinfo {year}
  {2012})}\BibitemShut {NoStop}%
\bibitem [{\citenamefont {Wallraff}\ \emph {et~al.}(2005)\citenamefont
  {Wallraff}, \citenamefont {Schuster}, \citenamefont {Blais}, \citenamefont
  {Frunzio}, \citenamefont {Majer}, \citenamefont {Devoret}, \citenamefont
  {Girvin},\ and\ \citenamefont {Schoelkopf}}]{wallraff2005approaching}%
  \BibitemOpen
  \bibfield  {author} {\bibinfo {author} {\bibfnamefont {Andreas}\ \bibnamefont
  {Wallraff}}, \bibinfo {author} {\bibfnamefont {David~I.}\ \bibnamefont
  {Schuster}}, \bibinfo {author} {\bibfnamefont {Alexandre}\ \bibnamefont
  {Blais}}, \bibinfo {author} {\bibfnamefont {Luigi}\ \bibnamefont {Frunzio}},
  \bibinfo {author} {\bibfnamefont {Johannes}\ \bibnamefont {Majer}}, \bibinfo
  {author} {\bibfnamefont {Michel~H.}\ \bibnamefont {Devoret}}, \bibinfo
  {author} {\bibfnamefont {Steven~M.}\ \bibnamefont {Girvin}}, \ and\ \bibinfo
  {author} {\bibfnamefont {Robert~J.}\ \bibnamefont {Schoelkopf}},\ }\bibfield
  {title} {\enquote {\bibinfo {title} {Approaching {unit} {visibility} for
  {control} of a {superconducting} {qubit} with {dispersive} {readout}},}\
  }\href {\doibase 10.1103/PhysRevLett.95.060501} {\bibfield  {journal}
  {\bibinfo  {journal} {Physical Review Letters}\ }\textbf {\bibinfo {volume}
  {95}},\ \bibinfo {pages} {060501} (\bibinfo {year} {2005})}\BibitemShut
  {NoStop}%
\bibitem [{\citenamefont {Blok}\ \emph {et~al.}(2021)\citenamefont {Blok},
  \citenamefont {Ramasesh}, \citenamefont {Schuster}, \citenamefont
  {O’Brien}, \citenamefont {Kreikebaum}, \citenamefont {Dahlen},
  \citenamefont {Morvan}, \citenamefont {Yoshida}, \citenamefont {Yao},\ and\
  \citenamefont {Siddiqi}}]{blok2021quantum}%
  \BibitemOpen
  \bibfield  {author} {\bibinfo {author} {\bibfnamefont {Machiel~S.}\
  \bibnamefont {Blok}}, \bibinfo {author} {\bibfnamefont {Vinay~V.}\
  \bibnamefont {Ramasesh}}, \bibinfo {author} {\bibfnamefont {Thomas}\
  \bibnamefont {Schuster}}, \bibinfo {author} {\bibfnamefont {Kevin}\
  \bibnamefont {O’Brien}}, \bibinfo {author} {\bibfnamefont {John-Mark}\
  \bibnamefont {Kreikebaum}}, \bibinfo {author} {\bibfnamefont {Dar}\
  \bibnamefont {Dahlen}}, \bibinfo {author} {\bibfnamefont {Alexis}\
  \bibnamefont {Morvan}}, \bibinfo {author} {\bibfnamefont {Beni}\ \bibnamefont
  {Yoshida}}, \bibinfo {author} {\bibfnamefont {Norman~Y.}\ \bibnamefont
  {Yao}}, \ and\ \bibinfo {author} {\bibfnamefont {Irfan}\ \bibnamefont
  {Siddiqi}},\ }\bibfield  {title} {\enquote {\bibinfo {title} {Quantum
  {information} {scrambling} on a {superconducting} {qutrit} {processor}},}\
  }\href {\doibase 10.1103/PhysRevX.11.021010} {\bibfield  {journal} {\bibinfo
  {journal} {Physical Review X}\ }\textbf {\bibinfo {volume} {11}},\ \bibinfo
  {pages} {021010} (\bibinfo {year} {2021})}\BibitemShut {NoStop}%
\bibitem [{\citenamefont {Motzoi}\ \emph {et~al.}(2009)\citenamefont {Motzoi},
  \citenamefont {Gambetta}, \citenamefont {Rebentrost},\ and\ \citenamefont
  {Wilhelm}}]{motzoi2009simple}%
  \BibitemOpen
  \bibfield  {author} {\bibinfo {author} {\bibfnamefont {Felix}\ \bibnamefont
  {Motzoi}}, \bibinfo {author} {\bibfnamefont {Jay~M.}\ \bibnamefont
  {Gambetta}}, \bibinfo {author} {\bibfnamefont {Patrick}\ \bibnamefont
  {Rebentrost}}, \ and\ \bibinfo {author} {\bibfnamefont {Frank~K.}\
  \bibnamefont {Wilhelm}},\ }\bibfield  {title} {\enquote {\bibinfo {title}
  {Simple {pulses} for {elimination} of {leakage} in {weakly} {nonlinear}
  {qubits}},}\ }\href {\doibase 10.1103/PhysRevLett.103.110501} {\bibfield
  {journal} {\bibinfo  {journal} {Physical Review Letters}\ }\textbf {\bibinfo
  {volume} {103}},\ \bibinfo {pages} {110501} (\bibinfo {year}
  {2009})}\BibitemShut {NoStop}%
\bibitem [{\citenamefont {Gambetta}\ \emph {et~al.}(2011)\citenamefont
  {Gambetta}, \citenamefont {Motzoi}, \citenamefont {Merkel},\ and\
  \citenamefont {Wilhelm}}]{gambetta2011analytic}%
  \BibitemOpen
  \bibfield  {author} {\bibinfo {author} {\bibfnamefont {Jay.~M.}\ \bibnamefont
  {Gambetta}}, \bibinfo {author} {\bibfnamefont {Felix}\ \bibnamefont
  {Motzoi}}, \bibinfo {author} {\bibfnamefont {Seth~T.}\ \bibnamefont
  {Merkel}}, \ and\ \bibinfo {author} {\bibfnamefont {Frank~K.}\ \bibnamefont
  {Wilhelm}},\ }\bibfield  {title} {\enquote {\bibinfo {title} {Analytic
  control methods for high-fidelity unitary operations in a weakly nonlinear
  oscillator},}\ }\href {\doibase 10.1103/PhysRevA.83.012308} {\bibfield
  {journal} {\bibinfo  {journal} {Physical Review A}\ }\textbf {\bibinfo
  {volume} {83}},\ \bibinfo {pages} {012308} (\bibinfo {year}
  {2011})}\BibitemShut {NoStop}%
\bibitem [{\citenamefont {Werninghaus}\ \emph {et~al.}(2021)\citenamefont
  {Werninghaus}, \citenamefont {Egger}, \citenamefont {Roy}, \citenamefont
  {Machnes}, \citenamefont {Wilhelm},\ and\ \citenamefont
  {Filipp}}]{werninghaus2021leakage}%
  \BibitemOpen
  \bibfield  {author} {\bibinfo {author} {\bibfnamefont {Max.}\ \bibnamefont
  {Werninghaus}}, \bibinfo {author} {\bibfnamefont {Daniel~J.}\ \bibnamefont
  {Egger}}, \bibinfo {author} {\bibfnamefont {Federico}\ \bibnamefont {Roy}},
  \bibinfo {author} {\bibfnamefont {Shai}\ \bibnamefont {Machnes}}, \bibinfo
  {author} {\bibfnamefont {Frank~K.}\ \bibnamefont {Wilhelm}}, \ and\ \bibinfo
  {author} {\bibfnamefont {Stefan}\ \bibnamefont {Filipp}},\ }\bibfield
  {title} {\enquote {\bibinfo {title} {Leakage reduction in fast
  superconducting qubit gates via optimal control},}\ }\href {\doibase
  10.1038/s41534-020-00346-2} {\bibfield  {journal} {\bibinfo  {journal} {npj
  Quantum Information}\ }\textbf {\bibinfo {volume} {7}},\ \bibinfo {pages}
  {14} (\bibinfo {year} {2021})}\BibitemShut {NoStop}%
\bibitem [{\citenamefont {Alexander}\ \emph {et~al.}(2020)\citenamefont
  {Alexander}, \citenamefont {Kanazawa}, \citenamefont {Egger}, \citenamefont
  {Capelluto}, \citenamefont {Wood}, \citenamefont {Javadi-Abhari},\ and\
  \citenamefont {C~McKay}}]{alexander2020qiskit}%
  \BibitemOpen
  \bibfield  {author} {\bibinfo {author} {\bibfnamefont {Thomas}\ \bibnamefont
  {Alexander}}, \bibinfo {author} {\bibfnamefont {Naoki}\ \bibnamefont
  {Kanazawa}}, \bibinfo {author} {\bibfnamefont {Daniel~J}\ \bibnamefont
  {Egger}}, \bibinfo {author} {\bibfnamefont {Lauren}\ \bibnamefont
  {Capelluto}}, \bibinfo {author} {\bibfnamefont {Christopher~J}\ \bibnamefont
  {Wood}}, \bibinfo {author} {\bibfnamefont {Ali}\ \bibnamefont
  {Javadi-Abhari}}, \ and\ \bibinfo {author} {\bibfnamefont {David}\
  \bibnamefont {C~McKay}},\ }\bibfield  {title} {\enquote {\bibinfo {title}
  {Qiskit pulse: programming quantum computers through the cloud with
  pulses},}\ }\href {\doibase 10.1088/2058-9565/aba404} {\bibfield  {journal}
  {\bibinfo  {journal} {Quantum Science and Technology}\ }\textbf {\bibinfo
  {volume} {5}},\ \bibinfo {pages} {044006} (\bibinfo {year}
  {2020})}\BibitemShut {NoStop}%
\bibitem [{\citenamefont {McKay}\ \emph {et~al.}(2018)\citenamefont {McKay},
  \citenamefont {Alexander}, \citenamefont {Bello}, \citenamefont {Biercuk},
  \citenamefont {Bishop}, \citenamefont {Chen}, \citenamefont {Chow},
  \citenamefont {Córcoles}, \citenamefont {Egger}, \citenamefont {Filipp},
  \citenamefont {Gomez}, \citenamefont {Hush}, \citenamefont {Javadi-Abhari},
  \citenamefont {Moreda}, \citenamefont {Nation}, \citenamefont {Paulovicks},
  \citenamefont {Winston}, \citenamefont {Wood}, \citenamefont {Wootton},\ and\
  \citenamefont {Gambetta}}]{mckay2018qiskit}%
  \BibitemOpen
  \bibfield  {author} {\bibinfo {author} {\bibfnamefont {David~C.}\
  \bibnamefont {McKay}}, \bibinfo {author} {\bibfnamefont {Thomas}\
  \bibnamefont {Alexander}}, \bibinfo {author} {\bibfnamefont {Luciano}\
  \bibnamefont {Bello}}, \bibinfo {author} {\bibfnamefont {Michael~J.}\
  \bibnamefont {Biercuk}}, \bibinfo {author} {\bibfnamefont {Lev}\ \bibnamefont
  {Bishop}}, \bibinfo {author} {\bibfnamefont {Jiayin}\ \bibnamefont {Chen}},
  \bibinfo {author} {\bibfnamefont {Jerry~M.}\ \bibnamefont {Chow}}, \bibinfo
  {author} {\bibfnamefont {Antonio~D.}\ \bibnamefont {Córcoles}}, \bibinfo
  {author} {\bibfnamefont {Daniel}\ \bibnamefont {Egger}}, \bibinfo {author}
  {\bibfnamefont {Stefan}\ \bibnamefont {Filipp}}, \bibinfo {author}
  {\bibfnamefont {Juan}\ \bibnamefont {Gomez}}, \bibinfo {author}
  {\bibfnamefont {Michael}\ \bibnamefont {Hush}}, \bibinfo {author}
  {\bibfnamefont {Ali}\ \bibnamefont {Javadi-Abhari}}, \bibinfo {author}
  {\bibfnamefont {Diego}\ \bibnamefont {Moreda}}, \bibinfo {author}
  {\bibfnamefont {Paul}\ \bibnamefont {Nation}}, \bibinfo {author}
  {\bibfnamefont {Brent}\ \bibnamefont {Paulovicks}}, \bibinfo {author}
  {\bibfnamefont {Erick}\ \bibnamefont {Winston}}, \bibinfo {author}
  {\bibfnamefont {Christopher~J.}\ \bibnamefont {Wood}}, \bibinfo {author}
  {\bibfnamefont {James}\ \bibnamefont {Wootton}}, \ and\ \bibinfo {author}
  {\bibfnamefont {Jay~M.}\ \bibnamefont {Gambetta}},\ }\bibfield  {title}
  {\enquote {\bibinfo {title} {Qiskit {backend} {specifications} for {OpenQASM}
  and {OpenPulse} {experiments}},}\ }\href {http://arxiv.org/abs/1809.03452}
  {\bibfield  {journal} {\bibinfo  {journal} {arXiv:1809.03452 [quant-ph]}\ }
  (\bibinfo {year} {2018})}\BibitemShut {NoStop}%
\bibitem [{\citenamefont {D’Ariano}\ \emph {et~al.}(2004)\citenamefont
  {D’Ariano}, \citenamefont {Maccone},\ and\ \citenamefont
  {Presti}}]{dariano2004quantum}%
  \BibitemOpen
  \bibfield  {author} {\bibinfo {author} {\bibfnamefont {Giacomo~Mauro}\
  \bibnamefont {D’Ariano}}, \bibinfo {author} {\bibfnamefont {Lorenzo}\
  \bibnamefont {Maccone}}, \ and\ \bibinfo {author} {\bibfnamefont
  {Paoloplacido~Lo}\ \bibnamefont {Presti}},\ }\bibfield  {title} {\enquote
  {\bibinfo {title} {Quantum {calibration} of {measurement}
  {instrumentation}},}\ }\href {\doibase 10.1103/PhysRevLett.93.250407}
  {\bibfield  {journal} {\bibinfo  {journal} {Physical Review Letters}\
  }\textbf {\bibinfo {volume} {93}},\ \bibinfo {pages} {250407} (\bibinfo
  {year} {2004})}\BibitemShut {NoStop}%
\bibitem [{\citenamefont {Lundeen}\ \emph {et~al.}(2009)\citenamefont
  {Lundeen}, \citenamefont {Feito}, \citenamefont {Coldenstrodt-Ronge},
  \citenamefont {Pregnell}, \citenamefont {Silberhorn}, \citenamefont {Ralph},
  \citenamefont {Eisert}, \citenamefont {Plenio},\ and\ \citenamefont
  {Walmsley}}]{lundeen2009tomography}%
  \BibitemOpen
  \bibfield  {author} {\bibinfo {author} {\bibfnamefont {Jeff~S.}\ \bibnamefont
  {Lundeen}}, \bibinfo {author} {\bibfnamefont {Alvaro}\ \bibnamefont {Feito}},
  \bibinfo {author} {\bibfnamefont {Hendrik}\ \bibnamefont
  {Coldenstrodt-Ronge}}, \bibinfo {author} {\bibfnamefont {Kenny~L.}\
  \bibnamefont {Pregnell}}, \bibinfo {author} {\bibfnamefont {Christine}\
  \bibnamefont {Silberhorn}}, \bibinfo {author} {\bibfnamefont {Timothy~C.}\
  \bibnamefont {Ralph}}, \bibinfo {author} {\bibfnamefont {Jens}\ \bibnamefont
  {Eisert}}, \bibinfo {author} {\bibfnamefont {Martin~B.}\ \bibnamefont
  {Plenio}}, \ and\ \bibinfo {author} {\bibfnamefont {Ian~A.}\ \bibnamefont
  {Walmsley}},\ }\bibfield  {title} {\enquote {\bibinfo {title} {Tomography of
  quantum detectors},}\ }\href {\doibase 10.1038/nphys1133} {\bibfield
  {journal} {\bibinfo  {journal} {Nature Physics}\ }\textbf {\bibinfo {volume}
  {5}},\ \bibinfo {pages} {27--30} (\bibinfo {year} {2009})}\BibitemShut
  {NoStop}%
\bibitem [{\citenamefont {Fiurášek}(2001)}]{fiurasek2001maximumlikelihood}%
  \BibitemOpen
  \bibfield  {author} {\bibinfo {author} {\bibfnamefont {Jaromír}\
  \bibnamefont {Fiurášek}},\ }\bibfield  {title} {\enquote {\bibinfo {title}
  {Maximum-likelihood estimation of quantum measurement},}\ }\href {\doibase
  10.1103/PhysRevA.64.024102} {\bibfield  {journal} {\bibinfo  {journal}
  {Physical Review A}\ }\textbf {\bibinfo {volume} {64}},\ \bibinfo {pages}
  {024102} (\bibinfo {year} {2001})}\BibitemShut {NoStop}%
\bibitem [{\citenamefont {Maciejewski}\ \emph {et~al.}(2020)\citenamefont
  {Maciejewski}, \citenamefont {Zimborás},\ and\ \citenamefont
  {Oszmaniec}}]{maciejewski2020mitigation}%
  \BibitemOpen
  \bibfield  {author} {\bibinfo {author} {\bibfnamefont {Filip~B.}\
  \bibnamefont {Maciejewski}}, \bibinfo {author} {\bibfnamefont {Zoltán}\
  \bibnamefont {Zimborás}}, \ and\ \bibinfo {author} {\bibfnamefont {Michał}\
  \bibnamefont {Oszmaniec}},\ }\bibfield  {title} {\enquote {\bibinfo {title}
  {Mitigation of readout noise in near-term quantum devices by classical
  post-processing based on detector tomography},}\ }\href {\doibase
  10.22331/q-2020-04-24-257} {\bibfield  {journal} {\bibinfo  {journal}
  {Quantum}\ }\textbf {\bibinfo {volume} {4}},\ \bibinfo {pages} {257}
  (\bibinfo {year} {2020})}\BibitemShut {NoStop}%
\bibitem [{\citenamefont {Puchała}\ \emph {et~al.}(2018)\citenamefont
  {Puchała}, \citenamefont {Pawela}, \citenamefont {Krawiec},\ and\
  \citenamefont {Kukulski}}]{puchala2018strategies}%
  \BibitemOpen
  \bibfield  {author} {\bibinfo {author} {\bibfnamefont {Zbigniew}\
  \bibnamefont {Puchała}}, \bibinfo {author} {\bibfnamefont {Łukasz}\
  \bibnamefont {Pawela}}, \bibinfo {author} {\bibfnamefont {Aleksandra}\
  \bibnamefont {Krawiec}}, \ and\ \bibinfo {author} {\bibfnamefont {Ryszard}\
  \bibnamefont {Kukulski}},\ }\bibfield  {title} {\enquote {\bibinfo {title}
  {Strategies for optimal single-shot discrimination of quantum
  measurements},}\ }\href {\doibase 10.1103/PhysRevA.98.042103} {\bibfield
  {journal} {\bibinfo  {journal} {Physical Review A}\ }\textbf {\bibinfo
  {volume} {98}},\ \bibinfo {pages} {042103} (\bibinfo {year}
  {2018})}\BibitemShut {NoStop}%
\bibitem [{\citenamefont {Carrasquilla}\ \emph {et~al.}(2019)\citenamefont
  {Carrasquilla}, \citenamefont {Torlai}, \citenamefont {Melko},\ and\
  \citenamefont {Aolita}}]{carrasquilla2019reconstructing}%
  \BibitemOpen
  \bibfield  {author} {\bibinfo {author} {\bibfnamefont {Juan}\ \bibnamefont
  {Carrasquilla}}, \bibinfo {author} {\bibfnamefont {Giacomo}\ \bibnamefont
  {Torlai}}, \bibinfo {author} {\bibfnamefont {Roger~G.}\ \bibnamefont
  {Melko}}, \ and\ \bibinfo {author} {\bibfnamefont {Leandro}\ \bibnamefont
  {Aolita}},\ }\bibfield  {title} {\enquote {\bibinfo {title} {Reconstructing
  quantum states with generative models},}\ }\href {\doibase
  10.1038/s42256-019-0028-1} {\bibfield  {journal} {\bibinfo  {journal} {Nature
  Machine Intelligence}\ }\textbf {\bibinfo {volume} {1}},\ \bibinfo {pages}
  {155--161} (\bibinfo {year} {2019})}\BibitemShut {NoStop}%
\bibitem [{\citenamefont {Acharya}\ \emph {et~al.}(2021)\citenamefont
  {Acharya}, \citenamefont {Saha},\ and\ \citenamefont
  {Sengupta}}]{acharya2021shadow}%
  \BibitemOpen
  \bibfield  {author} {\bibinfo {author} {\bibfnamefont {Atithi}\ \bibnamefont
  {Acharya}}, \bibinfo {author} {\bibfnamefont {Siddhartha}\ \bibnamefont
  {Saha}}, \ and\ \bibinfo {author} {\bibfnamefont {Anirvan~M.}\ \bibnamefont
  {Sengupta}},\ }\bibfield  {title} {\enquote {\bibinfo {title} {Shadow
  tomography based on informationally complete positive operator-valued
  measure},}\ }\href {\doibase 10.1103/PhysRevA.104.052418} {\bibfield
  {journal} {\bibinfo  {journal} {Physical Review A}\ }\textbf {\bibinfo
  {volume} {104}},\ \bibinfo {pages} {052418} (\bibinfo {year}
  {2021})}\BibitemShut {NoStop}%
\bibitem [{\citenamefont {Gambetta}(2013)}]{gambetta2013quantum}%
  \BibitemOpen
  \bibfield  {author} {\bibinfo {author} {\bibfnamefont {Jay~M.}\ \bibnamefont
  {Gambetta}},\ }\href@noop {} {\emph {\bibinfo {title} {Quantum {information}
  {processing} - {lecture} {notes} of the 44th {IFF} {spring} {school}
  2013}}},\ edited by\ \bibinfo {editor} {\bibfnamefont {David~P.}\
  \bibnamefont {DiVincenzo}}\ (\bibinfo  {publisher} {Forschungszentrum
  Jülich, Zentralbibliothek},\ \bibinfo {year} {2013})\ \bibinfo {note}
  {section: Control of Superconducting Qubits}\BibitemShut {NoStop}%
\bibitem [{\citenamefont {Schreier}\ \emph {et~al.}(2008)\citenamefont
  {Schreier}, \citenamefont {Houck}, \citenamefont {Koch}, \citenamefont
  {Schuster}, \citenamefont {Johnson}, \citenamefont {Chow}, \citenamefont
  {Gambetta}, \citenamefont {Majer}, \citenamefont {Frunzio}, \citenamefont
  {Devoret}, \citenamefont {Girvin},\ and\ \citenamefont
  {Schoelkopf}}]{schreier2008suppressing}%
  \BibitemOpen
  \bibfield  {author} {\bibinfo {author} {\bibfnamefont {Joseph~A.}\
  \bibnamefont {Schreier}}, \bibinfo {author} {\bibfnamefont {Andrew~A.}\
  \bibnamefont {Houck}}, \bibinfo {author} {\bibfnamefont {Jens}\ \bibnamefont
  {Koch}}, \bibinfo {author} {\bibfnamefont {David~I.}\ \bibnamefont
  {Schuster}}, \bibinfo {author} {\bibfnamefont {Bradley~R.}\ \bibnamefont
  {Johnson}}, \bibinfo {author} {\bibfnamefont {Jerry~M.}\ \bibnamefont
  {Chow}}, \bibinfo {author} {\bibfnamefont {Jay~M.}\ \bibnamefont {Gambetta}},
  \bibinfo {author} {\bibfnamefont {JJohannes}\ \bibnamefont {Majer}}, \bibinfo
  {author} {\bibfnamefont {Luigi}\ \bibnamefont {Frunzio}}, \bibinfo {author}
  {\bibfnamefont {Michel~H.}\ \bibnamefont {Devoret}}, \bibinfo {author}
  {\bibfnamefont {Steven~M.}\ \bibnamefont {Girvin}}, \ and\ \bibinfo {author}
  {\bibfnamefont {Robert~J.}\ \bibnamefont {Schoelkopf}},\ }\bibfield  {title}
  {\enquote {\bibinfo {title} {Suppressing charge noise decoherence in
  superconducting charge qubits},}\ }\href {\doibase
  10.1103/PhysRevB.77.180502} {\bibfield  {journal} {\bibinfo  {journal}
  {Physical Review B}\ }\textbf {\bibinfo {volume} {77}},\ \bibinfo {pages}
  {180502} (\bibinfo {year} {2008})}\BibitemShut {NoStop}%
\bibitem [{\citenamefont {Ristè}\ \emph {et~al.}(2013)\citenamefont {Ristè},
  \citenamefont {Bultink}, \citenamefont {Tiggelman}, \citenamefont {Schouten},
  \citenamefont {Lehnert},\ and\ \citenamefont
  {DiCarlo}}]{riste2013millisecond}%
  \BibitemOpen
  \bibfield  {author} {\bibinfo {author} {\bibfnamefont {Diego}\ \bibnamefont
  {Ristè}}, \bibinfo {author} {\bibfnamefont {Niels}\ \bibnamefont {Bultink}},
  \bibinfo {author} {\bibfnamefont {Marijn~J.}\ \bibnamefont {Tiggelman}},
  \bibinfo {author} {\bibfnamefont {Raymond~N.}\ \bibnamefont {Schouten}},
  \bibinfo {author} {\bibfnamefont {Konrad~W.}\ \bibnamefont {Lehnert}}, \ and\
  \bibinfo {author} {\bibfnamefont {Leonardo}\ \bibnamefont {DiCarlo}},\
  }\bibfield  {title} {\enquote {\bibinfo {title} {Millisecond charge-parity
  fluctuations and induced decoherence in a superconducting transmon qubit},}\
  }\href {\doibase 10.1038/ncomms2936} {\bibfield  {journal} {\bibinfo
  {journal} {Nature Communications}\ }\textbf {\bibinfo {volume} {4}},\
  \bibinfo {pages} {1913} (\bibinfo {year} {2013})}\BibitemShut {NoStop}%
\bibitem [{\citenamefont {Neugebauer}\ \emph {et~al.}(2020)\citenamefont
  {Neugebauer}, \citenamefont {Fischer}, \citenamefont {Jäger}, \citenamefont
  {Czischek}, \citenamefont {Jochim}, \citenamefont {Weidemüller},\ and\
  \citenamefont {Gärttner}}]{neugebauer2020neuralnetwork}%
  \BibitemOpen
  \bibfield  {author} {\bibinfo {author} {\bibfnamefont {Marcel}\ \bibnamefont
  {Neugebauer}}, \bibinfo {author} {\bibfnamefont {Laurin~E.}\ \bibnamefont
  {Fischer}}, \bibinfo {author} {\bibfnamefont {Alexander}\ \bibnamefont
  {Jäger}}, \bibinfo {author} {\bibfnamefont {Stefanie}\ \bibnamefont
  {Czischek}}, \bibinfo {author} {\bibfnamefont {Selim}\ \bibnamefont
  {Jochim}}, \bibinfo {author} {\bibfnamefont {Matthias}\ \bibnamefont
  {Weidemüller}}, \ and\ \bibinfo {author} {\bibfnamefont {Martin}\
  \bibnamefont {Gärttner}},\ }\bibfield  {title} {\enquote {\bibinfo {title}
  {Neural-network quantum state tomography in a two-qubit experiment},}\ }\href
  {\doibase 10.1103/PhysRevA.102.042604} {\bibfield  {journal} {\bibinfo
  {journal} {Physical Review A}\ }\textbf {\bibinfo {volume} {102}},\ \bibinfo
  {pages} {042604} (\bibinfo {year} {2020})}\BibitemShut {NoStop}%
\bibitem [{\citenamefont {Hradil}\ \emph {et~al.}(2004)\citenamefont {Hradil},
  \citenamefont {Řeháček}, \citenamefont {Fiurášek},\ and\ \citenamefont
  {Ježek}}]{hradil2004maximumlikelihood}%
  \BibitemOpen
  \bibfield  {author} {\bibinfo {author} {\bibfnamefont {Zdeněk}\ \bibnamefont
  {Hradil}}, \bibinfo {author} {\bibfnamefont {Jaroslav}\ \bibnamefont
  {Řeháček}}, \bibinfo {author} {\bibfnamefont {Jaromír}\ \bibnamefont
  {Fiurášek}}, \ and\ \bibinfo {author} {\bibfnamefont {Miroslav}\
  \bibnamefont {Ježek}},\ }\bibfield  {title} {\enquote {\bibinfo {title} {3
  {maximum}-{likelihood} {methods in} {quantum} {mechanics}},}\ }in\ \href
  {\doibase 10.1007/978-3-540-44481-7_3} {\emph {\bibinfo {booktitle} {Quantum
  {State} {Estimation}}}},\ Vol.\ \bibinfo {volume} {649},\ \bibinfo {editor}
  {edited by\ \bibinfo {editor} {\bibfnamefont {Matteo}\ \bibnamefont {Paris}}\
  and\ \bibinfo {editor} {\bibfnamefont {Jaroslav}\ \bibnamefont
  {Řeháček}}}\ (\bibinfo  {publisher} {Springer Berlin Heidelberg},\
  \bibinfo {address} {Berlin, Heidelberg},\ \bibinfo {year} {2004})\ pp.\
  \bibinfo {pages} {59--112},\ \bibinfo {note} {series Title: Lecture Notes in
  Physics}\BibitemShut {NoStop}%
\bibitem [{\citenamefont {Chen}\ \emph {et~al.}(2019)\citenamefont {Chen},
  \citenamefont {Farahzad}, \citenamefont {Yoo},\ and\ \citenamefont
  {Wei}}]{chen2019detector}%
  \BibitemOpen
  \bibfield  {author} {\bibinfo {author} {\bibfnamefont {Yanzhu}\ \bibnamefont
  {Chen}}, \bibinfo {author} {\bibfnamefont {Maziar}\ \bibnamefont {Farahzad}},
  \bibinfo {author} {\bibfnamefont {Shinjae}\ \bibnamefont {Yoo}}, \ and\
  \bibinfo {author} {\bibfnamefont {Tzu-Chieh}\ \bibnamefont {Wei}},\
  }\bibfield  {title} {\enquote {\bibinfo {title} {Detector tomography on {IBM}
  quantum computers and mitigation of an imperfect measurement},}\ }\href
  {\doibase 10.1103/PhysRevA.100.052315} {\bibfield  {journal} {\bibinfo
  {journal} {Physical Review A}\ }\textbf {\bibinfo {volume} {100}},\ \bibinfo
  {pages} {052315} (\bibinfo {year} {2019})}\BibitemShut {NoStop}%
\bibitem [{\citenamefont {Johansson}\ \emph {et~al.}(2012)\citenamefont
  {Johansson}, \citenamefont {Nation},\ and\ \citenamefont
  {Nori}}]{johansson2012qutip}%
  \BibitemOpen
  \bibfield  {author} {\bibinfo {author} {\bibfnamefont {Robert}\ \bibnamefont
  {Johansson}}, \bibinfo {author} {\bibfnamefont {Paul~D.}\ \bibnamefont
  {Nation}}, \ and\ \bibinfo {author} {\bibfnamefont {Franco}\ \bibnamefont
  {Nori}},\ }\bibfield  {title} {\enquote {\bibinfo {title} {{QuTiP}: {An}
  open-source {Python} framework for the dynamics of open quantum systems},}\
  }\href {\doibase 10.1016/j.cpc.2012.02.021} {\bibfield  {journal} {\bibinfo
  {journal} {Computer Physics Communications}\ }\textbf {\bibinfo {volume}
  {183}},\ \bibinfo {pages} {1760--1772} (\bibinfo {year} {2012})}\BibitemShut
  {NoStop}%
\bibitem [{\citenamefont {Nielsen}(2002)}]{nielsen2002simple}%
  \BibitemOpen
  \bibfield  {author} {\bibinfo {author} {\bibfnamefont {Michael~A.}\
  \bibnamefont {Nielsen}},\ }\bibfield  {title} {\enquote {\bibinfo {title} {A
  simple formula for the average gate fidelity of a quantum dynamical
  operation},}\ }\href {\doibase 10.1016/S0375-9601(02)01272-0} {\bibfield
  {journal} {\bibinfo  {journal} {Physics Letters A}\ }\textbf {\bibinfo
  {volume} {303}},\ \bibinfo {pages} {249--252} (\bibinfo {year}
  {2002})}\BibitemShut {NoStop}%
\end{thebibliography}%

\clearpage
\appendix

\section{Details on POVM implementations}
\label{app:povm_decomposition}
\subsection{Naimark construction for single-qubit POVMs}
\label{app_sec:POVM_ancilla_implementation}

Here, we detail the connection between a unitary $U$ applied to a four-dimensional extended Hilbert space $\mathcal{H}_\text{ext}$ and the POVM operators realized on the single-qubit space $\mathcal{H}_\text{S}$ through a Naimark dilation construction. 
In a tensor product extension (TPE), the four basis states of $\mathcal{H}_\text{ext}$ are formed with an ancilla qubit as
${\ket{0}_\text{ext} = \ket{0}_\text{S} \otimes \ket{0}_\text{A}}$, 
 $\ket{1}_\text{ext} = \ket{1}_\text{S} \otimes \ket{0}_\text{A}$, 
$\ket{2}_\text{ext} = \ket{0}_\text{S} \otimes \ket{1}_\text{A}$, and
$\ket{3}_\text{ext} = \ket{1}_\text{S} \otimes \ket{1}_\text{A}$.
In contrast, in a direct sum extension (DSE), the four states $\ket{i}_\text{ext}$ form a qudit space, where the qubit is encoded in the states $\ket{0}_\text{ext} \equiv \ket{0}_\text{S}$ and $\ket{1}_\text{ext} \equiv \ket{1}_\text{S}$.

For simplicity, we assume a pure state of the system qubit $\ket{\psi}_\text{S} = \alpha \ket{0}_\text{S} + \beta \ket{1}_\text{S}$.
A Naimark construction for both a TPE and a DSE applies a unitary $U$ to the initial state $\ket{\psi}_\text{init} = \alpha \ket{0}_\text{ext} + \beta \ket{1}_\text{ext}$, to create the final state
\begin{align}
U  \ket{\psi}_\text{init} 
= \sum_{m=0}^3 \left(\alpha U_{m, 0} + \beta U_{m, 1} \right) \ket{m}_\text{ext} .
\end{align}
Measuring $U  \ket{\psi}_\text{init} $ in $\mathcal{H}_\text{ext}$ produces an outcome $m \in \{0, 1, 2, 3\}$ with a probability $p_m = \left| U_{m, 0}\alpha + U_{m, 1} \beta \right|^2$. 
This is equal to the probabilities $p_m = \text{Tr}(\Pi^m \ket{\psi}_\text{S} \bra{\psi}_\text{S} )$ associated with a POVM of four rank-1 operators
\begin{align}
\label{eq_theo:POVM_ops_from_unitary}
\Pi^m = \Gamma_m \ket{\pi_m}\bra{\pi_m}
\end{align}
acting on $\mathcal{H}_\text{S}$, which are proportional to projectors along the states 
\begin{align}
\label{eq_theo:POVM_ops_state}
\ket{\pi_m} &=  \frac{1}{\sqrt{\Gamma_m}} \left( U_{m, 0}^* \ket{0}_\text{S} + U_{m, 1}^* \ket{1}_\text{S} \right)  
\end{align}
with normalization factors $\Gamma_m = | U_{m, 0}|^2 + |U_{m, 1}|^2$.
Through Eqs.~\eqref{eq_theo:POVM_ops_from_unitary} and~\eqref{eq_theo:POVM_ops_state}, the unitary $U$ applied to $\mathcal{H}_{\text{ext}}$ can emulate the measurement of any POVM with four rank-one operators on $\mathcal{H}_{\text{S}}$. 

Since the desired POVM only defines the first two columns of $U$, i.e., $U_{m,0}$ and $U_{m,1}$, we find the remaining columns with a Gram-Schmidt procedure to ensure $U$ is unitary.
Without loss of generality, for the decomposition algorithm presented in Sec.~\ref{app_sec:POVM_decomp_algorithm}, it is convenient to choose the top right element of $U$ to vanish, i.e., $U_{0, 3} = 0$. 

\subsection{Pulse decomposition for qudit-space POVMs}
\label{app_sec:POVM_decomp_algorithm}
Here, we review the decomposition algorithm that we use to realize the unitary $U$ with Givens rotations $\mathcal{G}$- and $\mathcal{Z}$-gates, as defined in the main text.
Since this algorithm decomposes special unitary operators, we first define the ${\rm SU}(4)$ operator $\mathcal{U}^{(0)}= U \det(U)^{-1/4}$, which encodes the same POVM as $U$.
The decomposition routine iteratively reduces $\mathcal{U}^{(0)}$ to the identity matrix through a sequence of gates. 
We denote the unitary after the $i$-th gate is applied to $\mathcal{U}^{(0)}$ by $\mathcal{U}^{(i+1)}$. 

The reduction to the identity matrix is accomplished by creating zeros in the off-diagonal entries starting from the top entry in the fourth column~\cite{schirmer2002constructive}.
Since by choice, the first entry $\mathcal{U}^{(0)}_{0,3}$ is already zero, we create a second zero in the fourth column of $\mathcal{U}^{(1)}$ with a Givens rotation $\mathcal{G}_{1\leftrightarrow 2}^{(0)}(\theta_1, \phi_1)$ that must satisfy $\mathcal{G}_{1\leftrightarrow 2}^{(0)} \big( 0, \mathcal{U}^{(0)}_{1, 3}, \mathcal{U}^{(0)}_{2, 3}, \mathcal{U}^{(0)}_{3, 3} \big)^T = \big(0, 0, \mathcal{U}^{(1)}_{2, 3}, \mathcal{U}^{(1)}_{3, 3}\big)^T $.
If $\mathcal{U}^{(0)}_{1, 3} = r_1 e^{i\delta_1}$ and $\mathcal{U}^{(0)}_{2, 3} = r_2 e^{i\delta_2}$ then the angles of the Givens rotation must be~\cite{schirmer2002constructive}
\begin{align}
    \label{eq_theo_decomp_algo_step_1}
    \theta_1 = 2 \arctan\left(\frac{r_1}{r_2}\right)\text{, and  } 
    \phi_1 = \dfrac{\pi}{2} - \delta_1 + \delta_2.
\end{align}
In the next iteration, we similarly apply another Givens rotation such that  $\mathcal{G}^{(1)}_{2\leftrightarrow 3} \big( 0, 0, \mathcal{U}^{(1)}_{2, 3}, \mathcal{U}^{(1)}_{3, 3} \big)^T = \big(0, 0, 0, \mathcal{U}^{(2)}_{2, 3}\big)^T$.
Due to unitarity, the remaining non-zero entry is a phase factor $\mathcal{U}^{(2)}_{2, 3} = e^{i\beta}$. 
A rotation $\mathcal{Z}^{(2)}_{2\leftrightarrow 3}$ with an angle $\varphi_z = -2\beta$ sets the phase $\beta$ to zero.
This finally results in the matrix
\begin{equation}
\mathcal{U}^{(3)} = \mathcal{Z}^{(2)}_{2\leftrightarrow 3} \mathcal{U}^{(2)} =
\left( 
\begin{array}{c | c} 
  \begin{array}{c c c} 
     \ast & \ast & \ast\\ 
     \ast & \ast & \ast\\ 
     \ast & \ast & \ast
  \end{array} &  
  \begin{array}{c} 
     0 \\ 
     0 \\ 
     0
  \end{array}
  \\ 
  \hline 
  \begin{array}{c c c} 
    0 & 0 & 0
  \end{array}
   & 1 
 \end{array} 
\right)
\end{equation}
that has been reduced to a $3\!\times\!3$ block. 
The above procedure is now repeated for the third column. 
This requires two Givens rotations and one $\mathcal{Z}$-rotation such that
\begin{align}
    \mathcal{Z}^{(5)}_{1\leftrightarrow 2} \mathcal{G}^{(4)}_{1\leftrightarrow 2} \mathcal{G}^{(3)}_{0\leftrightarrow 1} \big( \mathcal{U}^{(3)}_{0,2}, \mathcal{U}^{(3)}_{1, 2}, \mathcal{U}^{(3)}_{2, 2}, 0 \big)^T = (0, 0, 1, 0)^T.
\end{align}
Finally, applying the same strategy once more to the second column results in the identity matrix.
Our initial choice of an ${\rm SU}(4)$ matrix assures that the final phase of the top left entry vanishes. 

As a result, applying the inverse of all gates in reverse order gives a decomposition (up to an irrelevant global phase) of the target unitary $U$ into elementary operations of Givens rotations and (potentially virtual) $\mathcal{Z}$-gates:
\begin{align}
\label{eq_theo:U_final_decomposition_sequence}
U &= \mathcal{G}^{(0)\dagger}_{1\leftrightarrow 2} \,\,
    \mathcal{G}^{(1)\dagger}_{2\leftrightarrow 3}  \,\,
    \mathcal{Z}^{(2)\dagger}_{2\leftrightarrow 3}  \, \\
    & \hspace{15mm} \times 
    \mathcal{G}^{(3)\dagger}_{0\leftrightarrow 1}  \,\,
    \mathcal{G}^{(4)\dagger}_{1\leftrightarrow 2}  \,\,
    \mathcal{Z}^{(5)\dagger}_{1\leftrightarrow 2}  \,\,
    \mathcal{G}^{(6)\dagger}_{0\leftrightarrow 1}  \,\,
    \mathcal{Z}^{(7)\dagger}_{0\leftrightarrow 1}  \nonumber.
\end{align}

To simplify pulse calibrations, we restrict the gate set to virtual $\mathcal{Z}$-gates and the gates $\sqrt{\mathcal{X}} = \mathcal{G}(\theta\! =\! \frac{\pi}{2}, \phi\!=\!0)$ which describe a $\pi/2$-rotation around the $x$-axis between the states $\ket{n}$ and $\ket{n+1}$. A general Givens rotation $\mathcal{G}(\theta, \phi)$ can be exactly realized by a sequence of two $\sqrt{\mathcal{X}}$- and three $\mathcal{Z}$-gates
\begin{align}
\label{eq_theo:decomposition_into_SX}
\mathcal{G}(\theta, \phi) = 
 \mathcal{Z}_{}(\phi - \dfrac{\pi}{2})\,
\sqrt{\mathcal{X}}_{}\,
\mathcal{Z}_{}(\pi - \theta)\,
\sqrt{\mathcal{X}}_{}\,
\mathcal{Z}_{}(-\phi - \frac{\pi}{2})
\end{align}
where we have omitted the subscripts $n\!\leftrightarrow\! n+1$~\cite{mckay2017efficient}.
Replacing every $\mathcal{G}$-gate in Eq.~\eqref{eq_theo:U_final_decomposition_sequence} with the decomposition in Eq.~\eqref{eq_theo:decomposition_into_SX} results in a decomposition of $U$ that only contains ten $\sqrt{\mathcal{X}}$- and eleven $\mathcal{Z}$-gates.

For a qudit-space POVM realization in realistic hardware we need to apply the sequence of $\mathcal{G}$- or $\sqrt{\mathcal{X}}$- and $\mathcal{Z}$-gates through the hardware-native rotations $\mathcal{R}$ derived in Eq.~\eqref{eq:givens_rotation_hardware} of the main text.
It is sufficient to implement a unitary $\tilde{U}$ equivalent to the target unitary $U$ as long as the same measurement probabilities for any initial qubit state are recovered. 
In Sec.~\ref{app_sec:POVM_virtual_Z} and~\ref{app_sec:POVM_phases_Givens}, we  detail how to realize such an equivalent unitary $\tilde{U}$ with realistic $\mathcal{R}$-rotations.

\subsection{Generalized virtual Z-gates}
\label{app_sec:POVM_virtual_Z}

Each pulse as defined in Eq.~\eqref{eq_theo:pulse_drive_definition} in the main text is played in a \emph{frame} that consists of a carrier frequency $\omega$ and a phase $\phi$. 
For our implementation of the qudit-space unitary, three frames are relevant, which correspond to the three driven transitions,
i.e., ${0\!\leftrightarrow\!1}$, $1\!\leftrightarrow\!2$, and $2\!\leftrightarrow\!3$. 
While the frequencies of the drives in these frames always remain fixed to the transition energies of the system such that $\omega_{n} = E_{n+1} - E_{n}$, the phases of the frames need to be adjusted to account for phase advances during $\mathcal{R}$-rotations and to virtually implement $\mathcal{Z}$-gates.

As an example of a $\mathcal{Z}$-rotation in qudit space, consider the gate 
\begin{align}
\label{eq_app1:virtual_z_example}
\mathcal{Z}_{1\leftrightarrow 2}(\varphi) = \begin{pmatrix} 
    1 &  0 &  0& 0\\
     0 & e^{-i \frac{\varphi}{2}} & 0 & 0 \\
     0 & 0 & e^{i \frac{\varphi}{2}} & 0 \\
     0 & 0 & 0 & 1
    \end{pmatrix}
\end{align}
which applies a relative phase of $-\varphi$ between states $\ket{1}$ and $\ket{2}$. Therefore, an angle $\varphi$ needs to be  subtracted from the phase of all subsequent pulses played in the $1\!\leftrightarrow\!2$ frame. However, while the above gate leaves the levels $\ket{0}$ and $\ket{3}$ unchanged, it applies a relative phase of $\varphi/2$ between the levels $\ket{0}$ and $\ket{1}$, as well as between $\ket{2}$ and $\ket{3}$. Hence, in addition to affecting all following phases in the $1\!\leftrightarrow\!2$ frame, an angle $\varphi/2$ must be added to all drive phases in the $0\!\leftrightarrow\!1$ and $2\!\leftrightarrow\!3$ frames.     
In general, $\mathcal{Z}_{n\leftrightarrow n+1}(\varphi)$ gates can be virtually implemented by adding $\varphi/2$ to all subsequent pulses in the  $n+1 \leftrightarrow n+2$ and $n-1\leftrightarrow n$ frames while deducting a phase $\varphi$ from the following pulses in the $n \leftrightarrow n+1$ frame. 

\subsection{Correcting phase advances during $\mathcal{R}$-pulses}
\label{app_sec:POVM_phases_Givens}
While playing a pulse of a total duration $T$ in the $n\! \leftrightarrow\! n+1$ frame, the uncoupled levels acquire non-trivial phases, see Eq.~\eqref{eq:givens_rotation_hardware} of the main text.
It is instructive to look at an example of a drive in the $1\!\leftrightarrow\!2$ frame which implements the unitary
\begin{align}
\label{eq_app1:given_rotation_example_12}
\mathcal{R}_{1\leftrightarrow 2}(\theta, \phi) &=  e^{i( \omega_{1} - E_1)T} \\ 
& \times
\begin{pmatrix} 
    e^{-i (\omega_{1} - \omega_{0})T} &  \begin{matrix}0&\quad0\end{matrix} & 0\\
     \begin{matrix}0\\0\end{matrix} & G(\theta, \phi) & \begin{matrix}0\\0\end{matrix} \\
     0 &  \begin{matrix}0&\quad0\end{matrix}  & e^{-i (\omega_{2} - \omega_{1})T} \nonumber
\end{pmatrix}.
\end{align}
Let $\alpha_{n} = \omega_{n} - \omega_{n-1}$ for $n>0$, such that for an anharmonic oscillator, $\alpha_1$ would simply denote the anharmonicity. 
The above unitary results in a relative phase of $\Delta \phi_{0\leftrightarrow 1} = -\alpha_1 T$ between the states $\ket{0}$ and $\ket{1}$ and a relative phase of $\Delta \phi_{2\leftrightarrow 3}= \alpha_2 T$ between the states $\ket{2}$ and $\ket{3}$. 
To correct these phases, $\Delta \phi_{0\leftrightarrow 1} $  and $\Delta \phi_{2\leftrightarrow 3}$ have to be subtracted from the phases $\phi$ of all subsequent pulses in the $0\!\leftrightarrow\!1$ and $2\!\leftrightarrow\!3$ frames, respectively. 
Generalizing from this example, under a drive $\mathcal{R}_{n\leftrightarrow n+1}$, the $m$-th level acquires a phase (ignoring global phases) of $\phi_{m\leftrightarrow m+1} = \left((m-n)\omega_{n} + E_n - E_m\right)T$ which results in a phase difference of 
\begin{align}
\Delta\phi_{m \leftrightarrow m+1}= \left(\omega_{m} - \omega_{n} \right) T
\end{align}
This defines the necessary phase shift of all following pulses in the $m \leftrightarrow m+1$ frame.

In summary, a sequence of gate instructions consisting of Givens rotations $\mathcal{G}(\theta_{\mathcal{G}}, \phi_{\mathcal{G}}) $ and phase gates  $\mathcal{Z}(\varphi_\mathcal{Z})$ can be implemented in the qudit space through pulses $\mathcal{R}(\theta_{\mathcal{R}}, \phi_{\mathcal{R}})$ where the rotation angles remain unchanged ($\theta_{\mathcal{R}} = \theta_{\mathcal{G}}$) and the phases of the pulses $\phi_{\mathcal{R}} $ depend on the phases $\phi_{\mathcal{G}}$ and $\varphi_\mathcal{Z}$ of all previously implemented gates of the sequence. 
This procedure is summarized as a pseudocode algorithm in Alg.~\ref{alg_app1:summary_of_phase_shifts}.

\begin{algorithm}[H]
\caption{Implementation of a sequence of Givens rotations $\mathcal{G}$ and phase gates $\mathcal{Z}$ via hardware-native pulses $\mathcal{R}$ achieved by keeping track of all necessary phase shifts.}
\label{alg_app1:summary_of_phase_shifts}
\begin{algorithmic}
\State \co{levels} $\gets$ number of levels in qudit space
\State \co{phases} $\gets$ [$0, \dots, 0$]  \Comment{list of length \co{levels}$-1$}
\State \co{gates} $\gets$ sequence of $\mathcal{G}_{n\leftrightarrow n+1}(\theta, \phi)$ and $\mathcal{Z}_{n\leftrightarrow n+1}(\varphi)$ gates
\For{\co{gate} in \co{gates}}
    \If{\co{gate} is of type $\mathcal{G}_{n\leftrightarrow n+1}$}
        \State $\theta$ $\gets$ rotation angle of \co{gate}
        \State $\phi$ $\gets$ phase of \co{gate}
        \State $T$ $\gets$ duration of \co{gate}
        \For{$m$ in [0, \dots, $n-1$, $n+2$, \dots, \co{levels}]}
            \State \co{phases}[$m$] $\gets$ \co{phases}[$m$] -- $(\omega_{m} - \omega_{n})T$
        \EndFor
        \State play pulse $\mathcal{R}_{n\leftrightarrow n+1}(\theta,\, \co{phases}[n]+\phi)$
    \ElsIf{\co{gate} is of type $\mathcal{Z}_{n\leftrightarrow n+1}$}
        \State $\varphi$ $\gets$ rotation angle of \co{gate}
        \State \co{phases}[$n$] $\gets$ \co{phases}[$n$] -- $\varphi$
        \State \co{phases}[$n-1$] $\gets$ \co{phases}[$n-1$] + $\frac{\varphi}{2}$
        \State \co{phases}[$n+1$] $\gets$ \co{phases}[$n+1$] + $\frac{\varphi}{2}$
    \EndIf
\EndFor
\end{algorithmic}
\end{algorithm}

\clearpage
\section{Details on experiments in superconducting hardware}
\label{app:details_hardware}
\subsection{The transmon qubit}
\label{sec_sup:transmon_qubits}
A transmon qubit consists of a Josephson junction with Josephson energy $E_\text{J}$ shunted by a large capacitance whose single-electron charging energy is denoted as $E_\text{C}$, with $E_\text{C} \ll E_\text{J}$. The Hamiltonian of the circuit is
\begin{align}
\label{eq_sup:transmon_hamiltonian}
\hat H_{\text{TM}} = 4E_\text{C}\left( \hat{n} - n_g \right)^2 - E_\text{J} \cos(\hat{\phi}), 
\end{align}
where $\hat{n}$ and $\hat{\phi}$ are dimensionless conjugate variables describing the number of Cooper pairs on the capacitor and the superconducting phase across the Josephson junction, respectively~\cite{gambetta2013quantum}.
The \emph{offset charge} $n_g$ is a constant that results from capacitive coupling of undesired voltage sources due to imperfect isolation from the environment.

We denote the Hamiltonian in its eigenbasis by $\hat H_{\text{TM}} = \sum_{n} E_n \ket{n}\bra{n}$, where the qubit is encoded in the lowest-lying eigenstates $\ket{0}$ and $\ket{1}$. 
Through the expansion $\cos{\hat{\phi}} = 1 - \frac{\hat{\phi}^2}{2} + \frac{\hat{\phi}^4}{24} - \dots$, we see that, for small $\hat{\phi}$, the transmon resembles a harmonic oscillator. 
However, the higher powers of $\hat{\phi}$ create an anharmonic spectrum where the spacing of the eigenenergies is not equidistant, but decreases with higher levels. 
With the excitation energies $\omega_n = E_{n+1} - E_n$ between adjacent levels, we define the \emph{anharmonicity} $\alpha_n = \omega_n - \omega_{n-1}, \,\, n \geq 1$, as the difference in adjacent transition frequencies. 

\begin{figure}
\includegraphics[width=\columnwidth]{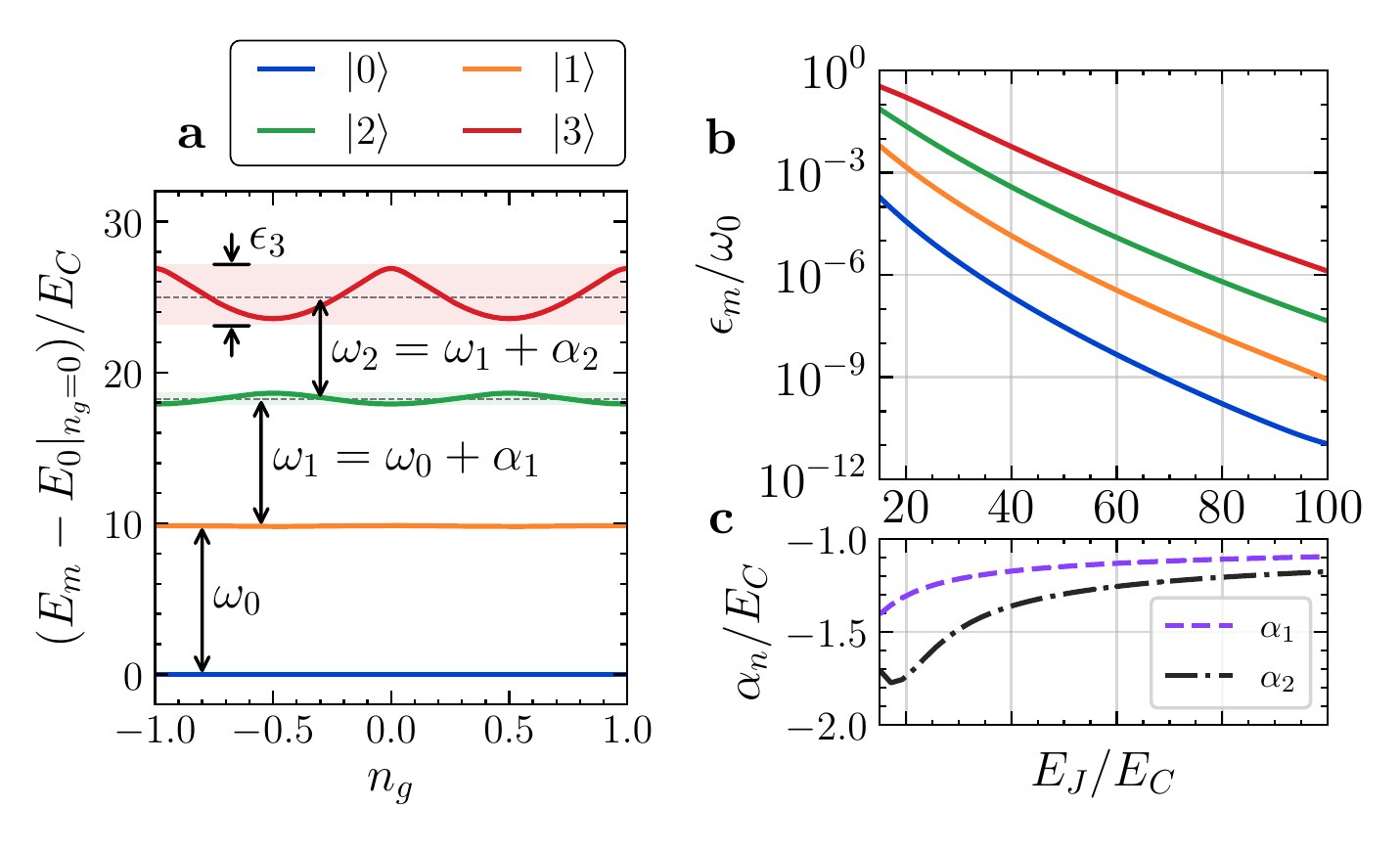}
\caption[]{Properties of the lowest energy eigenstates $\ket{m}$ of a transmon obtained from numerical diagonalization of the Hamiltonian in Eq.~\eqref{eq_sup:transmon_hamiltonian}. 
\textbf{a)} Fluctuations of the eigenenergies with the offset charge $n_g$ at $E_\text{J}/E_\text{C} = 15$, which become exponentially stronger for higher levels. The excitation energies $\omega_n = E_{n+1} - E_n$ and the anharmonicities $\alpha_n$ are defined with respect to the average value of $E_n$ over $n_g$. 
\textbf{b)} Reduction of the charge dispersion $\epsilon_n$ with increasing $E_\text{J}/E_\text{C}$.
{\textbf{c)} Anharmonicities} $\alpha_n$ as a function of $E_\text{J}/E_\text{C}$.
} 
\label{fig:transmon_properties}
\end{figure}

In a realistic experimental setting, $n_g$ is subject to fluctuations called \emph{charge noise}. 
This causes changes of the eigenenergies $E_n$ which are periodic in $n_g$~\cite{koch2007charge}, see Fig.~\ref{fig:transmon_properties}\textbf{a}. 
The maximal difference in eigenenergies of
\begin{align}
\label{eq_sup:charge_dispersion_definition}
\epsilon_n = \left|E_n(n_g=0) - E_n(n_g = 1/2) \right|,
\end{align}
is commonly called the \emph{charge dispersion}. 
Thus, under charge noise, the exact transition frequencies $\omega_n$ fluctuate, which creates phase errors~\cite{schreier2008suppressing}. 
Transmons mitigate this by increasing the ratio of $E_\text{J}/E_\text{C}$, which decreases charge dispersion, as shown in Fig.~\ref{fig:transmon_properties}\textbf{b}. 
However, the charge dispersion in the $\ket{2}$ and $\ket{3}$ state remain at least one and two orders of magnitude larger than in the $\ket{1}$ state, respectively.  
As $E_\text{J}/E_\text{C}$ increases, the  absolute value of the anharmonicity decreases (see Fig.~\ref{fig:transmon_properties}\textbf{c}) which complicates driving the individual transitions due to leakage into adjacent levels and phase errors. 
The transmon relies on the fact that the charge dispersion decreases exponentially with $E_\text{J}/E_\text{C}$ while the anharmonicity is only reduced with a weak power-law, making control at high $E_\text{J}/E_\text{C}$ favorable~\cite{koch2007charge}. 
Therefore, IBM Quantum devices currently employ transmon qubits with ($E_{\mathrm{J}}/E_{\mathrm{C}} \sim 35\,\text{--}\,45$), $\omega_0/(2\pi) \sim 5 \, \text{GHz}$, and $\alpha_1/(2\pi) \sim 300\,\text{MHz}$ \cite{IBMQuantum}.

\subsection{Decay of higher excited states}
\label{sec_app:decay_higher_states}
\begin{figure}
\includegraphics[width=0.95\columnwidth]{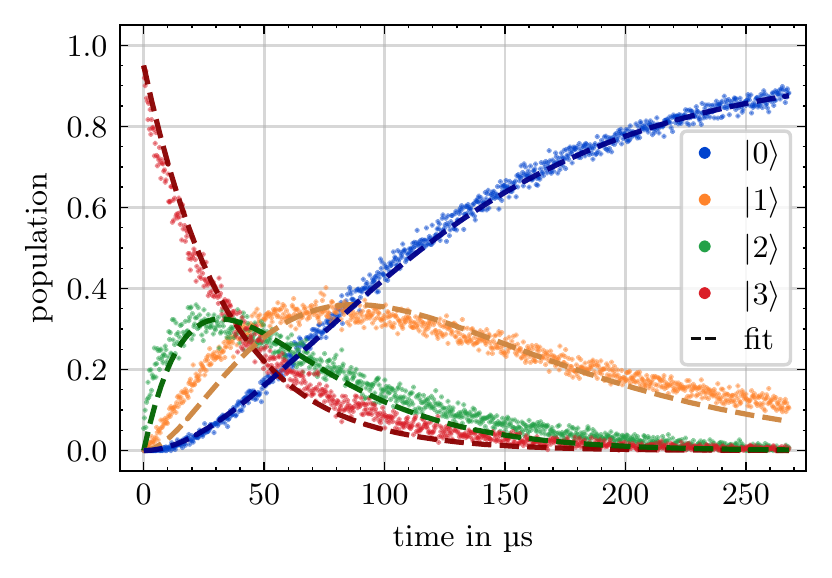}
\caption[]{Decay of state $\ket{3}$ in a transmon. 1000 measurements are taken at each time step which then undergo readout error mitigation to estimate the state populations.   
Fits are performed with a multi-channel exponential decay model.
Data taken on qubit $24$ of \textit{ibmq\_manhattan}.
}
\label{fig:app_T1_decay}
\end{figure}

Sufficient coherence of all involved states is required to perform qudit operations acting on higher-excited states. 
Here, we experimentally probe the $T_1$ times of the four lowest levels of a transmon in IBM Quantum hardware. 
We prepare the state $\ket{3}$ by a ladder sequence of $\pi$-pulses $\mathcal{X}_{0\leftrightarrow1}, \mathcal{X}_{1\leftrightarrow2}$, and $\mathcal{X}_{2\leftrightarrow 3}$. 
The system is left to decay for a time $t$ prior to a projective measurement which extracts the populations $\vec{p}(t)$ from 1000 measurements at each time step, see Fig.~\ref{fig:app_T1_decay}. 

To estimate the $T_1$ times, we fit a model based on a multi-channel rate equation  ${\text{d}\vec{p}(t)/\text{d}t = \Gamma^T \vec{p}(t)}$.
$\Gamma$ is a $4\times4$ matrix that contains the decay rates $\Gamma_{ij}$ associated with the decay from $\ket{i}$ to $\ket{j}$ and diagonal entries $\Gamma_{ii} = -\sum_{j=0}^{i-1} \Gamma_{ij}$. We consider only the possible ``downward'' transitions ${\ket{3}\!\shortrightarrow\!\ket{2}}$, ${\ket{3}\!\shortrightarrow\!\ket{1}}$, ${\ket{3}\!\shortrightarrow\! \ket{0}}$, ${\ket{2}\!\shortrightarrow\! \ket{1}}$, ${\ket{2}\!\shortrightarrow\! \ket{0}}$, and ${\ket{1}\!\shortrightarrow\! \ket{0}}$, so $\Gamma_{ij} = 0$ for $i<j$. 
The $T_1$ times arise from all possible decay channels, e.g., ${ T_1^{\ket{3}} = 1/(\Gamma_{32} + \Gamma_{31} + \Gamma_{30}})$. The obtained fit parameters are summarized in Tab.~\ref{tab_app:measured_decay_times}. We find that the non-sequential transitions are strongly suppressed and the qudit mainly decays sequentially, i.e., ${\ket{3}\!\shortrightarrow\!\ket{2}\!\shortrightarrow\!\ket{1}\!\shortrightarrow\!\ket{0}}$. Importantly, the $\sim\! 30\,\upmu \text{s}$ lifetimes of $\ket{2}$ and $\ket{3}$ leave plenty of coherence time to implement the POVM pulse schedule, which lasts $\sim 100\,\text{ns}$.

The fit in Fig.~\ref{fig:app_T1_decay}, accurately captures the population of the $\ket{0}$ state but deviates slightly for the other states. 
We attribute this to the significant misassignment errors present in the readout stage, which, even after readout error mitigation, remain significant, see Sec.~\ref{sec:experimental_demo}. 

\begin{table}
\caption{Experimentally measured decay constants $\Gamma_{ij}$ and $T_1$ times for a transmon qudit extracted from Fig.~\ref{fig:app_T1_decay}.}
\centering
\begin{ruledtabular}
\begin{tabular}{c|ccc|cc|c}
 & \multicolumn{3}{c|}{$\ket{3}$} & \multicolumn{2}{c|}{$\ket{2}$} & $\ket{1}$ \\
\hline
 $\Gamma_{ij}$ in & $\ket{3}\!\shortrightarrow\!\ket{2}$ &  $\ket{3}\!\shortrightarrow\!\ket{1}$ &  $\ket{3}\!\shortrightarrow\!\ket{0}$ &  $\ket{2}\!\shortrightarrow\!\ket{1}$ &  $\ket{2}\!\shortrightarrow\!\ket{0}$ &  $\ket{1}\!\shortrightarrow\!\ket{0}$  \\
 $(\upmu\text{s})^{-1}$ & 0.029 & 0.00 & 0.00 & 0.030 & 0.004 &  0.013 \\ 
\hline 
$T_1$ in $\upmu\text{s}$ & \multicolumn{3}{c|}{34.3} &  \multicolumn{2}{c|}{29.7} & \multicolumn{1}{c}{74.5} \\
\end{tabular}
\end{ruledtabular}
\label{tab_app:measured_decay_times}
\end{table}

\subsection{Measurement of charge dispersion in state $\ket{3}$}
\label{sec_app:meas_charge_disp}
\begin{figure}
\includegraphics[width=\columnwidth]{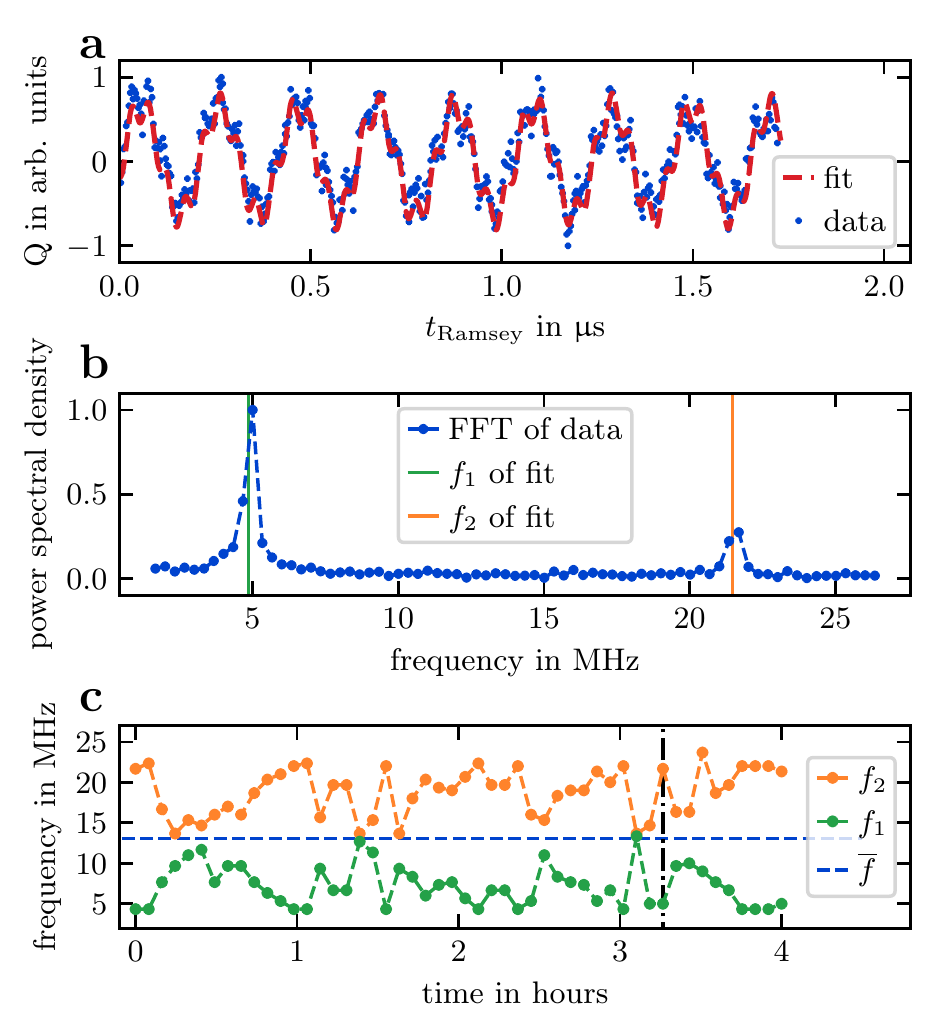}
\caption[]{Measurement of charge dispersion in the $\ket{3}$ state. \textbf{a)} Ramsey oscillations between states $\ket{2}$ and $\ket{3}$ with a fit of two sinusoidals forming a beating pattern. Each data point is averaged over 1000 shots.  
\textbf{b)} Fourier transform of the signal in \textbf{a} with two contributing frequencies. 
\textbf{c)} Symmetrical fluctuations of the fit frequencies around a mean frequency when repeating the procedure over time. The vertical black line indicates data from panel \textbf{b}. Data taken on qubit 0 of \textit{ibm\_lagos}.
} 
\label{fig:ramsey_measurement}
\end{figure}

Here, we present a direct measurement of the charge dispersion of the $\ket{3}$ state by performing a Ramsey interference experiment on the ${2\!\leftrightarrow\!3}$ transition. 
The experimental sequence consists of a preparation of the $\ket{2}$ state, followed by a $\pi/2$-pulse around the $x$-axis of the ${2\!\leftrightarrow\!3}$ transition, a delay time $t_{\text{Ramsey}}$, and finally a $-\pi/2$-pulse around the $x$-axis of the same transition. 
We measure the signal in the IQ-plane as $t_{\text{Ramsey}}$ is increased. 
This results in oscillations at the difference between the drive frequency and the true ${2\!\leftrightarrow\!3}$ transition frequency, see Fig.~\ref{fig:ramsey_measurement}\textbf{a}.
Transforming the signal into Fourier space reveals that the oscillation is a beating between two contributing frequencies $f_1$ and $f_2$, see Fig.~\ref{fig:ramsey_measurement}\textbf{b}. This can be attributed to quasiparticle tunneling across the qubit
junction~\cite{riste2013millisecond, peterer2015coherence}.
While repeating the Ramsey sequence 50 times, the two frequency components $f_1$ and $f_2$ fluctuate symmetrically around a center frequency $\overline{f}$ of $\sim 13\,\text{MHz}$, see Fig.~\ref{fig:ramsey_measurement}\textbf{c}. This represents the average detuning of the applied drive pulses. 
In total, this data suggests that the true frequency of the $2\!\leftrightarrow\!3$ transition fluctuates by as much as $15$-$20\,\text{MHz}$. 
For this particular qubit with a frequency of $\omega_0/(2\pi) = 5.2\,\text{GHz}$ and an anharmonicity of $\alpha_1/(2\pi)=-340\,\text{MHz}$, a direct diagonalization of the Hamiltonian from Eq.~\eqref{eq_sup:transmon_hamiltonian} predicts a charge dispersion of $\epsilon_3 = 13.9\,\text{MHz}$. 
Our measurements are thus in reasonable agreement with theory.

\subsection{Experimental POVM pulse sequence}
\label{sec:app_details_experiments_POVM}

In Sec.~\ref{sec:experimental_demo} of the main text, we present an experimental implementation of a single-qubit POVM measurement that consists of the operators given in Eq.~\eqref{eq:experimental_POVM_operators}.
Here, we motivate the choice of this POVM and provide further details on the corresponding pulse sequence. 

The average $2\!\leftrightarrow\!3$ transition frequency is difficult to calibrate due to the significant measurement misassignments between the involved states. 
This renders high-fidelity implementations of virtual $\mathcal{Z}_{2\leftrightarrow3}$-gates problematic, as the necessary phase updates to the drive frames depend on the transition frequency, see App.~\ref{app_sec:POVM_virtual_Z}.
We have thus chosen a POVM which does not require $\mathcal{Z}_{2\leftrightarrow3}$-gates. Instead, the qudit-space unitary $U$ that encodes our chosen POVM is built up from the  gate sequence
\begin{align}
\label{eq:U_decompositon_exp_POVM}
U &= \sqrt{\mathcal{X}}_{1\leftrightarrow2} \, \sqrt{\mathcal{X}}_{2\leftrightarrow3} \, \sqrt{\mathcal{X}}_{0\leftrightarrow1} \\
    & \hspace{20mm} \times 
\mathcal{Z}_{1\leftrightarrow2}(\pi/2)\, \sqrt{\mathcal{X}}_{1\leftrightarrow2}\,\sqrt{\mathcal{X}}_{0\leftrightarrow1}. \nonumber
\end{align}
The resulting POVM operators have a simple geometrical interpretation: three of the four operators points along the $x$-, $y$-, and $z$-axis of the Bloch sphere, Fig.~\ref{fig:experiment_results_POVM}\textbf{a}.

\begin{figure}
\includegraphics[width=0.95\columnwidth]{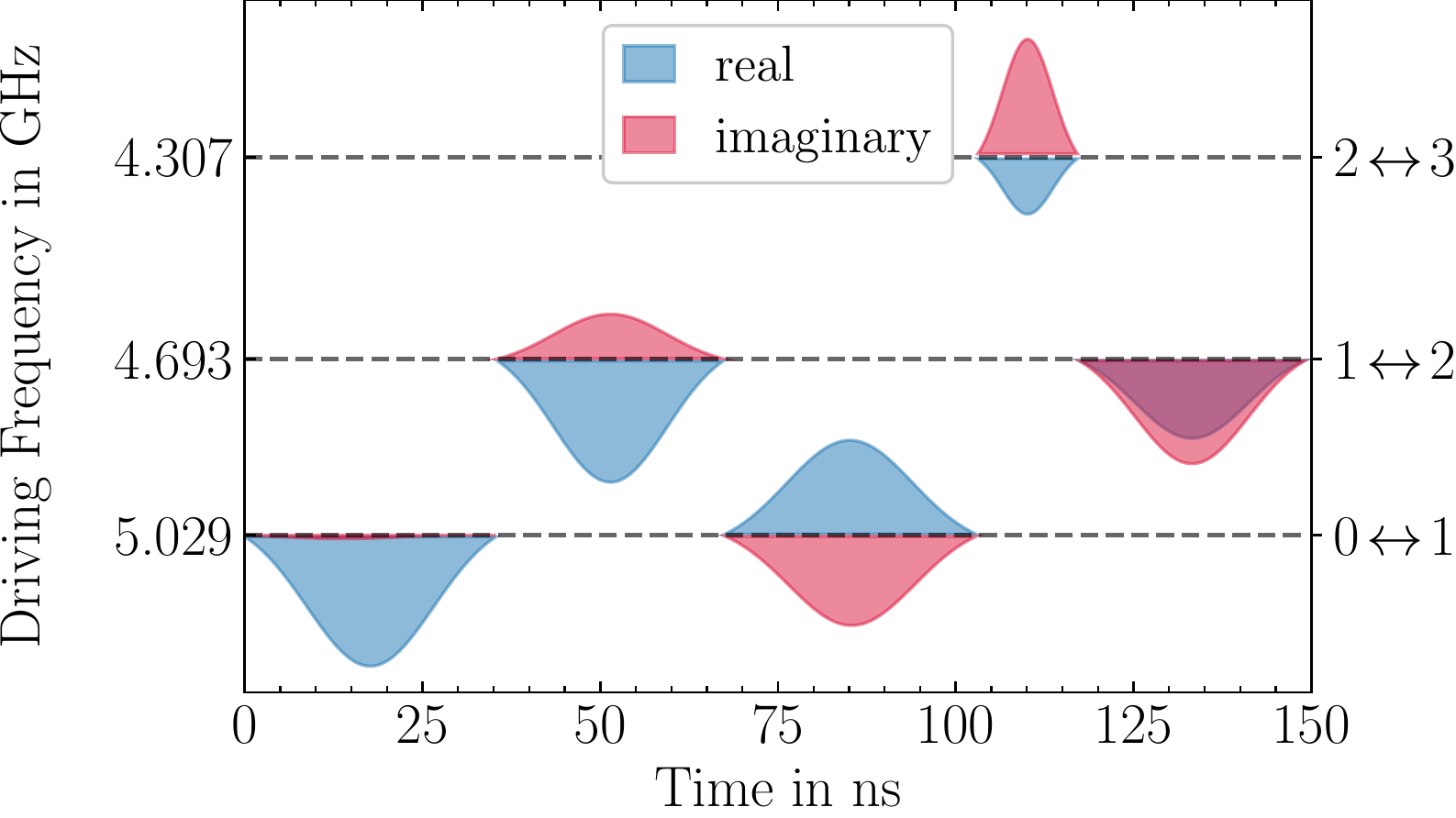}
\caption[]{Pulse sequence of five $\sqrt{\mathcal{X}}$-gates encoding the experimentally demonstrated POVM operators from Eq.~\eqref{eq:experimental_POVM_operators}.
The $0\!\leftrightarrow\!1$ drives are \textsc{Drag}-pulses, while the $1\!\leftrightarrow\!2$ and $2\!\leftrightarrow\!3$ transitions are driven with Gaussian pulses. 
Amplitudes of the real and imaginary parts of the pulse envelopes are depicted in arbitrary units. 
}
\label{fig:pulse_sequence_exp_povm}
\end{figure}

The pulse sequence that implements the unitary from Eq.~\eqref{eq:U_decompositon_exp_POVM} is shown in Fig.~\ref{fig:pulse_sequence_exp_povm}. 
The non-trivial phases of the pulses, manifested in non-zero imaginary parts, arise from both the $\mathcal{Z}_{1\leftrightarrow2}$-gate in the sequence as well as from phases acquired during frame changes between different transitions. 
With the lack of $\mathcal{Z}_{2\leftrightarrow3}$-gates in the sequence, this POVM does not represent the most general case from Eq.~\eqref{eq_theo:U_final_decomposition_sequence}. 
Besides this simplification, it exhibits all features of our proposed scheme, thus constituting a reasonable compromise between practical feasibility on hardware that is not tailored for qudit operation and generality of the proof of principle.

\section{Operational distance}
\label{app:operational_distance}
To compare the fidelity between two POVMs, such as the experimentally implemented POVM and the theoretical target POVM, a suitable distance measure is needed. 
In this work we use the \emph{operational distance} (OD)~\cite{maciejewski2020mitigation, puchala2018strategies}. 
For two $M$-outcome POVMs $\boldsymbol{\Pi}=\{\Pi^m \}$ and $\boldsymbol{\Sigma}=\{\Sigma^m\}$ the OD is defined as
\begin{align}
\label{eq_theo:Operational_distance_definition}
D_{\text{OD}}(\boldsymbol{\Pi}, \boldsymbol{\Sigma}) = \max_{\rho}
\frac{1}{2} \sum_{m=0}^{M-1} \left| \Tr(\rho \Pi^m) - \Tr(\rho \Sigma^m) \right|.
\end{align}
The OD is thus the worst-case \emph{total variation} between the probability distribution of measurement outcomes obtained with the two POVMs. Importantly, ${0\leq D_{\text{OD}}(\boldsymbol{\Pi}, \boldsymbol{\Gamma}) \leq 1}$ where  $D_{\text{OD}}(\boldsymbol{\Pi}, \boldsymbol{\Gamma})=0$ if and only if the two POVMs coincide. The OD can be calculated directly from the POVM operators through
\begin{align}
\label{eq_theo:Operational_distance_calculation}
D_{\text{OD}}(\boldsymbol{\Pi}, \boldsymbol{\Sigma}) = \max_{I'  \subset I } 
\left\|\sum_{m \in I'} \Pi^m - \Sigma^m \right\|_{\infty}, 
\end{align}
where $I$ is the set of all outcomes $I={\{0, \dots, M-1\}}$. 

\section{Quantum detector tomography}
\label{app:detector_tomography}
In an experiment that performs a quantum measurement, the implemented POVM operators can be characterized through \emph{quantum detector tomography} (QDT) \cite{fiurasek2001maximumlikelihood, dariano2004quantum}.
In combination with the better known quantum state tomography (QST) and quantum process tomography (QPT), QDT is required for a full specification of a quantum experiment \cite{lundeen2009tomography}. 
In QST an unknown state $\rho$ is estimated from measurements in a known set of reference POVM operators $\{\Pi^m_{\text{ref}}\}$. By contrast, in QDT the unknown POVM operators $\{\Pi^m \}$ are estimated from a known set of prepared reference quantum states $\{ \rho_{\text{ref}}^i \}$. 
As in QST, there is the concept of \emph{informational completeness}: For a full characterization of $\{\Pi^m\}$ through QDT, the  reference states $\{\rho_{\text{ref}}^i\}$ need to span the operator space of $\Pi^m$ \cite{lundeen2009tomography}. 
One possible set of such states for single-qubit POVMs are projectors on the six single-qubit stabilizer states $\{ \ket{0}, \ket{1}, \ket{+}, \ket{-}, \ket{+i}, \ket{-i} \}$ which are the eigenstates of $\sigma_z, \sigma_x$, and $\sigma_y$, respectively. 
This is a convenient choice since the initialization in $\ket{0}$ and subsequent single-qubit rotations to either of these states can be implemented with high fidelity on existing quantum processors. 
The POVM measurement is carried out on each such reference state, sampling from the probability distributions $p^{(j)}_m = \bra{\psi_j}\Pi^m \ket{\psi_j}$. 
Let $\mathcal{N}^{(j)}_m$ be the number of times outcome $m\in\{0, 1, 2, 3\}$ is recorded for initial state $\ket{\psi_j}$. 
One way to obtain an estimator for the underlying single-qubit POVM operators is to invert the system of linear equations
\begin{equation}
\label{eq_theo:POVM_linear_inversion}
\bra{\psi_j} \Pi^m \ket{\psi_j} \sim \frac{\mathcal{N}^{(j)}_m}{\sum\limits_{m'} \mathcal{N}^{(j)}_{m'}}
\end{equation}
to obtain the entries of $\Pi^m$. This approach suffers from the fact that the obtained POVM operators might be non-physical, as they are not necessarily positive. 
An analogous issue exists for QST through linear inversion of Eq.~\eqref{eq_theo:POVM_linear_inversion} \cite{neugebauer2020neuralnetwork}. 
Positivity can be enforced with a maximum-likelihood (ML) estimation by maximizing the likelihood functional 
\begin{equation}
\label{eq_theo:detector_tomog_likelihood}
\mathcal{L}\left(\Pi^0, \Pi^1, \Pi^2, \Pi^3\right) = \prod_{m, j} \left(\bra{\psi_j}\Pi^m \ket{\psi_j}\right)^{\mathcal{N}^{(j)}_m}
\end{equation}
under the constraint that the operators $\Pi^m$ form a valid POVM \cite{fiurasek2001maximumlikelihood}. 
As laid out in Ref.~\cite{hradil2004maximumlikelihood}, the optimization can be performed with an iterative algorithm that converges to the ML estimator. This procedure has recently been demonstrated experimentally on IBM hardware as a means to mitigate readout noise \cite{maciejewski2020mitigation, chen2019detector}. 
In this work, we make use of ML quantum detector tomography to reconstruct the implemented POVM operators both for the verification of the experimental proof-of-principle in Sec.~\ref{sec:experimental_demo} and for the simulations of our error mitigation scheme in Sec.~\ref{sec:bias_mitigation_tomography}. 

\section{Details on pulse-level simulations}
\label{app:details_simulation}
\begin{figure*}[t]
\centering
\includegraphics[width=0.99\linewidth]{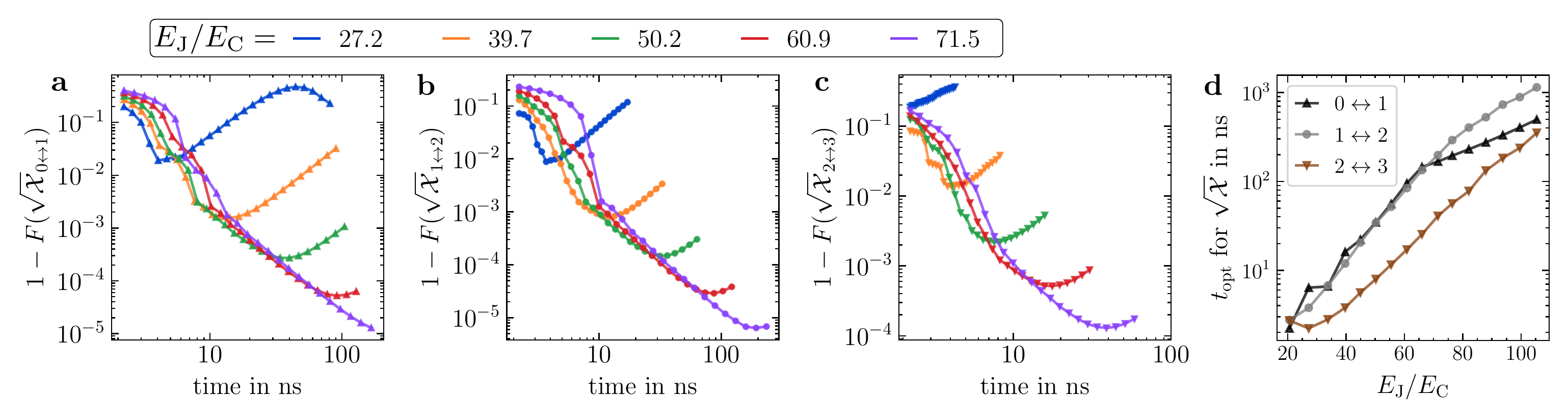}
\caption[]{Calibration of $\sqrt{\mathcal{X}}$-pulses through numerical simulations of the pulse-level dynamics. \textbf{a) -- c)} Average gate infidelities $1-F(\sqrt{\mathcal{X}})$ in the qudit space relevant for the POVM encoding as a function of the gate duration for selected values of $E_\text{J} / E_\text{C}$. \textbf{d)} Optimal times $t_{\text{opt}}$ that maximize $F(\sqrt{\mathcal{X}})$ for different $E_\text{J} / E_\text{C}$.}    \label{fig:app_simulation_SX_calibrations}
\end{figure*}

In this appendix, we summarize the technical details of the numerical simulations from Sec.~\ref{sec:optimal_transmon_regime} and Sec.~\ref{chap:applications}. 
Transmons are modeled by the Hamiltonian in Eq.~\eqref{eq_sup:transmon_hamiltonian}. 
This Hamiltonian is diagonalized in the charge representation after truncating to 20 Fourier modes in the superconducting phase $\hat{\phi}$ to obtain the low-energy spectrum~\cite{gambetta2013quantum}. 
The parameters $E_{\mathrm{J}}$ and $E_{\mathrm{C}}$ are then adjusted to fix the base frequency of the qubit (at $n_g=0$) at $5\,\text{GHz}$. 
The dynamics of the system under a drive as defined in  Eq.~\eqref{eq_theo:pulse_drive_definition} are modeled by the interaction Hamiltonian from Eq.~\eqref{eqn:theory_Hint_rot_frame}, where we assume the relative coupling of a harmonic oscillator, i.e., $g_n \propto \sqrt{n+1}$, and truncate the system at $d=5$ levels.

To implement a desired POVM, we decompose the corresponding target unitary into a sequence of $\sqrt{\mathcal{X}}$-pulses with virtual $\mathcal{Z}$-gates.
For simplicity, we employ Gaussian pulse envelopes with a standard deviation of one quarter of the pulse duration. 
The pulses feature a piece-wise constant envelope with a sample duration of $222\,\text{ps}$, matching IBM control hardware~\cite{IBMQuantum}. 
The transition frequencies depend on $n_g$, whose exact value fluctuates from one experimental run to another.
We model $n_g$ to be uniformly distributed as $p(n_g) \sim \text{Uni}([0, 1])$, which is sufficient due to the periodicity of the eigenenergies with $n_g$, see Fig.~\ref{fig:transmon_properties}.
Each pulse is played at the average transition frequency $\overline{\omega_n} = \int_0^1 p(n_g) \omega_n(n_g) \text{d}n_g$. 
The pulse amplitudes are chosen such that the resulting rotation angles for the average transition frequency are $\pi/2$. 

We model the quantum dynamics of a state $\rho$ under a pulse sequence by an effective channel 
\begin{align}
\label{eq_app:channel_charge_noise}
\mathcal{E}: \rho \longmapsto \int_0^1 p(n_g)  U(n_g) \rho U(n_g)^\dagger \text{d}n_g  
\end{align}
where $U(n_g)$ are the unitary dynamics for a fixed offset charge $n_g$. 
We obtain  $U(n_g)$ under a sequence of pulses with an integrator of the time-dependent Schr{\"o}dinger equation provided by QuTip~\cite{johansson2012qutip}.
The channel $\mathcal{E}$ is numerically approximated by computing $U(n_g)$ for 20 values of $n_g$ equally spaced between 0 and 1.

To calibrate $\sqrt{\mathcal{X}}$-pulses in the simulation, we keep the amplitude fixed while varying the duration of the pulses.
The target unitary of such a pulse is given by the implemented rotation $U^{\text{tar}} = \mathcal{R}(\varphi=\pi/2, \gamma=0)$, defined in Eq.~\eqref{eq:givens_rotation_hardware}, which includes phases that are accumulated in the idle levels. 
As a figure of merit, we compute the average gate fidelity $F(\mathcal{E}, U^{\text{tar}})$ between the target unitary and the channel of the simulated unitary under charge noise~\cite{nielsen2002simple}.
We hereby restrict the computation of $F(\mathcal{E}, U^{\text{tar}})$ to the subspace that is relevant for the POVM pulse sequence. 
Recall that the POVM-encoding unitary is always realized with pulses that couple adjacent levels in the order $0\!\leftrightarrow \!1$, $1\!\leftrightarrow\!2$, $0\!\leftrightarrow\!1$, $2\!\leftrightarrow\!3$, and finally $1\!\leftrightarrow\!2$.
Since the $\ket{3}$ state is only populated once prior to measurement, the phases acquired by $\ket{3}$ during a $0\!\leftrightarrow\!1$ and $1\!\leftrightarrow\!2$ gate do not affect the encoded POVM operators. 
The fidelities of $\sqrt{\mathcal{X}}_{0\leftrightarrow1}$ and $\sqrt{\mathcal{X}}_{1\leftrightarrow2}$ are thus only computed over the subspaces spanned by $\ket{0}, \ket{1}$, and $\ket{2}$. 
Similarly, only the $\ket{1}, \ket{2}, \ket{3}$ subspace is considered for the fidelities of $\sqrt{\mathcal{X}}_{2\leftrightarrow3}$ since no $0\!\leftrightarrow\!1$ pulses are applied after the $2\!\leftrightarrow\!3$ pulse.

The average gate fidelities for different hardware parameters as a function of the pulse duration are shown in Fig~\ref{fig:app_simulation_SX_calibrations}.
For short durations, the broad spectral range of the pulse leads to leakage errors. 
In contrast, for long pulse durations, the phases accumulated over time by the idle levels become difficult to track due to charge noise.
The infidelities $1-F(\sqrt{\mathcal{X}})$ thus typically show a distinct minimum where these two effects are traded off optimally.   
As $E_\text{J}/E_\text{C}$ increases, this optimum shifts towards longer gate durations, see Fig.~\ref{fig:app_simulation_SX_calibrations}\textbf{d}.
For reference, the default single-qubit $SX$-gate in current IBM Quantum hardware is carried out with \textsc{Drag} pulses of a duration of $36\,\text{ns}$. 
We find that it is important to employ much shorter pulses when including the phase uncertainty of a neighboring state, despite our use of simple Gaussian pulse envelopes, which are not specifically designed to correct for leakage errors (especially for the $\sqrt{\mathcal{X}}_{2\leftrightarrow3}$-gate). 
This suggests that phase uncertainties in higher excited states have an overall bigger impact on the qudit gate fidelities than leakage. 
The remaining leakage errors could be further reduced by a careful calibration of \textsc{Drag} pulses.
For current hardware ($E_{\mathrm{J}}/E_{\mathrm{C}} \sim 35\,\text{--}\,45$), our simulations suggest achievable gate fidelities in the relevant qudit spaces that reach up to $99.9\%$ for the $\sqrt{\mathcal{X}}_{0\leftrightarrow1}$- and $\sqrt{\mathcal{X}}_{1\leftrightarrow2}$-gates, and $98\%$ for $\sqrt{\mathcal{X}}_{2\leftrightarrow3}$. 
This can be improved by over an order of magnitude by tuning deeper into the transmon regime (e.g., $E_{\mathrm{J}}/E_{\mathrm{C}} \sim60$), at the expense of increased gate durations.  

For our simulation of the full POVM pulse sequences in Secs.~\ref{sec:optimal_transmon_regime} and \ref{chap:applications}, we employ the durations of the $\sqrt{\mathcal{X}}_{n\leftrightarrow n+1}$-pulses that maximize their respective fidelities. 
When limiting the total duration of the sequence as in Fig.~\ref{fig:transmon_parameter_regime}\textbf{a}, we incrementally shorten those pulses whose fidelity is affected the least. This is repeated until a pulse sequence is obtained which is at most as long as the desired total length.
From the implemented channel $\mathcal{E}$ of the pulse sequence, we finally obtain an effective POVM $\Pi_{\text{sim}}$ as the average over the POVM operators encoded by the unitaries $U(n_g)$.

\end{document}